\documentclass[rmp,twocolumn,groupedaddress,showpacs]{revtex4}

\def\Ref#1{\citeauthor{#1}, \citeyear{#1}}
\def\cit#1{\citeauthor{#1}, \citeyear{#1}}

\begin{document}
\bibliographystyle{apsrmp}


\title{
Doping a Mott Insulator: Physics of High Temperature Superconductivity
}


\author{Patrick A. Lee$^a$}
\author{Naoto Nagaosa$^b$}
\author{Xiao-Gang Wen$^a$}
\affiliation{$^a$ Department of Physics, Massachusetts Institute of Technology,
Cambridge, Massachusetts 02139\\
$^b$ CREST, Department of Applied Physics,
University of Tokyo,
7--3--1 Hongo, Bunkyo-ku, Tokyo 113, Japan}


\date{\today}

\begin{abstract}
This article reviews the effort to understand the physics of high temperature
superconductors from the point of view of doping a Mott insulator.  The basic
electronic structure of the cuprates is reviewed, emphasizing the physics of
strong correlation and establishing the model of a doped Mott insulator as a
starting point.  A variety of experiments are discussed, focusing on the
region of the phase diagram close to the Mott insulator (the underdoped
region) where the behavior is most anomalous.  The normal state in this region
exhibits the pseudogap phenomenon.  In contrast, the quasiparticles in the
superconducting state are well defined and behave according to theory.  We
introduce Anderson's idea of the resonating valence bond (RVB) and argue that
it gives a qualitative account of the data.  The importance of phase
fluctuation is discussed, leading to a theory of the transition temperature
which is driven by phase fluctuation and thermal excitation of quasiparticles.
However, we argue that phase fluctuation can only explain the pseudogap
phenomenology over a limited temperature range, and some additional physics is
needed to explain the onset of singlet formation at very high temperatures.
We then describe the numerical method of projected wavefunction which turns
out to be a very useful technique to implement the strong correlation
constraint, and leads to a number of predictions which are in agreement with
experiments.  The remainder of the paper deals with an analytic treatment of
the $t$-$J$ model, with the goal of putting the RVB idea on a more formal
footing.  The slave-boson is introduced to enforce the constraint of no double
occupation.  The implementation of the local constraint leads naturally to
gauge theories.  We follow the historical order and first review the $U(1)$
formulation of the gauge theory.  Some inadequacies of this formulation for
underdoping are discussed, leading to the $SU(2)$ formulation.  Here we
digress with a rather thorough discussion of the role of gauge theory in
describing the spin liquid phase of the undoped Mott insulator.  We emphasize
the difference between the high energy gauge group in the formulation of the
problem versus the low energy gauge group which is an emergent phenomenon.
Several possible routes to deconfinement based on different emergent gauge
groups are discussed, which lead to the physics of fractionalization and 
spin-charge separation.  We next describe the extension of the $SU(2)$ formulation to
nonzero doping.  We focus on a part of the mean field phase diagram called the
staggered flux liquid phase.  We show that inclusion of gauge fluctuation
provides a reasonable description of the pseudogap phase.  We emphasize that
$d$-wave superconductivity can be considered as evolving from a stable $U(1)$
spin liquid.  We apply these ideas to the high $T_c$ cuprates, and discuss
their implications for the vortex structure and the phase diagram.  A possible
test of the topological structure of the pseudogap phase is discussed.
\end{abstract}
\pacs{74.20.Mn, 71.27.+a}
\keywords{High $T_c$ superconductivity, spin liquid, slave-boson theory}

\maketitle


\tableofcontents

\section{Introduction}
The discovery of high temperature superconductivity in cuprates \cite{BM8689}
and the rapid raising of the transition temperature to well
above the melting point of nitrogen \cite{WAT8708}
ushered in an era
of great excitement for the condensed matter physics community.  For decades
prior to this discovery, the highest $T_c$ had been stuck at 23~K.  Not only
was the old record $T_c$ shattered, but the fact that high $T_c$
superconductivity was discovered in a rather unexpected material, a transition
metal oxide, made it clear that some novel mechanism must be at work.  The
intervening years have seen great strides in high  $T_c$ research.  First and
foremost, the growth and characterization of cuprate single crystals and thin
films have advanced to the point where sample quality and reproducibility
problems which plagued the field in the early days are no longer issues.  At
the same time, basically all conceivable experimental tools have been applied
to the cuprates.  Indeed, the need for more and more refined data has spurred
the development of experimental techniques such as angle resolved
photoemission spectroscopy (ARPES) and low temperature scanning tunneling
microscopy (STM).  Today the cuprate is arguably the best studied material
outside of the semiconductor family and a great deal of facts are known.  It
is also clear that many of the physical properties are unusual, particularly
in the metallic state above the superconductor.  Superconductivity is only one
aspect of a rich phase diagram which must be understood in its totality.
\begin{figure}[t]
\centerline{
\includegraphics[width=3.4in]{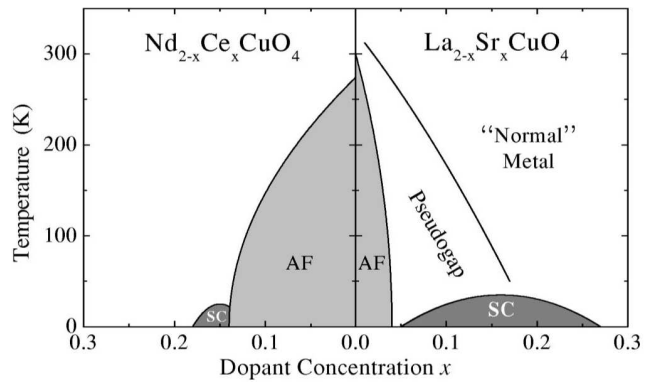}
}
\caption{Schematic phase diagram of high $T_c$ superconductors showing hole doping (right side) and
electron doping (left side). From 
\Ref{DHS0373}. }
\label{Fig.1}
\end{figure}

While there are hundreds of high $T_c$ compounds, they all share a layered
structure made up of one or more copper-oxygen planes.  They all fit into a
``universal'' phase diagram shown in Fig. 1.  We start with the so-called
``parent compound,'' in this case La$_2$CuO$_4$.  There is now general
agreement that the parent compound is an insulator, and should be classified
as a Mott insulator.  The concept of Mott insulation was introduced many years
ago \cite{M4916}
to describe a situation where a material should be metallic according to band
theory, but is insulating due to strong electron-electron repulsion.  In our
case, in the copper-oxygen layer there is an odd number of electrons per unit
cell.  More specifically, the copper ion is doubly ionized and is in a $d^9$
configuration, so that there is a single hole in the $d$ shell per unit cell.
According to band theory, the band is half-filled and must be metallic.
Nevertheless, there is a strong repulsive energy cost to put two electrons (or
holes) on the same ion, and when this energy (commonly called $U$) dominates
over the hopping energy $t$, the ground state is an insulator due to strong
correlation effects.  It also follows that the Mott insulator should be an
antiferromagnet (AF), because when neighboring spins are oppositely aligned,
one can gain an energy $4t^2/U$ by virtual hopping.  This is called the
exchange energy $J$.  The parent compound is indeed an antiferromagnetic
insulator.  The ordering temperature $T_N \approx 300$K shown in
Fig.~\ref{Fig.1} is in fact misleadingly low because it is governed by a small
interlayer coupling, which is furthermore frustrated in La$_2$CuO$_4$ (see
\Ref{KBS9897}).
The exchange energy $J$ is in fact extraordinarily high, of order 1500~K, and
the parent compound shows strong antiferromagnetic correlation much above
$T_N$.

The parent compound can be doped by substituting some of the trivalent La by
divalent Sr.  The result is that $x$ holes are added to the Cu-O plane in
La$_{2-x}$Sr$_x$CuO$_4$.  This is called hole doping.  In the compound
Nd$_{2-x}$Ce$_x$CuO$_4$, the reverse happens in that $x$ electrons are added
to the Cu-O plane.  This is called electron doping.  As we can see from Fig.
1, on the hole doping side the AF order is rapidly suppressed and is gone by 3
to 5\% hole concentration.  Almost immediately after the suppression of AF,
superconductivity appears, ranging from $x = 6$\% to 25\%.  The dome-shaped
$T_c$ is characteristic of all hole-doped cuprates, even though the maximum
$T_c$ varies from about 40~K in the La$_{2-x}$Sr$_x$CuO$_4$ (LSCO) family to
93~K and higher in other families such as YBa$_2$Cu$_3$O$_{6+y}$ (YBCO) and
Ba$_2$Sr$_2$CaCu$_2$O$_{8+y}$ (Bi-2212).  On the electron doped side, AF is
more robust and survives up to $x = 0.14$, beyond which a region of
superconductivity arises.  One view is that the carriers are more prone to be
localized on the electron doped side, so that electron doping to closer to
dilution by nonmagnetic ions, which is less effective in suppressing AF order
than itinerant carriers.  Another possibility is that
the next neighbor hopping term favors AF on the electron doped side
\cite{SG0214}.
It is as though a more robust AF region is covering
up the more interesting phase diagram revealed on the hole doped side.  In
this review we shall focus on the hole doped materials, even though we will
address the issue of the particle-hole asymmetry of the phase diagram from
time to time.

The region in the phase diagram with doping $x$ less than that of the maximum
$T_c$ is called the underdoped region.  The metallic state above $T_c$ has
been under intense study and exhibits many unusual properties not encountered
before in any other metal.  This region of the phase diagram has been called
the pseudogap phase.  It is not a well defined phase in that a definite finite
temperature phase boundary has never been found.  The line drawn in Fig.~1
should be regarded as a cross-over.  Since we view the high $T_c$ problem as
synonymous with that of doping a Mott insulator, the underdoped region is
where the battleground between Mott insulator and superconductivity is drawn
and this is where we shall concentrate on in this review.

The region of the normal state above the optimal $T_c$ also exhibits unusual
properties.  The resistivity is linear in $T$ and the Hall coefficient is
temperature dependent (see \Ref{CWO9188}).
These were cited as examples of non-Fermi liquid behavior since the early days
of high $T_c$.  Beyond optimal doping (called the overdoped region), sanity
gradually returns.  The normal state behaves more normally in that the
temperature dependence of the resistivity resembles $T^2$ over a temperature
range which increases with further overdoping.  The anomalous region above
optimal doping is sometimes referred to as the ``strange metal'' region.  We
offer a qualitative description of this region in section IX, but in our mind
the understanding of the ``strange metal'' is even more rudimentary that
of the ``pseudogap.''  A popular notion is that the strange metal is
characterized by a quantum critical point lying under the superconducting dome
(\cit{V9754}; \cit{CCG9737}; \cit{TL0053})
In our view, unless the nature of the ordered side of a quantum critical point
is classified, the simple statement of quantum criticality does not teach us
too much about the behavior in the critical region.  For this reason, we
prefer to concentrate on the underdoped region and leave the strange metal
phase to future studies.

Contrary to the experimental situation, the development of high $T_c$ theory
follows a rather tortuous path and people often have the impression that the
field is highly contentious and without a clear direction  or consensus.  We
do not agree with this assessment and would like to clearly state our point of
view from the outset.  Our starting point is, as already stated, that the
physics of high $T_c$ superconductivity is the physics of doping a Mott
insulator.  Strong correlation is the driving force behind the phase diagram.
We believe that there is a general consensus on this starting point.  The
simplest model which captures the strong correlation physics is the Hubbard
model and its strong coupling limit, the $t$-$J$ model.  Our view is that one
should focus on understanding these simple models before adding various
elaborations.  For example, further neighbor hopping certainly is significant
and as we shall discuss, plays an important role in understanding the
particle-hole asymmetry of the phase diagram.  Electron-phonon coupling can
generally be expected to be strong in transition metal oxides, and we shall
discuss their role in affecting spectral line shape.  However, these
discussions must be made in the context of strong correlation.  The logical
step is to first understand whether simple models such as the $t$-$J$ model
contains enough physics to explain the appearance of superconductivity and
pseudogaps in the phase diagram.

The strong correlation viewpoint was put forward by 
\Ref{A8796},
who revived his earlier work on a possible spin liquid state in a frustrated
antiferromagnet.  This state, called the resonating valence band (RVB), has no
long range AF order and is a unique spin singlet ground state.  It has spin
1/2 fermionic excitations which are called spinons.  The idea is that when
doped with holes the RVB is a singlet state with coherent mobile carriers, and
is indistinguishable in terms of symmetry from a singlet BCS superconductor.
The process of hole doping was further developed by \Ref{KRS8765}
who 
argue that the combination of the
doped hole with the spinon form a bosonic excitation. This excitation,
called the holon, carries charge but no spin whereas the
spinon carries spin 1/2 but no charge, and the notion of spin-charge
separation was born.  Meanwhile, a slave-boson theory was formulated by
\Ref{BZA8773}.
Many authors contributed to the development of the mean field theory,
culminating in the paper by 
\Ref{KL8842}
who found that the
superconducting state should have $d$-symmetry and that a state with spin gap
properties should exist above the superconducting temperature in the
underdoped region.  The 
possibility of d-wave superconductivity has been discussed in terms of the
exchange of spin fluctuations
(\cit{SLH8690}, \citeyear{SLH8794}; \cit{MSV8654}; \cit{E8377},
\citeyear{E8621}; \cit{MP9369}).
These discussions are either based on phenomenological coupling between spins
and fermions or via RPA treatment for the Harbbard model which is basically a
weak coupling expansion. In contrast, the salve-boson
theory was developed in the limit of strong repulsion. Details of the
mean field theory will be discussed
in section VIII.

At about the same time, the proposal by 
\Ref{A8796}
of using projected mean
field states as trial wavefunctions was implemented on the computer by 
\Ref{G8831}, \citeyear{G8853}.  
The idea is to remove by hand on a computer all components of the
mean field wavefunction with doubly occupied sites, and use this as a
variational wavefunction for the $t$-$J$ model.  
\Ref{G8831}, \citeyear{G8853} 
concluded that
the projected $d$-wave superconductor is the variational ground state for the
$t$-$J$ model over a range of doping.  The projected wavefunction method
remains one of the best numerical tools to tackle the $t$-$J$ or Hubbard model
and is reviewed in section VI.

It was soon realized that inclusion of fluctuations about the mean field
invariably leads to gauge theory 
\cite{BA8880,IL8988,NL9050}.
The gauge field fluctuations can be
treated at a Gaussian level and these early developments together with some of
the difficulties are reviewed in section IX.

In hindsight, the slave-boson mean field theory and the projected wavefunction
studies contain many of the qualitative aspects of the hole doped phase
diagram.  It is indeed quite remarkable that the main tools of treating the
$t$-$J$ model, \ie projected trial wavefunction, slave-boson mean field, and
gauge theory, were in place a couple of years after the discovery of high
$T_c$.  In some way the theory was ahead of its time, because the majority
view in the early days was that the pairing symmetry was $s$-wave, and the
pseudogap phenomenology remains to be discovered.  (The first hint came from
Knight shift measurements in 1989 shown in Fig. \ref{chi}(a).)  Some of the
early history and recent extensions are reviewed by 
\Ref{ARR0455}.

The gauge theory approach is a difficult one to pursue systematically because
it is a strong coupling problem.  One important development is the realization
that the original $U(1)$ gauge theory should be extended to $SU(2)$ in order
to make a smooth connection to the underdoped limit 
\cite{WLsu2}
This
is discussed in sections XI and XII.  More generally, it was gradually
realized that the concepts of confinement/deconfinement which are central to
QCD also play a key role here, except that the presence of matter field make
this problem even more complex.  Since gauge theories are not so familiar to
condensed matter physicists, these concepts are discussed in some detail in
section X.  One of the notable recent advances is that the notion of the spin
liquid and its relation to deconfinement in gauge theory has been greatly
clarified and several soluble models and candidates based on numerical exact
diagonalization have been proposed.  It remains true, however, that so far no
two-dimensional spin liquid has been convincingly realized experimentally.

Our overall philosophy is that the RVB idea of a spin liquid and its relation
to superconductivity contains the essence of the physics and gives a
qualitative description of the underdoped phase diagram.  The goal of our
research is to put these ideas on a more quantitative footing.  Given the
strong coupling nature of the problem, the only way progress can be made is
for theory to work in consort with experiment.  Our aim is to make as many
predictions as possible, beyond saying that the pseudogap is a RVB spin
liquid, and challenge the experimentalists to perform tests.  Ideas along
these lines are reviewed in section XII.

High $T_c$ research is an enormous field and we cannot hope to be complete in
our references.  Here we refer to a number of excellent review articles on
various aspects of the subject.  
\Ref{IFT9839} reviewed the general topic of metal-insulator transition.
\Ref{OM0068} and \Ref{NP0347}
have provided highly readable accounts of experiments and general
theoretical approaches.  
Early numerical work was reviewed by \Ref{D9463}.
\Ref{KBS9897} summarized the earlier
optical and magnetic neutron scattering data mainly on
La$_{2-x}$Sr$_x$CuO$_4$.  Major reviews of angle resolved photoemission data
(ARPES) have been provided by 
\Ref{CNR03} and \Ref{DHS0373}.  Optics measurements on underdoped materials are reviewed by
\Ref{TS9961}.  The volumes edited by 
\Ref{Gin89} contain
excellent reviews of early NMR work by C.P. Slichter and early transport
measurement by N.P. Ong among others.  Discussions of stripe physics are
recently given by 
\Ref{CEK03} and \Ref{KBF0301}.
A discussion of spin liquid states 
is given by
\Ref{S0313} with an emphasis on dimer order and by
\Ref{Wen04} with an emphasis on quantum order. 
For an account of experiments and early RVB theory, see the
book by 
\Ref{AndHtc97}.
 
\section{Basic electronic structure of the cuprates} 

\begin{figure}
\centerline{
\includegraphics[width=3.5in]{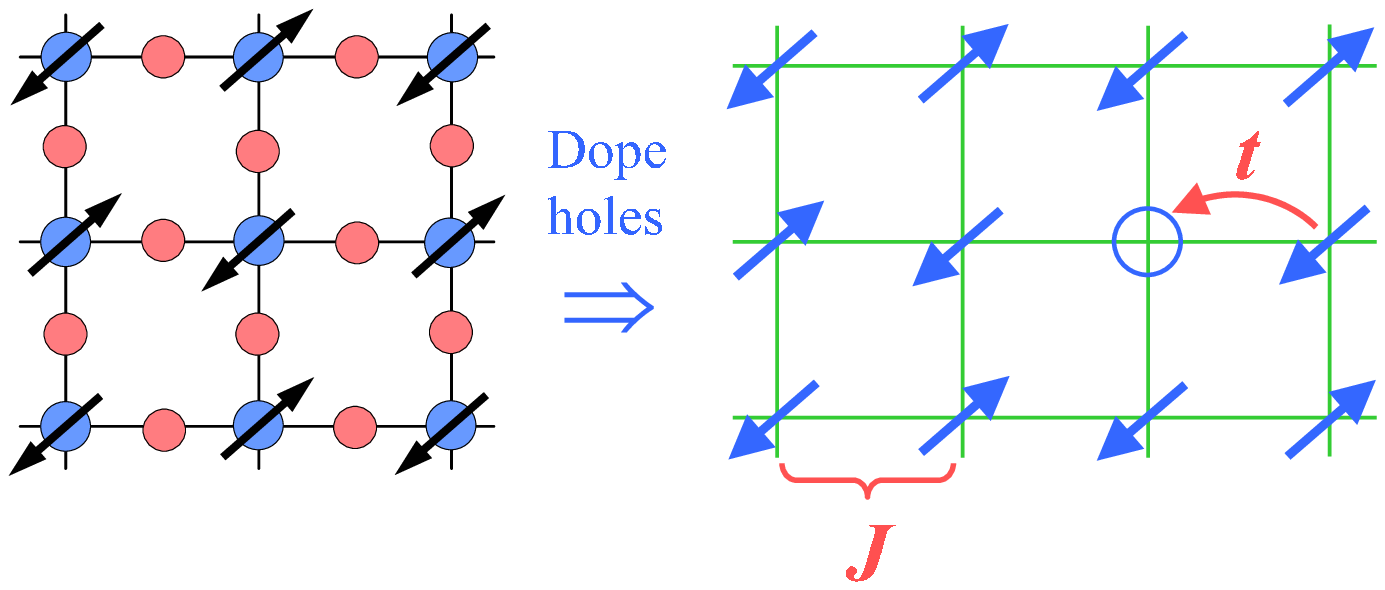}
}
\centerline{
\includegraphics[width=1.5in]{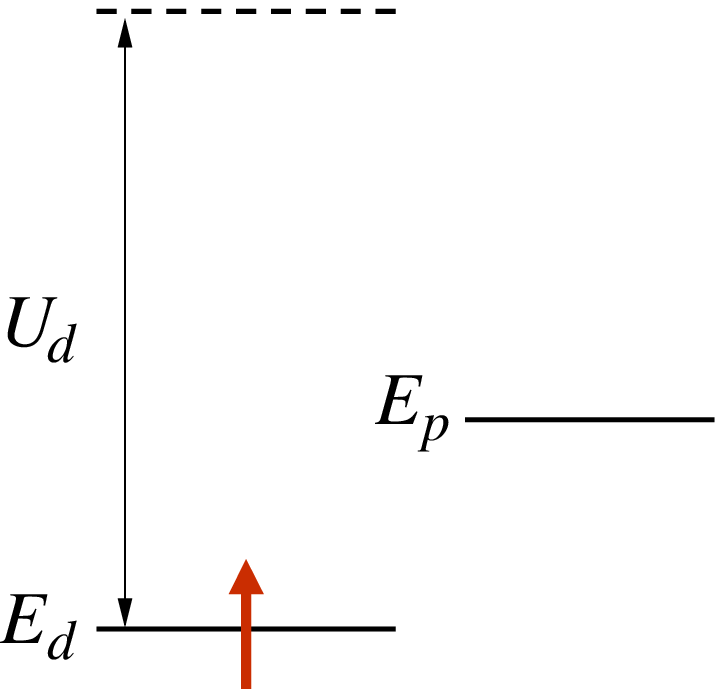}
}
\caption{The two-dimensional copper-oxygen layer (left) is simplified to the one-band model (right). 
Bottom figure shows the copper $d$ and oxygen $p$ orbitals in the hole picture.  A single hole with  $S = 1/2$ occupies the copper $d$ orbital in the insulator.
}
\label{fig.2}
\end{figure}

It is generally agreed that the physics of high $T_c$ superconductivity is
that of the copper oxygen layer, as shown in Fig. \ref{fig.2}.  In the parent
compound such as La$_2$CuO$_4$, the formal valence of Cu is $2+$, which means
that its electronic state is in the $d^9$ configuration.  The copper is
surrounded by six oxygens in an octahedral environment (the apical oxygen
lying above and below Cu are not shown in Fig. 2).  The distortion from a
perfect octahedron due to the shift 
\noindent of the apical oxygens splits the $e_g$ orbitals so that the highest
partially occupied $d$ orbital is $x^2 -y^2$.  The lobes of this orbital point
directly to the $p$ orbital of the neighboring oxygen, forming a strong
covalent bond with a large hopping integral $t_{pd}$.  As we shall see, the
strength of this covalent bonding is responsible for the unusually high energy
scale for the exchange interaction.  Thus the electronic state of the cuprates
can be described by the so-called 3 band model, where in each unit cell we
have the Cu $d_{x^2-y^2}$ orbital and two oxygen $p$ orbitals  
\cite{E8759,VSA8781}.
The Cu orbital is singly occupied while the $p$
orbitals are doubly occupied, but these are admixed by $t_{pd}$.  In addition,
admixtures between the oxygen orbitals may be included.  These tight-binding
parameters may be obtained by fits to band structure calculations 
\cite{M8728,YFX8735}.
However, the largest energy in the problem
is the correlation energy for doubly occupying the copper
orbital.  To describe these correlation energies, it is more convenient to go
to the hole picture. 
The Cu $d^9$ configuration is represented by energy level $E_d$ occupied by a
single hole with $S={1\over 2}$.  The oxygen $p$ orbital is empty of holes and
lies at energy $E_p$ which is higher than $E_d$. The energy to doubly occupy
$E_d$ (leading to a $d^8$ configuration) is $U_d$, which is very large and can
be considered infinity. The lowest energy excitation is the charge transfer
excitation where the hole hops from $d$ to $p$ with amplitude $-t_{pd}$.  If
$E_p-E_d$ is sufficiently large compared with $t_{pd}$, the hole will form a
local moment on Cu.  This is referred to as a charge transfer insulator in the
scheme of 
\Ref{ZSA8518}.  Essentially, $E_p-E_d$ plays
the role of the Hubbard $U$ in the one-band model of the Mott insulator.
Experimentally an energy gap of 2.0~eV is observed and interpreted as the
charge transfer excitation (see 
\Ref{KBS9897}).


Just as in the one-band Mott-Hubbard insulator, where virtual hopping to
doubly occupied states leads to an exchange interaction $J{\v{S}}_1\cdot
{\v{S}}_2$ where $J=4t^2/U$, in the charge-transfer insulator, the local
moments on nearest neighbor Cu prefer antiferromagnetic alignment because both
spins can virtually hop to the $E_p$ orbital.  Ignoring the $U_p$ for doubly
occupying the $p$ orbital with holes, the exchange integral is given by
\be
J = {t^4_{pd}\over (E_p-E_d)^3} .
\label{Eq.1}
\en
The relatively small size of the charge transfer gap means that we are not
deep in the insulating phase and the exchange term is expected to be large.
Indeed experimentally the insulator is found to be in an antiferromagnetic
ground state.  By fitting Raman scattering to two magnon excitations 
\cite{SFL9025},
the exchange energy is found to be $J = 0.13$~eV.   This
is one of the largest exchange energies known and is exceeded only by that of
the ladder compounds which involve the same Cu-O bonding.  This value of $J$
is confirmed by fitting spin wave energy to theory,  where an additional ring
exchange terms is found 
\cite{CHA0177}.

By substituting divalent Sr for trivalent La, the electron count on the Cu-O
layer can be changed in a process called doping. For example, in
La$_{2-x}$Sr$_x$CuO$_4$, $x$ holes per Cu is added to the layer. As seen in
Fig.~2, due to the large $U_d$, the hole will reside on the oxygen $p$
orbital.  The hole can hop via $t_{pd}$  and due to translational symmetry,
the holes are mobile and form a metal, unless localization due to disorder or
some other phase transition intervenes.  The full description of the hole
hopping in the three-band model is complicated, and a number of theories
consider this essential to the understanding of high T$_c$ superconductivity
\cite{E8759,VSA8781}.
On the other hand, there is strong
evidence that the low energy physics (on a scale small compared with $t_{pd}$
and $E_p-E_d$) can be understood in terms of an effective one-band model, and
we shall follow this route in this review.   The essential insight is that
the doped hole resonates on the four oxygen sites surrounding a Cu and the
spin of the doped hole combines with the spin on the Cu to form a spin
singlet.  This is known as the Zhang-Rice singlet \cite{ZR8859}.
This state is split off by an energy of order $t^2_{pd}/(E_p-E_d)$ because the
singlet gains energy by virtual hopping.  On the other hand, the Zhang-Rice
singlet can hop from site to site.  Since the hopping is a two step process,
the effective hopping integral $t$ is also of order $t^2_{pd}/(E_p-E_d)$.
Since $t$ is the same parametrically as the binding energy of the singlet, the
justification of this point of view relies on a large  numerical factor for
the binding energy which is obtained by studying small clusters.

By focusing on the low lying singlet, the hole doped three-band model
simplifies to a one-band tight binding model on the square lattice, with an
effective nearest neighbor hopping integral $t$ given earlier 
and with $E_p-E_d$ playing a role analogous to $U$.  In the large $E_p-E_d$ 
limit this maps onto the $t$-$J$ model
\be
H = &P&  \left(
-\sum_{\langle \v i\v j \rangle,\sigma} t_{\v i\v j}c^\dagger_{\v i\sigma}c_{\v i\sigma}  \right.  \nonumber  \\ 
&+& J\sum_{\langle \v i\v j \rangle}
\left. \rule{0in}{.3in}
\left(
\v{S}_{\v i} \cdot \v{S}_{\v j} - {1\over 2}n_{\v i}n_{\v j} \right)
\right) P
\label{Eq.2}
\en
where the projection operator $P$ restricts the Hilbert space to one which
excludes double occupation of any site.  $J$ is given by $4t^2/U$ and we can
see that it is the same functional form as that of the three-band model
described earlier.  It is also possible to dope with electrons rather than
holes. The typical electron doped system is Nd$_{2-x}$Ce$_x$CuO$_{4+\delta}$
(NCCO). The added electron corresponds to removal of a hole from the copper
site in the hole picture (Fig. \ref{fig.2}), \ie the Cu ion is in the
$d^{10}$ configuration.  This vacancy can hop with a $t_\text{eff}$ and the mapping
to the one-band model is more direct than the hole doped case.  Note that in
the full three-band model the object which is hopping is the Zhang-Rice
singlet for hole doping and the Cu $d^{10}$ configuration for electron doping.
These have rather different spatial structure and are physically quite
distinct.  For example, the strength of their coupling to lattice distortions
may be quite different.  When mapped to the one-band model, the nearest
neighbor hopping $t$ has the same parametric dependence, but could have a
different numerical constant.  As we shall see, the value of $t$ derived from
cluster calculations turn out to be surprisingly similar for electron and hole
doping.  For a bi-partate lattice, the $t$-$J$ model with nearest neighbor $t$
has particle-hole symmetry because the sign of $t$ can be absorbed by changing
the sign of the orbital on one sublattice.  Experimentally the phase diagram
exhibits strong particle-hole asymmetry. On the electron doped side, 
the antiferromagnetic insulator
survives up to much higher doping concentration (up to $x \approx 0.2$) and
the superconducting transition temperature is quite low (about 30~K).  Many of
the properties  of the superconductor resemble that of the overdoped region of
the hole doped side and the pseudogap phenomenon, which is so prominent in the
underdoped region, is not observed with electron doping.  It is as though the
greater stability of the antiferromagnet has covered up any anomalous regime
that might exist otherwise.  Precisely why is not clear at  the moment.  One
possibility is that polaron effects may be stronger on the electron doped
side, leading to carrier localization over a broader range of doping.  There
has been some success in modeling the contrast in the single hole spectrum by
introducing further neighbor coupling into the one-band model which breaks
the particle-hole symmetry 
\cite{SLE0402}.  This will be discussed
further below.

We conclude that
the electron correlation is strong enough to produce a Mott insulator at half
filling. Furthermore, the one band $t$-$J$
model captures the essence of the low
energy electronic excitations of the cuprates.  
Particle-hole
asymmetry may be accounted for by including further neighbor hopping
$t^\prime$. 
This point of view has been tested extensively by 
\Ref{HSS9068}
who used {\em ab initio} local density functional theory to generate
input parameters for the three-band Hubbard model and then solve the spectra
exactly on finite clusters.  The results are compared with the low energy
spectra of the one-band Hubbard model and the $t-t^\prime-J$ model.  They found
an excellent overlap of the low lying wavefunctions for both the one-band
Hubbard and the $t-t^\prime-J$ model and were able to extract effective
parameters.  They found $J$ to be $128 \pm 5$~meV, in excellent agreement with
experimental values.  Furthermore they found $t \approx 0.41$~eV and $0.44$~eV
for electron and hole doping, respectively.  The near particle-hole symmetry
in $t$ is surprising because the underlying electronic states are very
different in the two cases, as already discussed.  Based on their results, the
commonly used parameter $J/t$ for the $t$-$J$ model is 1/3.  They also found a
significant next nearest neighbor $t^\prime$ term, again almost the same for
electron and hole doping.

\begin{figure}[t]
\centerline{
\includegraphics[width=3in]{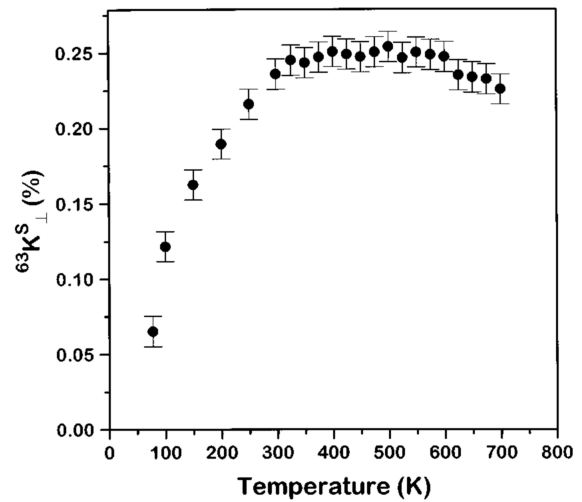}
}
\caption{
The Knight shift for YB$_2$Cu$_4$O$_8$.  It is an underdoped material with
$T_c = 79$K.  From 
\Ref{CIS9777}.}
\label{Knight}
\end{figure}

More recently, 
\Ref{Ao9685} have pointed out that in addition
to the three-band model, an additional Cu $4s$ orbital has a strong influence
on further neighbor hopping $t^\prime$ and $t^{\prime\prime}$, where
$t^\prime$ is the hopping across the diagonal and $t^{\prime\prime}$ is
hopping to the next-nearest neighbor along a straight line.  Recently 
\Ref{PDS0103}
emphasized the importance of the apical oxygen in
modulating the energy of the Cu $4s$ orbital and found a sensitive dependence
of $t^\prime/t$ on the apical oxygen distance.  They also pointed out an
empirical correlation between optimal $T_c$ and $t^\prime/t$.  As we will
discuss in sections VI.D and VII, $t^\prime$ may play an important role in
determining $T_c$ and in explaining the difference between electron and hole
doping.  However, in view of the fact that on-site repulsion is the largest
energy scale in the problem, it would make sense to begin our modeling of the
cuprates with the $t$-$J$ model and ask to what extent the phase diagram can
be accounted for.  As we shall see, even this is not a simple task and will
constitute the major thrust of this review.

\begin{figure}[t]
\centerline{
\includegraphics[width=3in]{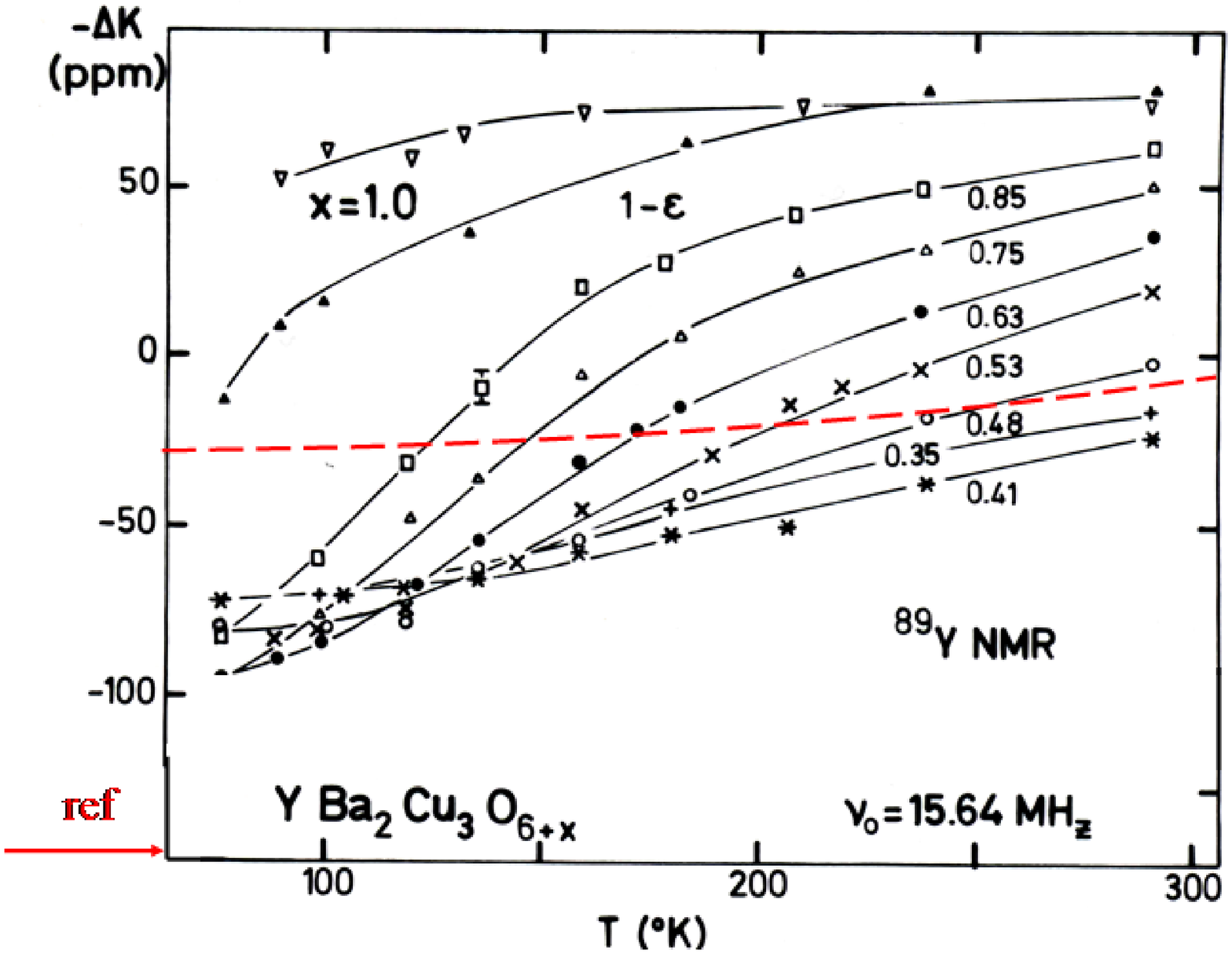}
}
\centerline{
\includegraphics[width=3in]{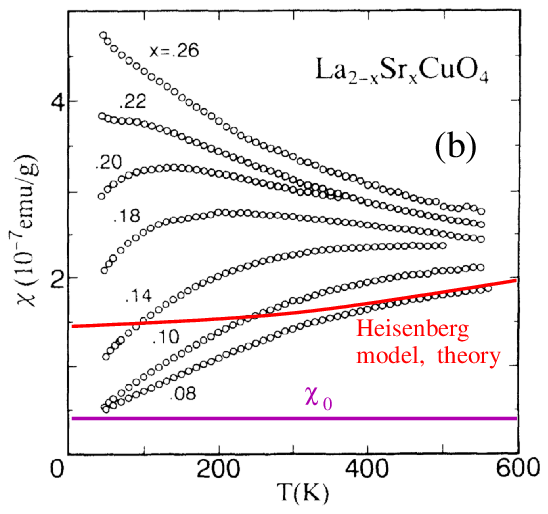}
}
\vspace{0.5cm}
\caption{
(a) Knight shift data of YBCO for a variety of doping (from 
\Ref{AOM8900}).  The zero reference level for the spin  contribution
is indicated by the arrow and the dashed line represents the prediction of the 2D $S = {1\over 2}$ Heisenberg model for
$J = 0.13$~eV. (b)~Uniform magnetic susceptibility for LSCO (from 
\Ref{NOM9400}). The orbital contribution $\chi_0$ is shown (see
text) and the solid line represents the Heisenberg model prediction.}
\label{chi}
\end{figure}

\begin{figure}[t]
\centerline{
\includegraphics[width=2.75in]{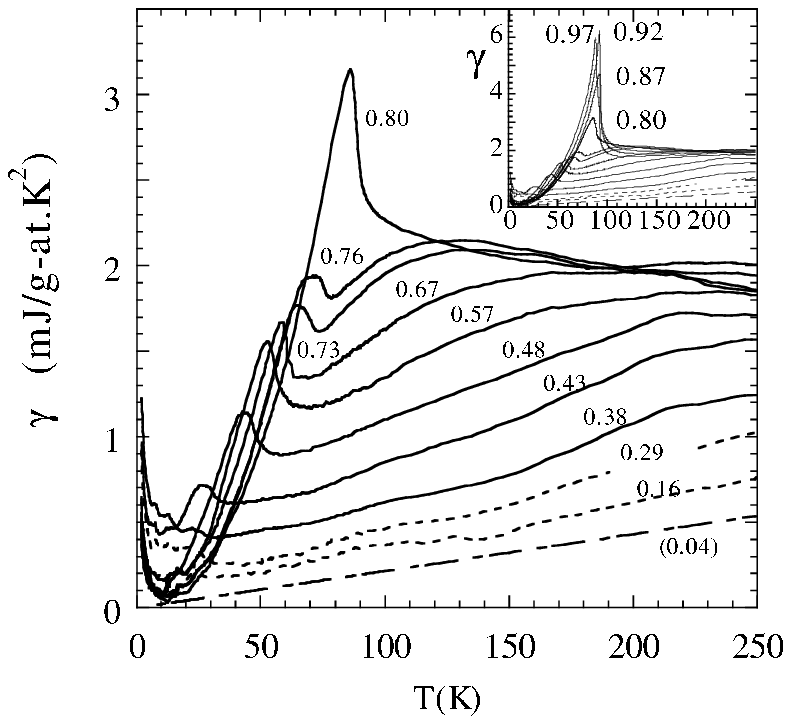}
}
\centerline{
\includegraphics[width=3.1in]{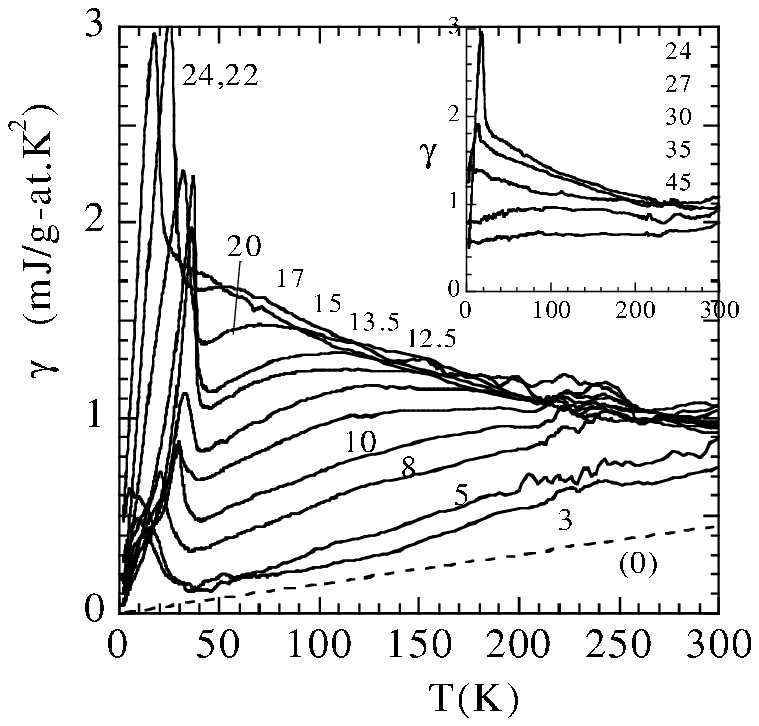}
}
\vspace{0.5cm}
\caption{ 
The specific heat coefficient $\gamma$ for YBa$_2$Cu$_3$O$_{6+y}$ (top) and
La$_{2-x}$Sr$_x$CuO$_4$ (bottom).  Curves are labeled by the oxygen content
$y$ in the top figure and by the hole concentration $x$ in the bottom figure.
Optimal and overdoped samples are shown in the inset.  The jump in $\gamma$
indicates the superconducting transition.  Note the reduction of the jump size
with underdoping.  (From 
\Ref{LMC9340} and \Ref{LLC0159}).
}
\label{gamma}
\end{figure}

\section{Phenomenology of the underdoped cuprates} The essence of the problem
of doping into a Mott insulator is readily seen from Fig. \ref{fig.2}.  When a
vacancy is introduced into an antiferromagnetic spin background, it would like
to hop with amplitude $t$ to lower its kinetic energy.  However, after one hop
its neighboring spin finds itself in a ferromagnetic environment, at an energy
cost of ${3\over 2} J$ if the spins are treated as classical $S = {1\over 2}$.
It is clear that the holes are very effective in destroying the
antiferromagnetic background.  This is particularly so when $t \gg J$ when the
hole is strongly delocalized.  The basic physics is the competition between
the exchange $J$ and the kinetic energy which is of order $t$ per hole or $xt$
per unit area.  When $xt \gg J$ we expect the kinetic energy to win and the
system should be a Fermi liquid metal with weak residual antiferromagnetic
correlation.  When $xt \leq J$, however, the outcome is much less clear
because the system would like to maintain the antiferromagnetic correlation
while allowing the hole to move as freely as possible.  Experimentally we know
that N\'{e}el order is destroyed with 3\% hole doping, after which $d$-wave
superconducting state emerges as the ground state up to 30\% doping. Exactly how and
why superconductivity emerges as the best compromise is the centerpiece of the
high T$_c$ puzzle but we already see that the simple competition between $J$
and $xt$ sets the correct scale $x = J/t = {1\over 3}$ for the appearance of
nontrivial ground states. We shall focus our attention on the so-called
underdoped region, where this competition rages most fiercely.  Indeed it is
known experimentally that the ``normal'' state above the superconducting T$_c$
behaves differently from any other metallic state that we have known about up
to now.  Essentially an energy gap appears in some properties and not others.
This region of the phase diagram is referred to as the pseudogap region and is
well documented experimentally.  We review below some of the key properties.

\subsection{The pseudogap phenomenon in the normal state}

As seen in Fig. \ref{Knight} Knight shift measurement in the YBCO 124 compound
shows that while the spin susceptibility $\chi_s$ is almost temperature
independent between 700~K and 300~K, as in an ordinary metal, it decreases
below 300~K and by the time the T$_c$ of 80~K is reached, the system has lost
80\% of the spin susceptibility 
\cite{CIS9777}.  To emphasize the
universality of this phenomenon, we reproduce in Fig. \ref{chi} some old data
on YBCO and LSCO.  Figure \ref{chi}(a) shows the Knight shift data from 
\Ref{AOM8900}. We have subtracted the orbital contribution, which is
generally agreed to be 150~ppm 
\cite{THS9350}, and drawn in the
zero line to highlight the spin contribution to the Knight shift which is
proportional to $\chi_s$.  The proportionality constant is known, which allows
us to draw in the Knight shift which corresponds to the 2D square $S = {1\over
2}$ Heisenberg antiferromagnet with $J = 0.13$~eV 
\cite{DM9162,SDS9701}.   The point of this exercise is to show that in the
underdoped region, the spin susceptibility drops {\it below} that of the
Heisenberg model at low temperatures before the onset of superconductivity.
This trend continues even in the severely underdoped limit (O$_{0.53}$ to
O$_{0.41}$), showing that the $\chi_s$ reduction cannot simply be understood
as fluctuations towards the antiferromagnet.  Note that the discrepancy is
worse if $J$ were replaced by a smaller $J_\text{eff}$ due to doping, since
$\chi_s \sim J_\text{eff}^{-1}$.  The data seen in this light strongly point
to singlet formation as the origin of the pseudogap seen in the uniform spin
susceptibility.

It is worth noting that the trend shown in Fig. \ref{chi}(a) is not so
apparent if one looks at the measured spin susceptibility directly 
\cite{TMG8877}. This is because the van~Vleck part of the spin
susceptibility is doping dependent, due to the changing chain contribution.
This problem does not arise for LSCO, and in Fig. 4(b) we show the uniform
susceptibility data 
\cite{NOM9400}.  The zero of the spin part is
determined by comparing susceptibility measurements to $^{17}$O Knight shift
data 
\cite{IKZ9116}.  
\Ref{NOM9400}  find an
excellent fit for the $x = 0.15$ sample (see Fig. 9 of this reference) and
determine the orbital contribution for this sample to be $\chi_0 \sim 0.4
\times 10^{-7}$~emu/g.  This again allows us to plot the theoretical
prediction for the Heisenberg model.  Just as for YBCO, $\chi_s$ for the
underdoped samples ($x = $0.1 and 0.08) drops below that of the Heisenberg
model.  In fact, the behavior of $\chi_s$ for the two systems is remarkably
similar, especially in the underdoped region.\footnote{ We note that a
comparison of $\chi_s$ for YBCO and LSCO was made by 
\Ref{MM9310}.
Their YBCO analysis is similar to ours.  However, for LSCO they find  a rather
different  $\chi_0$ by matching the measurement above 600~K to that of the
Heisenberg model.  Consequently, their $\chi_s$ looks different for YBCO and
LSCO.  We believe their procedure is not really justified.}

\begin{figure}
\centerline{
\includegraphics[width=3in]{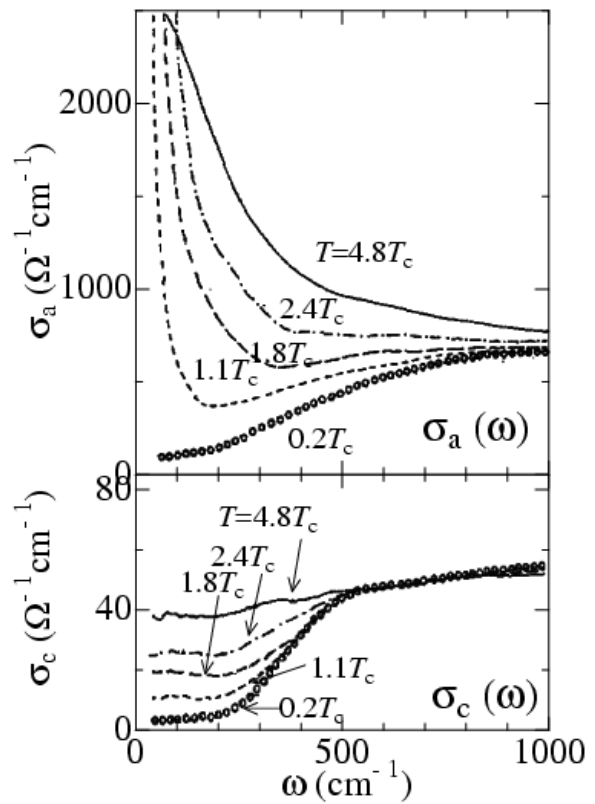}
}
\caption{ 
The frequency dependent conductivity with electric field parallel to the plane ($\sigma_a(\omega)$, top figure) and perpendicular to the plane ($\sigma_c(\omega)$ bottom figure) in an underdoped YBCO crystal.  From 
\Ref{U9712}.
}
\label{sigma}
\end{figure}

A second indication of the pseudogap comes from the linear $T$ coefficient of
the specific heat, which shows a marked decrease below room temperature
(see Fig. \ref{gamma}).  
Furthermore, the specific heat jump at T$_c$ is greatly reduced with
decreasing doping.  It is apparent that the spins are forming into singlets
and the spin entropy is gradually lost.  On the other hand, as shown in Fig.
\ref{sigma} the frequency dependent conductivity behaves very differently
depending on whether the electric field is in the $ab$ plane $(\sigma_{ab})$
or perpendicular to it $(\sigma_c)$.

\begin{figure}
\centerline{
\includegraphics[width=3.1in]{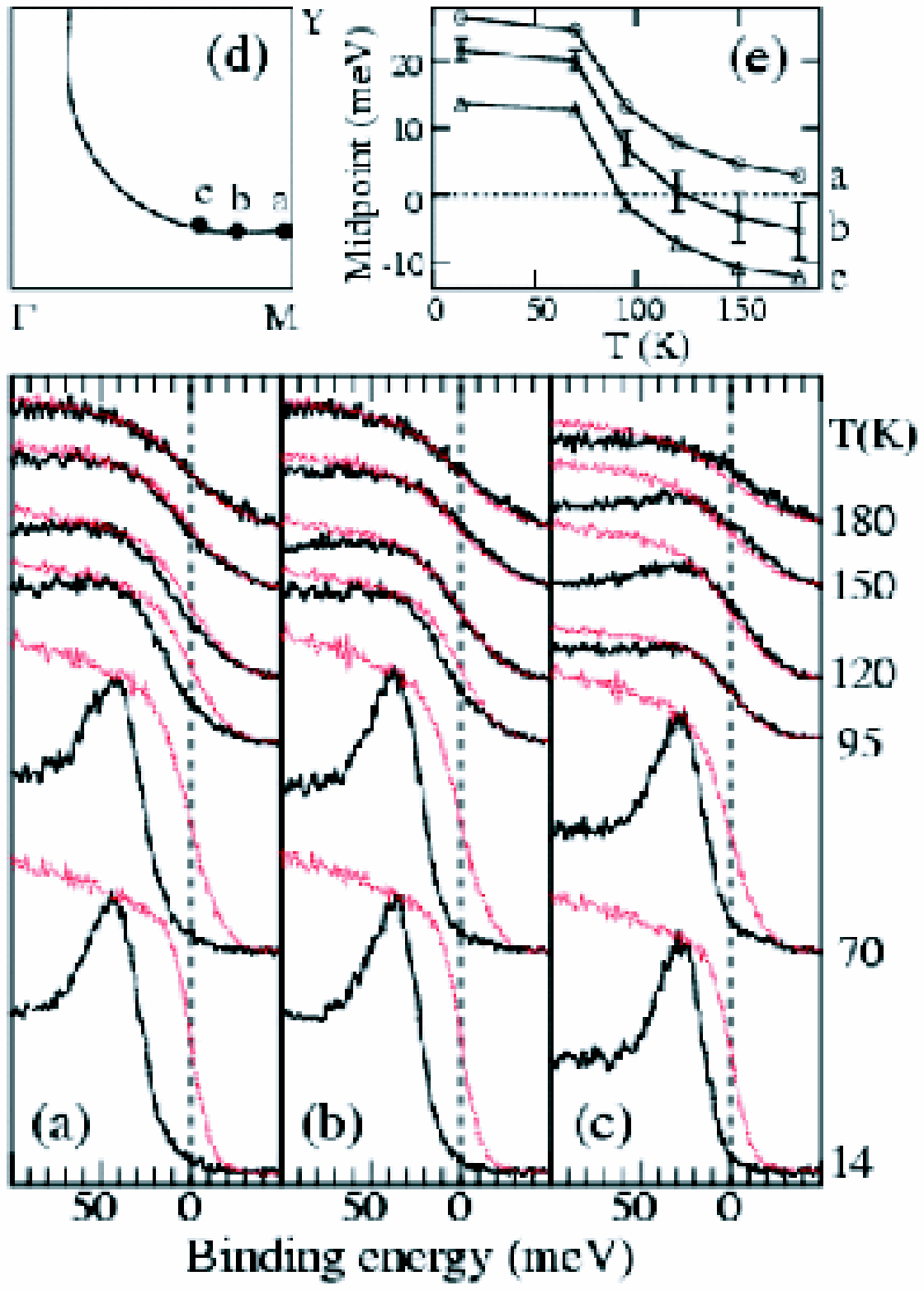}
}
\vskip 0.05in
\centerline{
\includegraphics[width=2.3in]{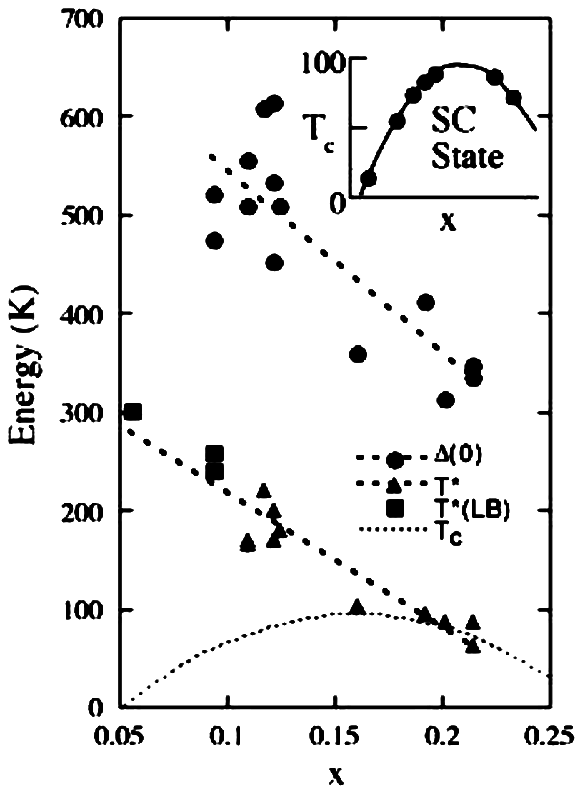}
}
\caption{
(a--c)~Spectra from underdoped Bi-2212 ($T_c = 85$K) taken at different $k$ points along the Fermi surface shown in (d).  Note the pullback of the spectrum from the Fermi surface as determined by the Pt reference (red lines) for $T > T_c$. (e)~Temperature dependence of the leading-edge midpoints. (from 
\Ref{NDR9857})  Bottom figure shows the temperature $T^\ast$
where the pseudogap determined from the leading edge first appears plotted as a function of doping for
Bi-2212 samples.  Triangles are determined from data such as shown in Fig.~7(a) and squares are lower
bound estimates. Circles show the energy gap $\Delta$ measured at $(0,\pi)$ at low temperatures. (from
\Ref{CNR03}).}
\label{edge}
\end{figure}

At low frequencies (below 500~cm$^{-1}$) $\sigma_{ab}$ shows a typical
Drude-like behavior for a metal with a width which decreases with temperature,
but an area (spectral weight) which is independent of temperature
\cite{SLB0205}. Thus there is no sign of the pseudogap in
the spectral weight.  This is surprising because in other examples where an
energy gap appears in a metal, such as the onset of charge or spin density
waves, there is a redistribution of the spectral weight from the Drude part to
higher frequencies.  An important observation concerning the spectral weight
is that the integrated area under the Drude peak is found to be proportional
to $x$ 
(\cit{OTM9042}; \cit{CRK9333}; \cit{UIT9142}; \cit{PLD04}).  In the superconducting state this
weight collapses to form the delta function peak, with the result that the
superfluid density $n_s/m$ is also proportional to $x$.  It is as though only
the doped holes contribute to charge transport in the plane.  In contrast,
angle-resolved photoemission shows a Fermi surface at optimal doping very
similar to that predicted by band theory, with an area corresponding to
$(1-x)$ electrons (see Fig. \ref{edge}(d)).  With underdoping, this Fermi
surface is partially gapped in an unusual manner which we shall next discuss.

In contrast to the metallic behavior of $\sigma_{ab}$, it was discovered by
\Ref{HTL9310} that  below 300~K $\sigma_c(\omega)$ is gradually
reduced for frequencies below 500~cm$^{-1}$ and a deep hole is carved out of
$\sigma_c(\omega)$ by the time T$_c$ is reached.
This is clearly seen in the lower panel of Fig.
\ref{sigma}.

Finally, angle-resolved photoemission shows that an energy gap (in the form of
a pulling back of the leading edge of the electronic spectrum from the Fermi
energy) is observed near momentum $(0,\pi)$. Note that the lineshape is
extremely broad and completely incoherent. The onset of superconductivity is
marked by the appearance of a small coherent peak at this gap edge (Fig.
\ref{edge}).  The size of the pull back of the leading edge is the same as the
energy gap of the superconducting state as measured by the location of the
coherence peak.  As shown in Fig. \ref{edge} this gap energy increases with
decreasing doping, while the superconducting T$_c$ decreases. This trend is
also seen in tunneling data.

It  is possible to map out the Fermi surface by tracking the momentum of the
minimum excitation energy in the superconducting state for each momentum
direction.  Along the Fermi surface the energy gap does exactly what is
expected for a $d$-wave superconductor.  It is maximal near $(0,\pi)$ and
vanishes along the line connection $(0,0)$ and $(\pi,\pi)$ where the
excitation is often referred to as nodal quasiparticles.  Above T$_c$ the
gapless region expands to cover a finite region near the nodal point, beyond
which the pseudogap gradually opens as one moves towards $(0,\pi)$. This
unusual behavior is sometimes referred to as the Fermi arc 
(\cit{LSD9625}; \cit{MDL9641}; \cit{DYC9651}).  It is
worth noting that unlike the anti-nodal direction (near $(0,\pi)$) the
lineshape is relatively sharp along the nodal direction even above T$_c$.
From the width in  momentum space, a lifetime which is linear  in temperature
has been extracted for a sample near optimal doping 
\cite{VFJ9910}.  
A narrow lineshape in the nodel direction has also been observed in LSCO
\cite{Yo0301} and in $Na$ doped $Ca_2CuO_2Cl_2$
\cite{RSK0301}. So the notion of relatively well defined nodal excitations in
the normal state is most likely a universal feature.

\begin{figure}[t]
\centerline{
\includegraphics[width=3.5in]{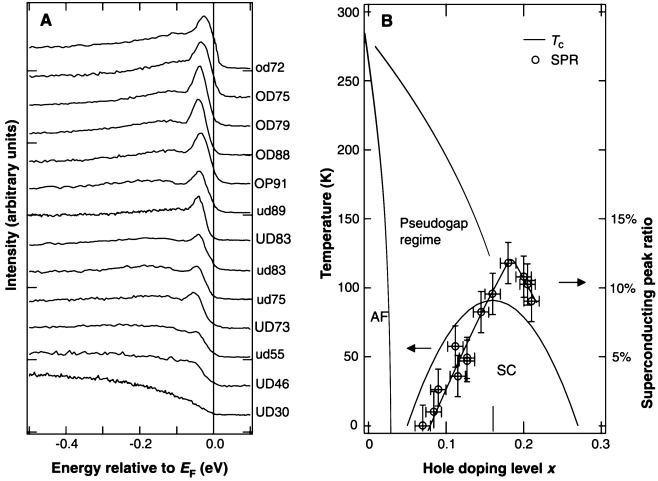}
}
\caption{
(A) Doping dependence of the ARPES spectra at $(0,\pi)$ at $T \ll T_c$ for
overdoped (OD), optimally doped (OP), and underdoped (UD) materials labeled by
their $T_c$'s.  (B)~The spectral weight of the coherent peak in Fig.~8(a)
normalized to the background is plotted vs. doping $x$.  From 
\Ref{Fo0077}.}
\label{ARPES}
\end{figure}

As mentioned earlier, the onset of superconductivity is marked by the
appearance of a sharp coherence peak near $(0,\pi)$.  The spectral weight of
this peak is small and gets even smaller with decreasing doping, as shown in
Fig. \ref{ARPES}(b).  Note that this behavior is totally different from
conventional superconductors.  There the quasiparticles are well defined in
the normal state and according to BCS theory, the sharp peak pulls back from
the Fermi energy and opens an energy gap in the superconducting state.

\begin{figure}[t]
\centerline{
\includegraphics[width=3.5in]{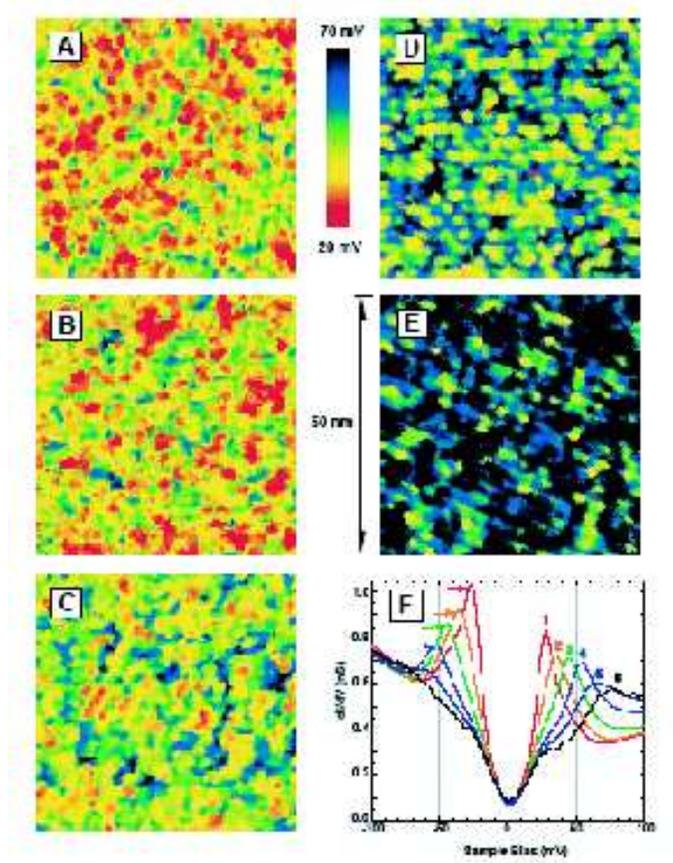}
}
\caption{
From 
\Ref{MLH0405}.  STM images showing the spatial distribution of energy gaps for
a variety of samples which are progressively more underdoped from A to E.
Panel F shows the average spectrum for a given energy gap.
}
\label{STM}
\end{figure} 

In the past few years, low temperature STM data have become available, mainly
on Bi-2212 samples.  STM provides a measurement of the local density of states
$\rho(E, \v{r})$ with atomic resolution.  It is complementary to ARPES in that
it provides real space information but no direct momentum space information.
One important outcome is that STM reveals spatial inhomogeneity of the Bi-2212
on roughly 50 to 100~\AA \hspace{.002in} length scale, which becomes more and
more significant with underdoping.  As shown in Fig.~\ref{STM}(f) spectra with
different energy gaps are associated with different patches and with
progressively more underdoping, patches with large gaps become more and more
pre-dominant.  Since ARPES is measuring the same surface, it becomes necessary
to reinterpret the ARPES data with inhomogeneity in mind.  In particular, the
decrease of the weight of the coherent peak shown in Fig.~\ref{ARPES}(b) may
simply be due to a reduction of the fraction of the sample which has sharp
coherent peaks.

A second remarkable observation by STM is that the low lying density of states
($\rho (E, \v{r})$ for $E \lesssim 10$ to 15 meV) is remarkably homogeneous.
This is clearly seen in Fig. 9(f).  It is reasonable to associate this low
energy excitation with the quasiparticles near the nodes.  Indeed, the low
lying quasiparticles exhibit interference effects due to scattering by
impurities, which is direct evidence for their spatial coherence over long
distances.  Then the combined STM and ARPES data suggest a kind of phase
separation in momentum space, \ie the spectra in the anti-nodal region (near
$0,\pi$) is highly inhomogeneous in space whereas the quasiparticles near the
nodal region are homogeneous and coherent.  The nodal quasiparticles must be
extended and capable of averaging over the spatial homogeneity, while the
anti-nodal quasiparticles appear more localized.  In this picture the
pseudogap phenomenon mainly has to do with the anti-nodal region.

\Ref{MLH0405} argued that there is a limiting spectrum (the
broadest curve in Fig. \ref{STM}(f)) which characterizes the extreme
underdoped region at zero temperature.  It has no coherent peak at all, but
shows a reduction of spectral weight up to a very high energy of 100 to 200
meV.  Very recently, 
\Ref{HLK04} provided support of this
point of view in their study of Na doped Ca$_2$CuO$_2$Cl$_2$.  In this
material the apical oxygen in the CuO$_4$ cage is replaced by Cl and the
crystal cleaves easily.  For Na doping ranging from $x = 0.08$ to $0.12$, a
tunneling spectrum very similar to the limiting spectrum for Bi-2212 is
observed.  This material appears free of the inhomogeneity which plagues the
Bi-2212 surface.  ARPES experiments on these crystals are becoming available
\cite{RSK0301}
and the combination of STM and ARPES should yield much information on the real
and momentum space dependence of  the electron spectrum.  There is much
excitement concerning the discovery of a static $4 \times 4$ pattern in this
material, and their relation to the incommensurate pattern seen in the vortex
core of Bi-2212 
\cite{HHL0266} and reported also in the absence
of magnetic field, albeit in a much weaker form 
\cite{HEK0333,VSO0495}.  How this spatial modulation is related to the
pseudogap spectrum is a topic of current debate.

In the literature, the pseudogap behavior is often associated with anomalous
behavior of the nuclear spin relaxation rate ${1\over T_1}$.  In normal metals
the nuclear spin relaxes by exciting low energy particle-hole excitations,
leading to the Koringa behavior, \ie ${1\over T_{1}T}$ is temperature
independent.  In high T$_c$ materials, it is rather ${1\over T_1}$ which is
temperature independent, and the enhanced relaxation (relative to Koringa) as
the temperature is reduced is ascribed to antiferromagnetic spin fluctuations.
It was found that in underdoped YBCO, the nuclear spin relaxation rate at the
copper site reaches a peak at a temperature $T^\ast_1$ and decreases rapidly
below this temperature 
(\cit{WWB8993}; \cit{YIS8954}; \cit{TRH9147}).  
The resistivity also shows a decrease below $T^\ast_1$.  In some
literature $T^\ast_1$ is referred to as the pseudogap scale.  However, we note
that $T^\ast_1$ is lower than the energy scale we have been discussing so far,
especially compared with that for the uniform spin susceptibility and the
$c$-axis conductivity.  Furthermore, the gap in ${1\over T_1}$ is not
universally observed in cuprates. It is  not seen in LSCO.  In
YBa$_2$Cu$_4$O$_8$, which is naturally underdoped, the gap in ${1\over T_1T}$
is wiped out by 1\% Zn doping, while the Knight shift remains unaffected
\cite{ZOM0391}.  It is known from neutron scattering that the low
lying spin excitations near $(\pi,\pi)$ is sensitive to disorder.  Since
${1\over T_1}$ at the copper site is dominated by these fluctuations, it is
reasonable that ${1\over T_1}$ is sensitive as well.  In contrast, the
gap-like behavior we described thus far in a variety of physical properties is
universally observed across different families of cuprates (wherever data
exist) and are robust.  Thus we prefer not to consider $T^\ast_1$ as the
pseudogap temperature scale.

\subsection{Neutron scattering, resonance and stripes} 
\begin{figure}[t]
\centerline{
\includegraphics[width=3in]{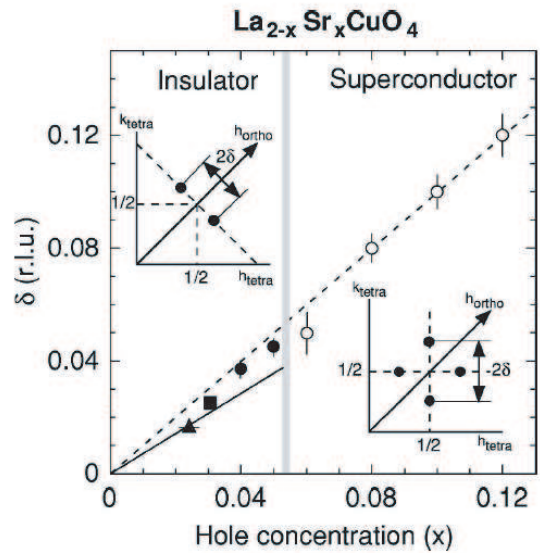}
}
\caption{
From 
\Ref{MFY0048}.  Plot of the incommensurability $\delta$ vs.
hole concentration $x$.  In the superconducting state, the open circles denote
the position of the fluctuating spin density wave observed by neutron
scattering.  (Data from 
\Ref{Yo9865}.)  In the insulator the spin density wave becomes static at low
temperatures and its orientation is rotated by 45$^\circ$.  The dashed line
$(\delta = x)$ is the prediction of the stripe model which assumes a fixed
density of holes along the stripe.
}
\label{incomm}
\end{figure}

Neutron scattering
provides a direct measure of the spin excitation spectrum.  Early work (see
\Ref{KBS9897}) has shown that the long range N\'{e}el order gives
way to short range order with progressively shorter correlation length with
doping, so that at optimal doping, the static spin correlation length is no
more than 2 or 3 lattice spacings.  Much of the early work was focused on the
La$_{2-x}$Sr$_x$CuO$_4$ family, because of the availability of large single
crystals.  It was found that there is enhanced spin scattering at low
energies, centered around the incommensurate positions $\v{q}_0 = \left( \pm
{\pi \over 2}, \pm \delta  \right)$, 
\cite{CAM9191}.  
\Ref{Yo9865}
found that $\delta$ increases systematically with doping, as
shown in Fig. \ref{incomm}.  
Meanwhile it was noted that in the La$_2$CuO$_4$ family, there is a marked
suppression of $T_c$ near $x = {1 \over 8}$.  This suppression is particularly
strong with Ba doping, and $T_c$ is completely destroyed if some Nd is
substituted for La, as in La$_{1.6-x}$Nd$_{0.4}$Sr$_x$CuO$_4$ for $x= {1 \over
8}$.  
\Ref{TSA9561}
discovered static spin density wave and
charge density wave order in this system, which onsets below about 50~K.  The
period of the spin and charge density waves are 8 and 4 lattice constants,
respectively.  The static order is modeled by a stripe picture where the holes
are concentrated in period 4 charge stripes separated by spin ordered regions
with anti-phase domain walls.  Recently, the same kind of stripe order was
observed in La$_{1.875}$Ba$_{0.125}$CuO$_4$ 
\cite{FGY0496}.  Note
that in this model there is one hole per two sites along the charge stripe.
It is tempting to interpret the low energy spin density wave observed in LSCO
as a slowly fluctuating form of stripe order, even though the associated
charge order (presumably dynamical also) has not yet been seen.  The most
convincing argument for this interpretation comes from the observation that
over a range of doping $x = 0.06$ to $x = 0.125$, the observed
incommensurability $\delta$ is given precisely by the stripe picture, \ie
$\delta = x$, while $\delta$ saturates at approximately ${1 \over 8}$ for $x
\gtrsim 0.125$ (see Fig.\ref{incomm}).  However, it must be noted that in this
interpretation, the charge stripe must be incompressible, \ie they behave as
charge insulators.  Upon changing $x$, it is energetically more favorable to
add or remove stripes and change the average stripe spacing, rather than
changing the hole density on each stripe, which is pinned at ${1 \over 4}$
filling.  It is difficult to reconcile this picture with the fact that LSCO is
metallic and superconducting in the same doping range.  An alternative
interpretation of the incommensurate spin scattering is that it is due to
Fermi surface nesting 
\cite{LZA9387,SZL9355,TKF9355}.
However, in this case the $x$ dependence of $\delta$ requires some fine
tuning.  Regardless of interpretation, it is clear that in the LSCO family,
there are low lying spin density wave fluctuations which are almost ready to
condense.  At low temperatures, static SDW order is stabilized by Zn doping
\cite{KHY9917}, near $x = {1 \over 8}$ 
\cite{WSE9969}, and in oxygen doped systems 
\cite{LBK9943}.  However, in the
latter case, there is evidence from $\mu$SR 
\cite{SFG0224} that
there may be microscopic phase separations in this material (not too
surprising in view of the STM data on Bi-2202).  It was also found that SDW
order is stabilized in the vicinity of vortex cores 
(\cit{KYS0077}; \cit{LAC0159}; \cit{KLE0228}).

\begin{figure}[t]
\centerline{
\includegraphics[width=3in]{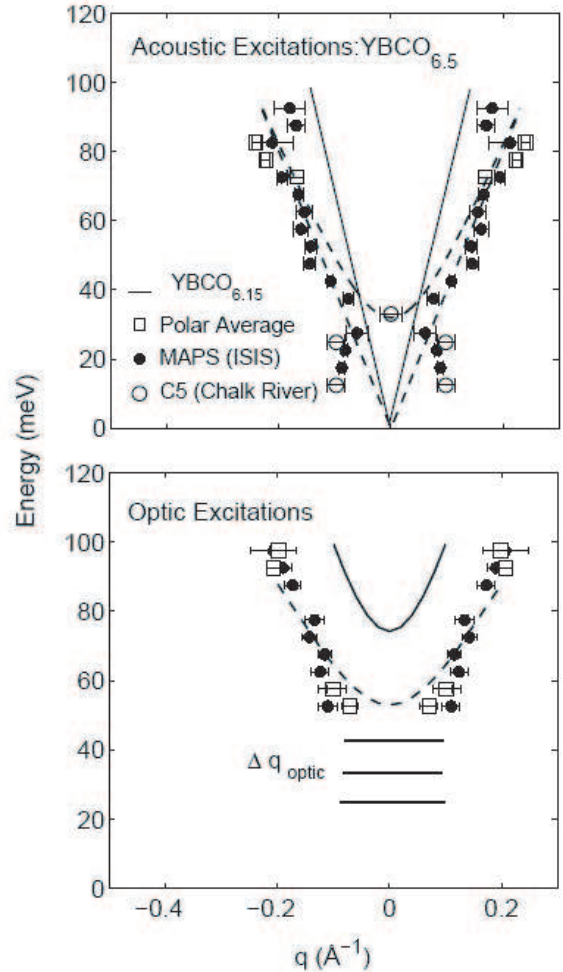}
}
\caption{
Neutron scattering from YBCO$_{6.5}$.  This sample has $T_c = 59$~K and the
experiment was performed at 6~K (from 
\Ref{SBC0471}).  Top panel
refers to in-phase fluctuations between the bilayer which shows a resonance
located at $(\pi,\pi)$ ($q = 0$ in the figure) and at energy 33~meV.
Incommensurate peaks disperse down from the resonance.  Broad peaks also
disperse upward from the resonance, forming the hourglass pattern.  Solid line
is the spin wave spectrum of the insulating parent compound.  Bottom panel
denotes out-of-phase fluctuations between the bilayers.
 }
 \label{hourglass}
\end{figure}

The key question is then whether the fluctuating stripe picture is special to
the LSCO family or plays a significant role in all the cuprates.  Outside of
the LSCO family, the spin response is dominated by a narrow resonance at
$(\pi,\pi)$.  The resonance was first discovered at 41~meV for optimally doped
YBCO 
(\cit{RRV9186}; \cit{MYA9390}).  Careful subtraction of an accidentally degenerate
phonon line reveals that the resonance appears only below $T_c$ at optimal
doping 
\cite{FKA9516}.  Now it is known that with underdoping, the resonance moves
down in energy and survives into the pseudogap regime above $T_c$.  The
resonance moves smoothly to almost zero energy at the edge of the transition
to N\'{e}el order in YBa$_2$Cu$_3$O$_{6.35}$ 
\cite{Bo04} and
clearly plays the role of a soft mode at that transition.

The resonance was interpreted as a spin triplet exciton bound below
$2\Delta_0$ 
\cite{FKA9516}.  This idea was elaborated upon by a number of RPA calculations 
(\cit{LZL9530}; \cit{BS9649}; \cit{N0051}; \cit{N0109}; \cit{KSL0098};
\cit{OP0215}; \cit{BL9915}; \cit{BL0202}; \cit{ACE0202}).  An alternative
picture making use of the particle-particle channel was proposed
\cite{DZ9526}.  However, as explained by 
\Ref{TNC0107} and by 
\Ref{NP0347}, this theory predicts an anti-bound resonance above the
two-particle continuum, which is not in accord with experiments.

Further support of the triplet exciton idea comes from the observation that
incommensurate branches extend below the resonance energy 
\cite{BSF0034}.  This behavior is predicted by RPA-type theories 
(\cit{N0051}; \cit{OP0215}; \cit{BL0202})
 in that the gap in the
particle-hole continuum extends over a region near $(\pi,\pi)$, where the
resonance can be formed.  With further underdoping this incommensurate branch
extends to lower energies (see Fig.~\ref{hourglass}).  Now it becomes clear
that the low energy incommensurate scattering previously reported for
underdoped YBCO 
\cite{MDD0004} is part of this downward dispersing
branch 
(\cit{SBL0402}; \cit{PSB0409}).  

It should be
noted that while the resonance is prominent due to its sharpness, its spectral
weight is actually quite small, of order 2\% of the total spin moment sum rule
for optimal doping and increasing somewhat with underdoping.  There is thus
considerable controversy over its significance in terms of its contribution to
the electron self-energy and towards pairing (see 
\Ref{NP0347}).
The transfer of this spectral weight from above to below $T_c$ has been
studied in detail by 
\Ref{SBL0402}.  These authors emphasized
that in the pseudogap state above $T_c$ in YBa$_2$Cu$_3$O$_{6.5}$, the
scattering below the resonance is gapless and in fact increases in strength
with decreasing temperature.  This is in contrast with the sharp drop seen in
${1 \over T_1T}$ below 150~K.  Either a gap open up at very low energy (below
4~meV) or the $(\pi,\pi)$ spins fluctuating seen by neutrons are not the
dominant contribution to the nuclear spin relaxation, \ie the latter may be
due to excitations which are smeared out in momentum space and undetected by
neutrons.  We note that a similar discrepancy between neutron scattering
spectral weight and ${1\over T_1T}$ was noted for LSCO 
\cite{AMH9511}.  This reinforces our view that the decrease in ${1\over T_1T}$ should
not be considered a signature of the pseudogap.  We also note that an enhanced
$(\pi,\pi)$ scattering together with singlet formation is just what is
predicted by the $SU(2)$ theory in section XI.D.

\begin{figure}[t]
\centerline{
\includegraphics[width=3.4in]{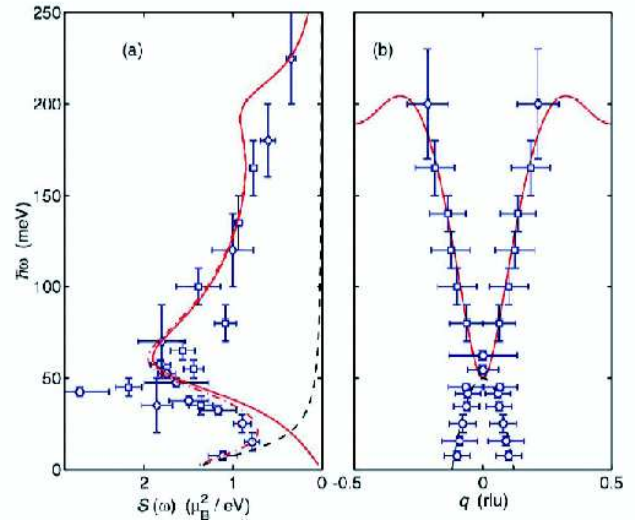}
}
\caption{
Neutron scattering from La$_{1.875}$Ba$_{0.125}$CuO$_4$ at 12~K $(> T_c)$
(from 
\Ref{TWP0434}).  Right panel shows the hourglass pattern
of the excitation spectrum (cf Fig. \ref{hourglass}).  Solid line is a fit to
a two-leg ladder spin model.  Left panel shows the momentum integrated
scattering intensity.  Dashed line is a Lorenztian fit to the rising intensity
at the incommensurate positions. Sharp peak at 40~meV could be a phonon.
 }
 \label{tranquada}
\end{figure}

Recently, neutron scattering has been extended to energies much above the
resonance.  It is found that very broad features disperse upward from the
resonance, resulting in the ``hourglass'' structure shown in
Fig.~\ref{hourglass} which was first proposed by 
\Ref{BSF0034}
\cite{HMD0431,SBL0402}.  Interestingly, there
has also been a significant evolution of the understanding of the neutron
scattering in the LSCO family.  For a long time it has been thought that the
LSCO family does not exhibit the resonance which shows up prominently below
$T_c$ in YBCO and other compounds.  However, neutron scattering does show a
broad peak around 50~meV which is temperature independent.  
\Ref{TWP0434} studied La$_{1.875}$Ba$_{0.125}$CuO$_4$ which exhibits static
charge and spin stripes below 50~K, and a greatly suppressed $T_c$.  Their
data also exhibits an ``hourglass''-type dispersion, remarkably similar to
that of underdoped YBCO.  In particular, the incommensurate scattering which
was previously believed to be dispersionless now exhibits downward dispersion
\cite{FGY0496}.  The same phenomenon is also seen in optimally
doped La$_{2-x}$Sr$_x$CuO$_4$ 
\cite{CMR0439}.  It is
remarkable that in these materials known to have static or dynamic stripes,
the incommensurate low energy excitations are not spin waves emanating  from $
\left( {\pi \over 2} \pm \delta \right) $ as one might have expected, but
instead are connected to the peak at $(\pi, \pi)$ in the hourglass fashion.
\Ref{TWP0434} fit the $\v{k}$ integrated intensity to a model
of a two-leg ladder.  It is not clear how unique this fit is because one may
expect high energy excitations to be relatively insensitive to details of the
model.  What is emerging though is a picture of a universal hourglass shaped
spectrum which is common to LSCO and YBCO families.  The high energy
excitations appear common while the major difference seems to be in the
re-arrangement of spectral weight at low energy.  In LBCO, significant weight
has been transferred to the low energy incommensurate scattering, as shown in
Fig.~\ref{tranquada}, and is associated with stripes.  In our view the
universality supports the picture that 
all the cuprates share the same short distance and high energy physics, which
include the pseudogap behavior.  Stripe formation is a competing state which
becomes prominent in the LSCO family, especially near $x = {1 \over 8}$, and
may dominate the low energy and low temperature (below 50~K) physics.  There
is a school of thought which holds the opposite view (see 
\Ref{CEK03}), that fluctuating stripes are responsible for the pseudogap
behavior and the appearance of superconductivity.  From this point of view the
same data have been interpreted as an indication that stripe fluctuations are
also important in the YBCO family 
\cite{TWP0434}.  Clearly,
this is a topic of much current debate.

\subsection{Quasiparticles in the superconducting state} In contrast with the
anomalous properties of the normal state, the low temperature properties of
the superconductor seem relatively normal.  There are two major differences
with conventional BCS superconductors, however.  First, due to the proximity
to the Mott insulator, the superfluid density of the superconductor is small,
and vanishes with decreasing hole concentration.  Second, because the pairing
is $d$-wave, the gap vanishes on four points on the Fermi surface (called gap
nodes), so that the quasiparticle excitations are gapless and affect the
physical properties even at the lowest temperatures.  We will focus on these
nodal quasiparticles in this sub-section.

The nodal quasiparticles clearly contribute to the thermal dynamical
quantities such as the specific heat.  Because their density of states  vanish
linearly in energy, they give rise to a $T^2$ term which dominates the low
temperature specific heat.  In practice, disorder rounds off the linear
density of states, giving instead an  $\alpha T + \beta T^3$  behavior.  An
interesting effect in the presence of a magnetic field
was proposed by 
\Ref{V9369}.  He argued that in the presence of a vector potential or superfluid flow,
the quasiparticle dispersion $E(\v{k}) = \sqrt{(\epsilon_{\v{k}}-\mu)^2 + \Delta_{\v k}^2}$ is shifted by
\be
E_{\v{A}}(\v{k}) = E(\v{k}) + \left(
{1\over 2e} \v{\nabla}\theta - \v{A}
\right) \cdot \v{j}_{\v k}
\label{Eq.3}
\en
where $\v{j}_{\v k}$ is the current carried by ``normal state'' quasiparticles with
momentum $\v{k}$ and is usually taken to be $-e {\partial \epsilon_{\v k} \over
\partial\v{k}}$.  Note that since the BCS quasiparticle is a superposition of
a particle and a hole, the charge is not a good quantum number.  However, the
particle component with momentum $\v{k}$ and the hole component with momentum
$-\v{k}$ each carry the same electrical current $\v{j}_{\v k} = -e {\partial
\epsilon_{\v k} \over \partial\v{k}}$ and it makes sense to consider this to be the
current carried by the quasiparticle.  Note that $\v{j}_{\v{k}}/e$ is very
different from the group velocity ${\partial E(\v{k}) \over \partial\v{k}}$.

In a magnetic field which exceeds $H_{c1}$, vortices enter the sample. The
superfluid flow $\v{\nabla}\theta \sim {2\pi \over r}$ where $r$ is the
distance to the vortex core.  On average, ${1 \over 2}|\v{\nabla}\theta|
\approx \pi/R$ where $R = (\phi_0/H)^{1/2}$ is the average spacing between
vortices and $\phi_0 =  hc/2e$ is the flux quantum.  Volovik then predicts a
shift of the quasiparticle spectra by $\approx ev_F (H/\phi_0)^{1/2}$ which in
turn gives a contribution to the specific heat proportional to $\sqrt{H}$.
This contribution has been observed experimentally
\cite{MBU9444}.

The quasiparticles contribute to the low temperature transport properties as
well. 
\Ref{L9387} considered the frequency-dependent conductivity
$\sigma(\omega)$ due to quasiparticle excitations.  In the low temperature
limit, he found that the low frequency limit of the conductivity is universal
in the sense that it does not depend on impurity strength, but only on the
ratio $v_F/v_\Delta$ where $v_\Delta$ is the velocity of the nodal
quasiparticle in the direction of the maximum gap $\Delta_0$, \ie 
$\sigma(\omega \rightarrow 0) = {e^2 \pi v_F \over h v_\Delta}$, if $v_F \gg
v_\Delta$.  This result was derived within the self-consistent $t$-matrix
approximation and can easily be understood as follows.  In the presence of
impurity scattering, the density of states at zero energy becomes finite.  At
the same time, the scattering rate is proportional to the self-consistent
density of states.  Since the conductivity is proportional to the density of
states and inversely to the scattering rate, the impurity dependence cancels.

\begin{figure}[t]
\centerline{
\includegraphics[width=3.5in]{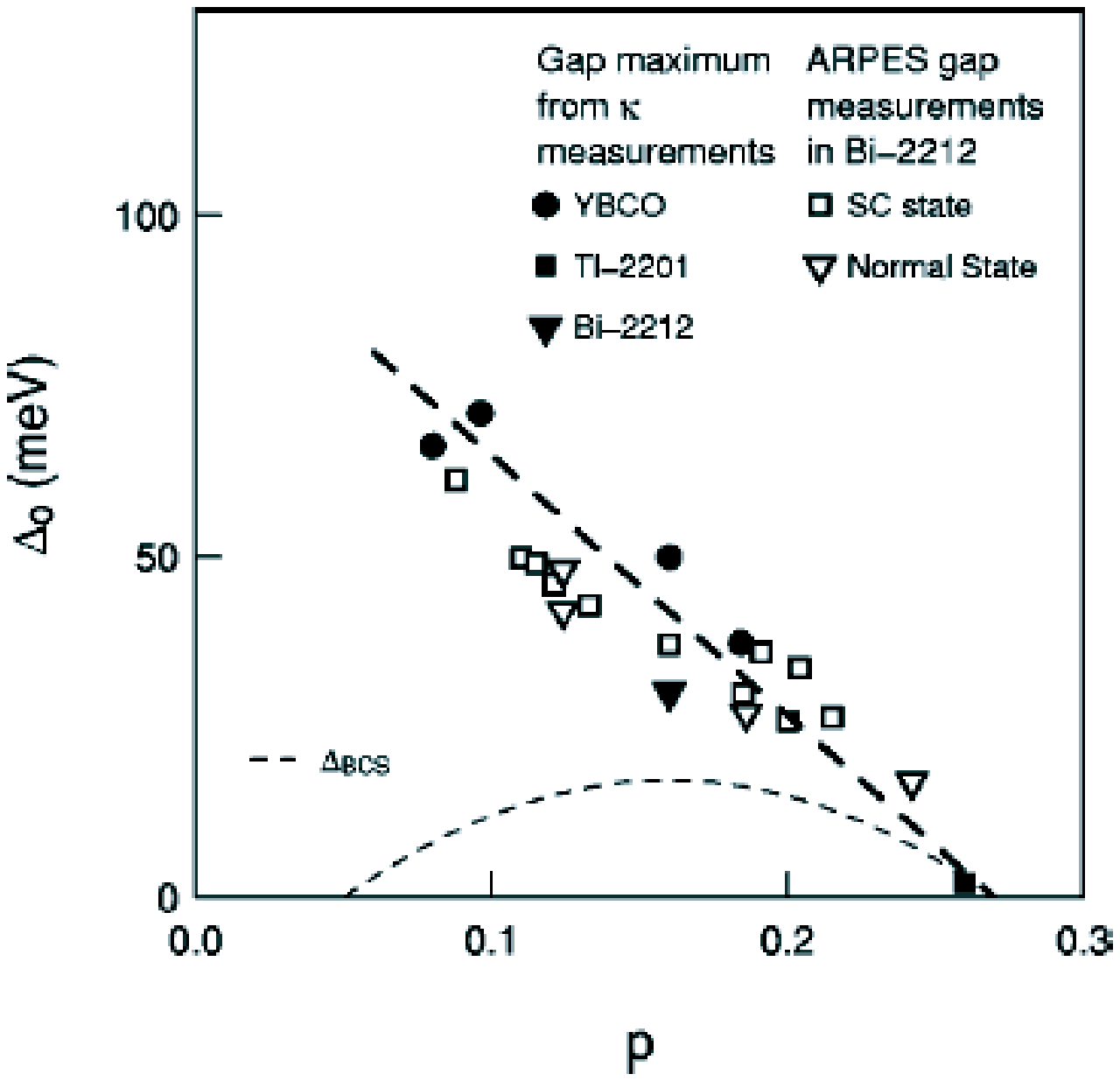}
}
\caption{
Figure from 
\Ref{SHH0320}. Doping dependence of the
superconducting gap $\Delta_0$ obtained from the quasiparticle velocity
$v_\Delta$ using eq. (\ref{Eq.4}) (filled symbols).  Here we assume $\Delta = \Delta_0
\cos2\phi$, so that $\Delta_0 = \hbar k_Fv_\Delta/2$, and we plot data for
YBCO alongside Bi-2212 
\cite{CHL0054} and Tl-2201 
\cite{PBH0203}.  For comparison, a BCS gap of the form $\Delta_{\bf BCS} = 2.14
k_BT_c$ is also plotted.  The value of the energy gap in Bi-2212, as
determined by ARPES, is shown as measured in the superconducting state
\cite{CDN9909} and the normal state 
\cite{NDR9857} (open symbols).  The thick dashed line is a guide to the eye.
 }
 \label{gap}
\end{figure} 

The frequency-dependent $\sigma(\omega)$ is difficult to measure and it was
realized that thermal conductivity $\kappa$ may provide a better test of the
theory because according to the Wiedemann-Franz law, $\kappa/T$ is
proportional to the conductivity and should be universal.  Unlike
$\sigma(\omega)$, thermal conductivity does not have a superfluid contribution
and can be measured at DC.  More detailed considerations by 
\Ref{DL0070}
 show that $\sigma(\omega)$ has two non-universal corrections: one due
to backscattering effects, which distinguishes the transport rate from the
impurity rate which enters the density of state; and a second one due to Fermi
liquid corrections.  On the other hand, these corrections do not exist for
thermal conductivity.  Consequently, the  Wiedemann-Franz law is violated, but
the thermal conductivity per layer is truly universal and is given by
\be
{\kappa \over T} = {k^2_B \over 3\hbar c}  \left(
{v_F \over v_\Delta }+ {v_\Delta \over v_F}
\right)
\label{Eq.4}
\en
We note that this result is obtained within the self-consistent $t$-matrix
approximation which is expected to break down if the impurity scattering is
strong, leading to localization effects.  The localization of nodal
quasiparticles is a complex subject.  Due to particle-hole mixing in the
superconductor, zero energy is a special point and quasiparticle localization
belongs to a different universality class 
\cite{SF9993}  from the standard ones.  Senthil and Fisher also pointed out
that since quasiparticles carry well defined spin, the Wiedemann-Franz law for
spin conductivity should hold and spin conductivity should be universal.  We
note that 
\Ref{DL0070} argued that Fermi liquid corrections enter the spin conductivity,
but we now believe their argument on this point is faulty.

Thermal conductivity has been measured to mK temperatures in a variety of YBCO
and BCCSO samples.  The universal nature of $\kappa/T$ has been demonstrated
by studying samples with different Zn  doping and showing that $\kappa/T$
extrapolates to the same constant at low temperatures 
\cite{TLG9783}
A magnetic field dependence analogous to the Volovik effect for the
specific heat has also been observed 
\cite{CHL0054}  Using
eq.~(\ref{Eq.4}), the experimental data can be used to extract the ratio
$v_F/v_\Delta$.  In the case of BCCSO where photoemission data for $v_F$ and
the energy gap is available, the extracted ratio $v_F/v_\Delta$ is in
excellent agreement with ARPES results, assuming a simple $d$-wave
extrapolation of the energy gap from the node to the maximum gap $\Delta_0$.
In particular, the trend that $\Delta_0$ increases with decreasing doping $x$
is directly observed as a decrease of $v_F/v_\Delta$ extracted from
$\kappa/T$.  A summary of the data is shown in Fig.~\ref{gap} 
\cite{SHH0320}.    Results of such systematic studies strongly support the
notion that in clean samples the nodal quasiparticles behave exactly as one
expects for well defined quasiparticles in a $d$-wave superconductor.  We
should add that in LSCO the ratio $v_F/v_\Delta$ extracted from $\kappa/T$
seems anomalously small, suggesting that strong disorder may be playing a role
here to invalidate eq.~(\ref{Eq.4}).

\begin{figure}[tb]
\centerline{
\includegraphics[width=3in]{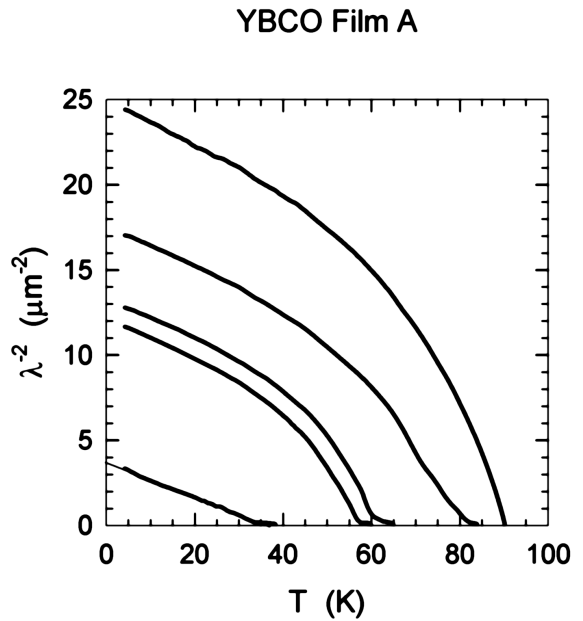}
}
\caption{
The London penetration depth measured in a series of YBCO film 
with different oxygen concentration and T$_c$'s.  The plot shows
$\lambda^{-2}$ plotted vs.  temperature.  Data provided by T.R. Lemberger and
published in 
\Ref{BSL0061}. 
}
\label{Lemberger}
\end{figure}

\Ref{LW9711}
pointed out that the nodal quasiparticles also manifest
themselves in the linear $T$ dependence of the superfluid density.  They
showed that by treating them as well defined quasiparticles in  the sense of
Landau, a general expression of the linear $T$ coefficient can be written
down, independent of the microscopic origin of the superconductivity.  We have
\be
{n_s(T) \over m}  = {n_s(0) \over m} - {2\ln 2 \over \pi} \alpha^2 \left( {v_F  \over v_\Delta}
\right) T
\label{Eq.5}
\en
The only assumption made is that the quasiparticles carry an electric current
\be
\v{j}(\v{k}) = -e \alpha \v{v}_F
\label{Eq.6}
\en
where $\alpha$ is a phenomenological Landau parameter which was left out in
the original Lee-Wen paper but added in by 
\Ref{MGI9842}.  While
the linear $T$ dependence is well known in the conventional BCS theory of a
$d$-wave superconductor, the same theory gives $n_s/m$ of order unity.  It is
therefore useful to write $n_s$ in this phenomenological way, and choose
$n_s(T=0)$ to be of order $x$ as we discussed in section III.A.  The key
question raised by eq.~(\ref{Eq.6}) is whether $\alpha$ depends on $x$ or not.  There
is experimental evidence that the linear $T$ coefficient of
$n_s(T)/m$ which is directly related to London penetration depth measurements,
is almost independent of $x$ for $x$ less than optimal doping.
Figure~\ref{Lemberger} shows data obtained for a series of thin films of YBCO
\cite{BSL0061,SIL0320}
The thin film data are in full agreement with earlier  but less extensive data
on bulk crystals 
\cite{Bo9695}.
However, we note that very recent data on severely underdoped YBCO crystal
($T_c<20$K)  show that $\frac{d(n_s/m)}{dT}$ is roughly linear in $T_c$
\cite{BTO04}.

Since $v_F/v_\Delta$ is known to go to a constant for small $x$ (and, indeed,
decreases with decreasing $x$), the independence of the linear $T$ term in
$n_s/m$ on $x$ means that $\alpha$ approaches a constant for small $x$.  By
combining with $v_F/v_\Delta$ extrapolated from thermal conductivity,
$\alpha^2$ has been estimated to be 0.5 (see 
\Ref{IM0259} for an
excellent summary).  This is an important result because it states that
despite the proximity to the Mott insulator, the nodal quasiparticles carry a
current which is similar to that of the tight-binding Fermi liquid band.  We
note that the simplest microscopic theory which gives correctly $n_s(T=0)$ to
be proportional to $x$ is the slave-boson  mean-field theory to be discussed
in section IX.B.  That theory predicts $\alpha$ to be proportional to $x$ and
the resulting $x^2T$ term is in strong disagreement with experiment.  The
search for a microscopic theory which gives correctly both $n_s(T=0)$ and the
linear $T$ term is one of the open problems that faces us today.

The unusual combination of a small $n_s(T=0)$ and a large linear $T$ reduction
due to quasiparticles has a number of immediate  consequences.  Simply by
extrapolating the linear $T$ dependence, we can conclude that $n_s$ vanishes
at the temperature scale proportional to $x$ and T$_c$ must be bound by it.
Furthermore, at $T_c$ the number of quasiparticles which are thermally excited
is still small, and not sufficient to close the gap as in standard BCS theory.
Thus the transition must not be thought of as a gap-closing transition, and
the effect of an energy gap must persist considerably above $T_c$.  This can
potentially explain at least part of the pseudogap phenomenon.  As we shall
see in the next section, when combined with phase fluctuations, the
quasiparticle excitations explain the magnitude of $T_c$ in the underdoped
cuprates and account for a wide phase fluctuation region above $T_c$, but not
the full pseudogap phenomenon.

The disconnect between the gap energy $\Delta_0$ and $kT_c$ introduces two
length scales, $\xi_0 = \hbar v_F/\Delta_0$ and $R_2 = \hbar v_F/kT_c$, where
$kT_c$ is proportional to $x$.  Around a vortex, the supercurrent induces a
population of quasiparticles by the Volovik effect, and in analogy to eq.
(\ref{Eq.5})
causes a reduction in $n_s$.  
\Ref{LW9711} show that at a radius of
$R_2$ the circulating supercurrent exceeds the critical current and inside
that radius the superconductor looses its phase stiffness.  They suggest that
the system becomes normal once the large core radius $R_2$ overlaps and
$H_{c2} \approx \phi_0/R_2^2$, in contrast with $H_{c2}^\ast \approx
\phi_0/\xi_0^2$ as in conventional BCS theory.  Note that $H_{c2}$ decreases
while $H_{c2}^\ast$ increases with underdoping.  Experimentally the resistive
transition to the normal state indeed takes place at an $H_{c2}$ which
decreases with decreasing $T_c$.  However, there are signs that vortices
survive above this magnetic field up to $H_{c2}^\ast$, as will be discussed in
section. V.B.

Finally, we comment on suggestions in the literature that classical
fluctuations of the superconducting phase can lead to a linear reduction of
$n_s$ at low temperatures 
\cite{CKE9912}.  Just as in the case of lattice displacements, such
fluctuations must be treated quantum mechanically at low temperatures (as
phonons in that case) to avoid the $3 k_B$ low temperature limit for the
specific heat.  In the case of phonons, the characteristic temperature scale
is the phonon frequency.  In the case of the superconductor, the phase mode is
pushed up to the plasma frequency by long-range  Coulomb interaction.
Nevertheless, due to the coupling to the low-lying particle-hole excitations,
the cross-over from classical to quantum fluctuations must be treated with
some care.  
\Ref{PRR0086}, \citeyear{P0221} and \Ref{BCC0113}
 have calculated that the cross-over happens at quite a high temperature
scale and we believe the low-temperature linear reduction of $n_s$ is entirely
due to thermal excitations of quasiparticles.
 
\section{Introduction to RVB and a simple explanation of the pseudogap} 

We explained in the last section that the N\'{e}el spin order is incompatible
with hole hopping.  The question is whether there is another arrangement of
the spin which achieves a better compromise between exchange energy and the
kinetic energy of the hole.  For $S={1\over 2}$ it appears possible to take
advantage of the special stability of the singlet state. The ground state of
two spins $S$ coupled with antiferromagnetic Heisenberg exchange is a spin
singlet with energy $-S(S+1)J$.  Compared with the classical large  spin
limit, we see that quantum mechanics provides an additional stability in the
term unity in $(S+1)$ and this contribution is strongest for $S={1\over 2}$.
Let us consider a one-dimensional spin chain.  A N\'{e}el ground state with
$S_z = \pm {1\over 2}$ gives an energy of $-{1\over 4}J$ per site.  On the
other hand, a simple trial wavefunction of singlet dimers already gives a
lower energy of $-{3\over 8}J$ per site.  This trial wavefunction breaks
translational symmetry and the exact ground state can be considered to be a
linear superposition of singlet pairs which are not limited to nearest
neighbors, resulting in a ground state energy of 0.443~J.  In a square and
cubic lattice the N\'{e}el energy is $-{1\over 2}J$ and $-\frac34J$ per site,
respectively, while the dimer variational energy stays at $-{3\over 8}J$.  It
is clear that in a 3D cubic lattice, the N\'{e}el state is a far superior
starting  point, and in two dimensions the singlet state may present a serious
competition.  Historically, the notion of a linear superposition of spin
singlet pairs spanning different ranges, called the resonating valence bond
(RVB), was introduced by 
\Ref{A7353} and 
\Ref{FA7432} as a
possible ground state for the $S={1\over 2}$ antiferromagnetic Heisenberg
model on a triangular lattice.  The triangular lattice is of special interest
because an Ising-like ordering of the spins is frustrated.  Subsequently, it
was decided that the ground state forms a $\sqrt{3} \times \sqrt{3}$
superlattice where the moments lie on the same plane and form $120^\circ$
angles between neighboring sites 
\cite{HE8831}.  Up to now there is no known spin Hamiltonian with full $S(U2)$
spin rotational symmetry outside of one dimension which is known to have an
RVB ground state.  However, see section X.H for examples which either violate
spin rotation or which permit charge fluctuations.

The N\'{e}el state has long range order of the staggered magnetization and an
infinite degeneracy of ground states leading to Goldstone modes which are
magnons.  In contrast, the RVB state is a unique singlet ground state with
either short range or power law decay of antiferromagnetic order.  This state
of affairs is sometimes referred to as a spin liquid.  However, the term spin
liquid is often used more generally to denote any kind of short range or power
law decay, \ie the absence of long range order, even when the unit cell is
doubled, either spontaneously or explicitly. For example, the ladder system
has two states per unit cell and in the limit of strong coupling across the
rung, the ground state is naturally a spin singlet with short range
antiferromagnetic order.  Another example is the spontaneously dimerized
ground state for the frustrated spin chains when the next-nearest neighbors
exchange $J^\prime$ is sufficiently large.  This kind of ground state is more
properly called a valence band solid and is smoothly connected to spin
singlet ground states often observed for systems with an even number of
electrons per unit cell, the extreme example being Si.  Thus we think it is
better to reserve the term spin liquid to cases where there is an odd number
of electrons per unit cell.

Soon after the discovery of high T$_c$ superconductors, 
\Ref{A8796}
revived the RVB idea and proposed that with the introduction of holes the
N\'{e}el state is destroyed and the spins form a superposition of singlets.
The vacancy can hop in the background of what he envisioned as a liquid of
singlets and a better compromise between the hole kinetic energy and the spin
exchange energy may be achieved.  Many elaborations of this idea followed,
but here we argue that the basic physical picture described above gives a
simple account of the pseudogap phenomenon.  The singlet formation explains
the decrease of the uniform spin susceptibility and the reduction of the
specific heat $\gamma$. The vacancies are responsible for transport in the
plane.  The conductivity spectral weight in the $ab$ plane is given by the
hole concentration $x$ and is unaffected by the singlet formation.  On the
other hand, for $c$-axis conductivity, an electron is transported between
planes.  Since an electron  carries spin ${1\over 2}$, it is necessary to
break a singlet.  This explains the gap formation in $\sigma_c(\omega)$ and
the energy scale of this gap should be correlated with that of the uniform
susceptibility.  In photoemission, an electron leaves the solid and reaches
the detector, the pull back of the leading edge simply reflects the energy
cost to break a singlet.

A second concept associated with the RVB idea is the notion of spinons and
holons, and spin charge separations.  Anderson  postulated that the spin
excitations in an RVB state are $S={1\over 2}$ fermions which he called
spinons.  This is in contrast with excitations in a N\'{e}el state which are
$S = 1$ magnons or $S = 0$ gapped singlet excitations.

 \begin{figure}[t]
\centerline{
\includegraphics[width=3in]{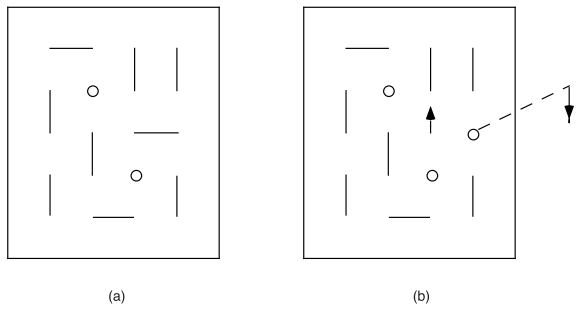}
}
\caption{
A cartoon representation of the RVB liquid or singlets.  Solid bond represents a spin singlet
configuration and circle represents a vacancy.  In (b) an electron is removed from the plane in
photoemission or $c$-axis conductivity experiment.  This necessitates the breaking of a singlet. }
 \label{RVB}
\end{figure}

Initially the spinons are  suggested to form a Fermi surface, with Fermi
volume equal to that of $1-x$ fermions.  Later it was proposed that the Fermi
surface is gapped to form $d$-wave type structure, with maximum gap near
$(0,\pi)$.  This $\v{k}$ dependence of the energy gap is needed to explain the
momentum dependence observed in photoemission.

The concept of spinons is a familiar one in one-dimensional spin chains where
they are well understood to be domain walls.  In two dimensions the concept is
a novel one  which does not involve domain walls.  Instead, a rough physical
picture is as follows.  If we assume a background of short range singlet
bonds, forming the so-called short-range RVB state, a cartoon of the spinon is
shown in Fig. \ref{RVB}.  If the singlet bonds are ``liquid,'' two $S={1\over
2}$ formed by breaking a single bond can drift apart, with the liquid of
singlet bonds filling in the space between them.  They behave as free
particles and are called spinons.  The concept of holons follows naturally
\cite{KRS8765}
as the vacancy left over by removing a
spinon.  A holon carries charge $e$ but no spin.

\section{Phase fluctuation vs. competing order} 

One of the hallmarks of doping
a Mott insulator is that the spectral weight of the frequency dependent
conductivity $\sigma(\omega)$ should go to zero in the limit of small doping.
Indeed, $\sigma(\omega)$ shows a Drude-like peak at low frequencies and its
area was shown to be proportional to the hole concentration 
(\cit{OTM9042}; \cit{CRK9333}; \cit{UIT9142}; \cit{PLD04}).
Results from exact diagonalization of small samples  are
consistent with a Drude weight of order $xt$ 
\cite{DMO9241}.
When the metal becomes superconducting, all the spectral weight collapses into
a $\delta$-function if the sample is in the clean limit. The London
penetration depth for field penetration perpendicular to the $ab$ plane is
given by
\be
\lambda_\perp^{-2}
= {4\pi n_s^{3d}e^2\over m^\ast c^2} ,
\label{Eq.7}
\en
where $n_s^{3d}/m^\ast$ is the spectral weight and $n_s^{3d}$ is the $3d$
superfluid charge density.  As an example, if we take $\lambda_\perp =
1600$~Angstrom for YBa$_2$Cu$_3$O$_{6.9}$,  and take $n_s^{3d}$ to be the hole
density, we find from eq. (\ref{Eq.7}) $m^\ast \approx 2m_e$ which corresponds
to an effective hopping $t^\ast = {1\over 3} t$.  The notion that
$\lambda_\perp^{-2}$ is proportional to $xt$ is also predicted by slave-boson
theory, as will be discussed in section IX.B.

 \begin{figure}[t]
\centerline{
\includegraphics[width=3.3in]{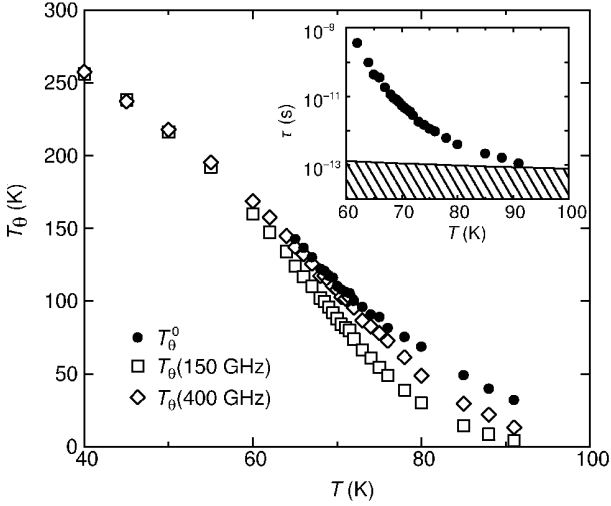}
}
\caption{
The phase stiffness $T_\theta$ measured at different frequencies $(T_\theta =
\hbar^2 n_s/m^\ast)$.  The solid dots give the bare stiffness obtained by
extrapolation to infinite frequency.  $T_c$ of this sample is 74~K.This is
where the phase stiffness measured at low frequency would vanish according to
BKT theory.  Note the linear decrease of the bare stiffness with $T$ which
extends considerably above $T_c$.  This decrease is due to thermal excitations
of nodal quasiparticles.  Inset shows the time scale of the phase fluctuation.
Hatched region denotes ${\hbar \over \tau} = kT_c$. From 
\Ref{CMO9921}. 
}
\label{Corson}
\end{figure}

\Ref{ULS8917}
discovered empirically a linear relation between
$\lambda_\perp^{-2}$ measured by $\mu$SR and the superconducting T$_c$.   He
interpreted this relation as indicative of Bose condensation of holes, since in
two dimensions the Bose-Einstein condensation temeprature is proportional to
the areal density.  Since $\lambda_{\perp}^{-2}$ is proportional to the 3d
density, in principle, some adjustment for the layer spacing should be made.
Furthermore, $\lambda_{\perp}^{-2}$ is highly sensitive to disorder, and it is
now known that in many systems, not all the spectral weight collapses to the
$\delta$-function, \ie  some residual normal conductivity is left, presumably
due to inhomogeneity 
\cite{BPH9465,COO0069}.
Thus the Uemura plot should be viewed as providing a qualitative trend, rather
than a quantitative relation.  Nevertheless, it is important in that it draws
a relationship between T$_c$ and carrier density.

\begin{figure}[t]
\centerline{
\includegraphics[width=2.5in]{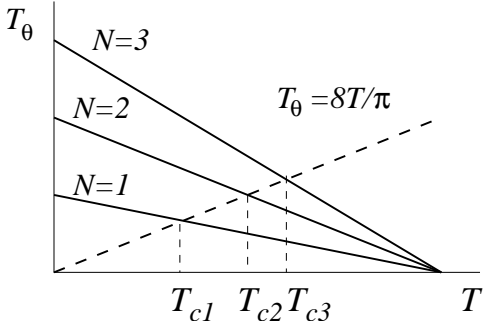}
}
\vspace{0.5cm}
\caption{ 
Schematic plot of the phase stiffness $T_\theta = \hbar^2 n_s/m^\ast$ for superconductors with $N$ coupled layers.  The linear decrease with temperature is due to the thermal excitation of quasiparticles.  The transition temperatures $T_{cN}, N=1, 2, 3$ are estimated by the interception with the BKT line $T_\theta = 8 T/\pi$.
}
\label{Tc}
\end{figure}

\subsection{A theory of $T_c$} 

The next important step was taken by 
\Ref{EK9534}, who noted that it is the superfluid density which controls
the phase stiffness of the superconducting order parameter $\Delta =
|\Delta|e^\theta$, \ie the energy density cost of a phase twist is
\be
H = {1\over 2} K_s^0 (\nabla\theta)^2
\label{Eq.8}
\en
Here the superscript on $K^0_s$ denotes the bare stiffness on a short distance
scale.  For two-dimensional layers the stiffness $K_s =
\hbar^2(n_s/2)/2m^\ast$, \ie the kinetic energy of Cooper pairs.  The
spectral weight $n_s/m^\ast$ and the stiffness are given by
\be
K_s = {1 \over 4}{\hbar^2n_s \over m^\ast}
= {1\over 4} {\hbar^2n_s^{3d}c_0 \over m^\ast} 
\label{Eq.9}
\en
where $c_0$ is the spacing between the layers 
and using Eqs. (\ref{Eq.7}) and (\ref{Eq.9}), can be directly measured in terms of
$\lambda_\perp$.  If $K_s^0$ is small due to the proximity to the Mott
insulator, then phase fluctuation is strong and the T$_c$ in the underdoped
cuprates may be governed by phase fluctuations.  The theory of phase
fluctuations in two dimensions is well understood due to the work of
\Ref{B7144}
and 
\Ref{KT7381}.  The BKT transition is
described by the thermal unbinding of vortex anti-vortex pairs.  The energy of
a single vortex is given by
\be
E_\text{vortex} = E_c + 2\pi K_s^0 \ln (L/\xi_0)
\label{Eq.10}
\en
where $L$ is the sample size, $\xi_0$ is the BCS coherence length which serves
as a short distance cut-off, and $E_c$ is the core energy.  For vortex
anti-vortex pairs, the sample size $L$ is replaced by the separation of the
pairs.  The vortex unbinding transition is driven by the balance between this
energy and the entropy which also scales logarithmically with the vortex
separation.  At T$_c$, $K_s$ is predicted to jump between zero and a finite
value $K_s(T_c)$ given by a universal relation 
\be
kT_c = (\pi/2)K_s(T_c) =
{\pi \over 8} {\hbar^2 n_s \over m^\ast}
\label{Eq.11}
\en
\cite{NK7701}.  The precise value of T$_c$ depends on $K_s(T=0)$ and weakly on
the core energy.  In  the limit of very large core energy, $kT_c = 1.5 K_s^0$,
whereas for an  XY model on a square lattice 
$E_c$ is basically zero if $\xi_0$ in eq. (\ref{Eq.10}) is replaced by the lattice
constant and $T_c = 0.95 K_s^0$.  Thus $K_s^0$ should give a reasonable guide
to T$_c$ in the phase fluctuation scenario.  Emery and Kivelson estimated
$K^0_s$ from $\lambda_\perp$ data for a variety of materials and concluded that
$K_s^0$ is indeed on the scale of T$_c$.  However, they assumed that each
layer is fluctuating independently, even for systems with strongly Josephson
coupled bi-layers. Subsequent work using microwave conductivity has confirmed
the BKT nature of the phase transition, but concluded that in BSCCO, it is the
bi-layer which should be considered as a unit, \ie the superconducting phase
is strongly correlated between the two layers of a bi-layer 
\cite{CMO9921}.  This increases the $K_s^0$ estimate by a factor of 2.  For
example, for $\lambda_\perp = 1600$ Angstroms, Emery and Kivelson quoted
$K_s^0$ to be 145~K for YBCO.  This should really be replaced by 290~K, a
factor of 3 higher than T$_c$.

We can get around this difficulty by realizing that $K_s$ is reduced by
thermal excitation of quasiparticles and the bare $K_s^0$ in the BKT  theory
should include this effect.  In Section III.B we showed empirical evidence
that the linear $T$ coefficient of $n_s(T)$ is relatively independent of $x$.
The bare $K_s^0$ is measured as the high frequency limit in a microwave
experiment 
\cite{CMO9921}.  As seen in Fig. \ref{Corson}, the
bare phase stiffness $T_\theta^0 \equiv \hbar^2 n_s^0/m^\ast$ continues to
decrease linearly with $T$ above T$_c$ = 74~K.  Given the universal relation
eq.~(\ref{Eq.11}), an estimate of T$_c$ can be obtained by the interception of
the straight line $T_\theta = (8/\pi)kT$ with the bare stiffness.  This yields
an estimate of the BKT transition temperature of $\approx 60$~K.  The somewhat
higher actual T$_c$ of 74~K is due to three dimensional ordering effects
between bi-layers.  Now we can extend this procedure to a multi-layer
superconductor.  In Fig.\ref{Tc} we show schematically $T_\theta^0 = \hbar^2
n_s^0/m^\ast$ plot of single-layer, bi-layer and tri-layer systems $(N=1,2,3)$
assuming that the layers are identical.  We expect $n_s^0(T=0)$, which is the
areal density per $N$ layers, to scale linearly with $N$.  On the other hand,
the linear $T$ slope also scales with $N$, because the number of thermally
excited quasiparticles per area scale with $N$.  The extrapolated ``T$_c$'s''
are therefore the same.  Now we may estimate $T_c(N)$ from the interception of
the line $T_\theta^0 = (8/\pi)kT$.  We see that $T_c$ increases monotonically
with $N$, but much slower than linear.  This trend is in agreement with what
is seen experimentally, notably in the Tl and Hg compounds.  As $N$ increases
further, the assumption that the layers are identical breaks down as the
charge density of each layer begins to differ.  We therefore conclude that the
combination of phase fluctuations and the thermal excitation of $d$-wave
quasiparticles can account for $T_c$ in underdoped cuprates, including the
qualitative trend as a function of the number of layers within a unit cell.

 This theory of $T_c$ receives confirmation from measurement of the oxygen
isotope effect of $T_c$ and on the penetration depth.  It is found that there
is substantial isotope effect on the $n_s/m^\ast$ for both underdoped and
optimally doped YBCO films.
On the other hand, there is significant isotope effect on
$T_c$ in underdoped YBCO 
\cite{KSC0317}, but no effect on
optimally doped samples 
\cite{Ko0402}.  Setting aside the
origin of the isotope effect on $n/m^\ast$, the remarkable doping dependence
of the isotope effect on $T_c$ is readily explained in our theory, since $T_c$
is controlled by $n_s/m^\ast$ in the underdoped but not in the overdoped
region.  In fact, a more detailed examination of the data for two underdoped
samples show that $n_s(T)/m^\ast$ appears to be shifted down by a constant
when O$^{16}$ is replaced by O$^{18}$.  This suggests that there is no isotope
effect on the temperature dependent term in eq.~\ref{Eq.5} which depends on
$v_F$.  This is consistent with direct ARPES measurements 
\cite{GSZ0487}.  Thus the data is consistent with an isotope effect only on
the zero temperature spectral weight $n_s(0)/m^\ast$.  The latter is a
complicated many body property of the ground state which is not simply related
to the effective mass of the quasiparticles in the naive manner.

\subsection{Cheap vortices and the Nernst effect} 

\Ref{EK9534},
also suggested that the notion of strong phase fluctuations may provide an
explanation of the pseudogap phenomenon.  They proposed that the pairing
amplitude is formed at a temperature $T_{MF}$ which is much higher than T$_c$
and the region between $T_{MF}$ and T$_c$ is characterized by robust pairing
amplitude and energy gap.  

This leaves open the microscopic origin of the robust pairing amplitude and
high $T_{MF}$ but we shall argue that even as phenomenology, phase
fluctuations alone cannot be the full explanation of the pseudogap.  Since
T$_c$ is driven by the unbinding of vortices, let us examine the vortex energy
more carefully.  As an extreme example, let us suppose $T_{MF}$ is described
by the standard BCS theory.  The vortex core energy in BCS theory is estimated
as $E_c \approx {\Delta_0^2 \over E_Fa^2} \xi_0^2$ where $\Delta_0$ is the
energy gap, $\Delta_0^2/(E_Fa^2)$ is the condensation energy per area, and
$\xi_0^2$ is the core size.  Using $\xi_o = {v_F \over \Delta_0}$, we conclude
that $E_c \approx E_F$ in BCS theory, an enormous energy compared with T$_c$.
Even if we assume $E_c$ to be of order of the exchange energy $J$ or the mean
field energy $T_{MF}$, it is still much larger than T$_c$.  We already note
that in BKT theory, T$_c$ is relatively insensitive to the core energy. Now we
emphasize that despite the insensitivity of $T_c$ to $E_c$, the physical
properties above $T_c$ are very sensitive to the core energy. This is because
BKT theory is an asymptotic long-distance theory which becomes simple in the
limit of dilute vortex or large $E_c$.  The typical vortex spacing which is
$n_v^{-{1\over 2}}$ where the vortex density $n_v$ goes as $e^{-E_c/kT}$.
Vortex unbinding happens on a renormalized length scale, \ie the typical
spacing between {\em free} vortices, which is much larger than $n_V^{-{1\over
2}}$.  As a result, the physics of the system above T$_c$ is very sensitive to
$E_c$.  If $E_c \gg kT_c$, vortices are dilute and the system will behave like
a superconductor for all measurements performed on a reasonable spatial or
temporal scale.  However, except for the close vicinity of  T$_c$, the
pseudogap region is not characterized by strong superconducting fluctuations,
but rather behaves like a metal.  Thus a large vortex core energy can be ruled
out.  The core energy must be small, of the order T$_c$, \ie it is comparable
to the second term in eq.~(\ref{Eq.10}).  The notion of ``cheap'' vortices has
two important consequences.  First, it is clear that the amplitude fluctuation
and phase fluctuation are controlled by the same energy scale, $kT_c$.  This
is because the vortex core is a region where the pairing amplitude vanishes
and, in addition, the phase $\theta$ winds by $2\pi$.  If we do away with the
phase winding and retain the amplitude fluctuation, this should cost even less
energy.  Thus the temperature scale where vortices proliferate is also the
scale where amplitude fluctuation proliferates.  Then the notion of strong
phase fluctuations is applicable only on a temperature scale of say 2~T$_c$
and this scale must become small as $x$ becomes small.  Thus phase fluctuation
cannot explain a pseudogap phenomenon which extends to finite $T$ in the small
$x$ limit.

Second, the notion of a cheap vortex means that there is a non-superconducting
state which is very close in energy.  In an ordinary superconductor, the core
can be thought of as a  patch of normal metals with a finite density of states
at the Fermi level.  The reason the core energy is large is because the energy
gained by opening up an energy gap is lost.  In underdoped and in slightly
overdoped cuprates there is experimental evidence from STM tunneling into the
core that the energy gap is retained inside the core 
\cite{MRE9554,PHG0036}.  The large peak in the density state predicted for
$d$-wave BCS theory 
\cite{WM9576} is simply not there.  The nature of the state in the core, which
one can think of as a competing state to the superconductor, is highly
nontrivial and is a topic of current debate.

\begin{figure}
\centerline{
\includegraphics[scale=1.2]{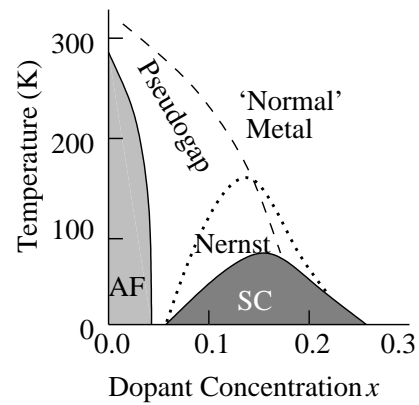}
}
\caption{
Schematic phase diagram showing the phase fluctuation regime where the Nernst effect is large. 
Note that this regime is a small part of the pseudogap region for small doping. }
\label{Nernst}
\end{figure}

The above discussion is summarized by a schematic phase diagram shown in Fig.
\ref{Nernst}.  A temperature scale of about 2~T$_c$ in the underdoped region
marks the range of phase fluctuation.  This is the region where the picture
envisioned by 
\Ref{EK9534} may be valid.  Here the phase is
locally well defined and vortices are identifiable objects.  Indeed, this is
the region where a large Nernst effect has been measured 
\cite{WYK0119,WOX0203,WOO0386}.
The Nernst effect is the voltage transverse to a thermal
gradient in the presence of a magnetic field perpendicular to the plane.  It
is exquisitely sensitive to the presence of vortices, because vortices drift
along the thermal gradient and produce the phase winding which supports a
transverse voltage by the Josephson effect.  A large Nernst signal has been
taken to be strong evidence for the presence of well-defined vortices above
T$_c$ 
\cite{WYK0119,WOX0203,WOO0386}.   At higher temperatures,
vortices overlap and the Nernst signal smoothly crosses over to that
describable by Gaussian fluctuation of superconducting amplitude and phase
\cite{USH0201}.  Very recently, the identification of the
Nernst region with fluctuating superconductivity was confirmed by the
observation of diamagnetic fluctuations which persist up to the same
temperature as the onset of the Nernst signal 
\cite{WLN04}.

It remains necessary to explain  why the resistivity looks metallic-like in
this temperature range and does not show the strong magnetic field dependence
one ordinarily expects for flux flow resistivity in the presence of thermally
excited vortices. The explanation may lie in the breakdown of the standard
Bardeen-Stephen model of flux flow resistivity.  Here the vortices have
anomalously low dissipation because in contrast to BCS superconductors, there
are no states inside the core to dissipate.  
\Ref{IM0213}
proposed
that the vortices are fast and yield a large flux flow resistivity.  In the
two fluid model, the conductivity is the sum of the flux flow conductivity
(the superfluid part) and the quasiparticle conductivity (the normal part).
The small flux flow conductivity is quickly shorted out by the nodal
quasiparticle contributions, and the system behaves like a metal, but with
carriers only in the nodal region.  This is also reminiscent of the Fermi arc
picture.  Unfortunately, a more detailed modeling requires an understanding
of the state inside the large core radius $R_2$ introduced in section III.C
which is not available up to now.

Instead of generating vortices thermally, one can also generate them by
applying a magnetic field.  
\Ref{WOO0386} have applied fields up to
45~T and found evidence that the Nernst signal remains large beyond that field
in the underdoped samples.  They estimate that the field needed to suppress
the Nernst signal to be of order $H^\ast_{c2} \approx \phi_0/\xi_0^2$ where
$\phi_0 \approx \hbar v_F/\Delta_0$.  This is the core size consistent with
what is reported by STM tunneling experiments.  At the same time, the field
needed for a resistive transition is much lower.  Recently 
\Ref{SHH04}
showed that in YBCO$_{6.35}$ superconductivity is destroyed 
by annealing or by applying a modest magnetic field.
Beyond this point the material is a thermal metal, with a
thermal conductivity which is unchanged from the superconducting side, where
it is presumably due to nodal quasiparticles and described by
eq.~(\ref{Eq.4}). Thus this field induced metal may be coexisting with pairing
amplitude and may be a very interesting new metallic state.

What is the nature of the gapped state inside the vortex core as revealed by
STM tunneling and how is it related to the pseudogap region?  A popular
notion is that the vortex core state is characterized by a competing order.  A
variety of competing order has been proposed in the literature.  An early
suggestion was that the core has antiferromagnetic order and an explicit model
was constructed based on the $SU(5)$ model of 
\Ref{Z9789}
\cite{ABK9771}.  However, this particular version has been criticized for its
failure to take into account the strong Coulomb repulsion and the proximity to
the Mott insulator 
(\cit{G9798}; \cit{BA9880}).  Recently,
more phenomenological version based on Landau theory have been proposed
(\cit{DSZ0102}; \cit{CCA0416}) where the
antiferromagnetism may be incommensurate.  As the temperature is raised into
the pseudogap regime, vortices proliferate and their cores overlap and,
according to this view, the pseudogap is characterized by fluctuating
competing order.  The dynamic stripe picture 
\cite{CEK03} is an example of this point of view.  Another proposal for
competing order is for orbital currents 
(\cit{V9754}; \cit{CLM0203}).  In this case the competing order is proposed to persist
in the pseudogap region but is ``hidden'' from detection because of
difficulties of coupling to the order.  Finally, as discussed earlier, the
recent observation of checkerboard patterns in the vortex core and in some
underdoped cuprates has inspired various proposals of charge density ordering. 

Most of the proposals for competing order are phenomenological in nature.  For
example, the proximity of $d$-wave superconductivity to antiferromagnetism is
simply assumed as an experimental fact.  However, from a microscopic point of
view, the surprise is that $d$-wave superconductivity turns out to be the
winner of this competition.  Our goal is a microscopic explanation of both the
superconducting and the pseudogap states.  We shall give a detailed proposal
for the vortex core in section XII.C.  Here we mention that while our proposal
also calls for slowly varying orbital currents in the core, this fluctuating
order is simply one manifestation of a quantum state.  For example, enhanced
antiferromagnetic fluctuation is another manifestation.  As discussed in
section VI.C, this picture is fundamentally different from competing states
described by Landau theory.  In the pseudogap phase, vortices proliferate and
overlap and all orders become very sort range.  Apart from characterizing this
state as a spin liquid (or RVB), the only possibility of order is a subtle
one, called topological/quantum order.  These concepts are described in
section X and a possible experimental consequence is described in section
XII.E.

\subsection{Two kinds of pseudogaps} Since the pseudogap is fundamentally a
cross-over phenomenon, there is a lot of confusion about the size of the
pseudogap and the temperature scale where it is observed.  Upon surveying the
experimental literature, it seems to us that we should distinguish between two
kinds of pseudogaps.  The first is clearly due to superconducting
fluctuations.  The energy scale of the pseudogap 
is the same as the low temperature superconducting gap
and it extends over a surprisingly large range of temperatures above $T_c$.
This is what we called the Nernst region in the last section.  This kind of
pseudogap has been observed in STM tunneling, where it is found that a
reduction of the density of states persists above $T_c$ even in overdoped
samples 
\cite{KFR0111}.  We believe the pull-back of the leading
edge observed in ARPES shown in Fig.~\ref{edge}(a) should be understood along these
lines.  There is another kind of pseudogap which is associated with singlet
formation.  A clear signature of this phenomenon is the downturn in uniform
spin susceptibility shown in Figs. \ref{Knight} and \ref{chi}.  The
temperature scale of the onset is high and increases up to 300 to 400~K with
underdoping.  The energy scale associated with this pseudogap is also very
large, and can extend up to 100~meV or beyond.  For example, the onset of the
reduction of the $c$-axis conductivity (which one may interpret as twice the
gap) has been reported to exceed 1000~cm$^{-1}$.  This is also the energy
scale one associates with the limiting STM tunneling spectrum observed in
highly underdoped Bi-2212 (Fig. \ref{STM}(f)) and in Na doped
Ca$_2$CuO$_2$Cl$_2$ 
\cite{HLK04}.  The gap in these spectra
is very broad and ill defined.  In the ARPES literature it is described as the
``high-energy pseudogap'' (see 
\Ref{DHS0373}) or the ``hump''
energy.  These spectra evolve smoothly into that  of the insulating parent.
This is most clearly demonstrated in Na-doped Ca$_2$CuO$_2$Cl$_2$ and the
ARPES spectrum near the antinodal point looks remarkably similar to that seen
by STM 
\cite{RSK0301}.  Examples of this kind of a spectrum can
be seen in the samples UD46 and UD30 shown in Fig. \ref{ARPES}(a).  In
contrast to the low energy pseudogap, a coherent quasiparticle peak is never
seen at these very high energies when the system enters the superconducting
state.  Instead, weak peaks may appear at lower energies, but judging from the
STM data, these may be associated with inhomogeneity.
In this connection, we point out that the often quoted $T^\ast$ line shown in
Fig. \ref{edge} is actually a combination of the two kinds of pseudogaps. The
solid triangles marking the onset of the leading edge refer to the fluctuating
superconductor gap, while the solid squares are lower bounds based on the
observation of the ``hump.''  Another example of this difference is that in
LSCO, the superconducting gap is believed to be much smaller and the pull back
of the leading edge is not seen by ARPES.  On the other hand, the singlet
formation is clearly seen in Fig. \ref{Knight}(b) and the broad hump-like
spectra is also seen by ARPES 
\cite{ZYL0481}.

We note that in contrast to superconducting fluctuations which extend across
the entire doping range but are substantially reduced for overdoped samples,
the onset of singlet formation seems to end rather abruptly near optimal
doping.  The Knight shift is basically temperature independent just above
$T_c$ in optimally doped and certainly in slightly overdoped samples 
(\cit{THS9350}; \cit{HBB9348}).  For this reason, we propose
that the pseudogap line and the Nernst line may cross in the vicinity of
optimal doping, as sketched in Fig. \ref{Nernst}.  In this connection it is
interesting to note that the pseudogap has also been seen inside the vortex
core 
\cite{MRE9554,PHG0036}.  By
definition, this is where the superconducting amplitude is suppressed to zero
and the gap is surely not associated with the pairing amplitude.  We have
argued that the gap offers a glimpse of the state which lies behind the
pseudogap associated with singlet formation.  It is interesting to note that
the gap in the vortex core has been reported in a somewhat overdoped sample
\cite{HKR0101}.  It is as though at zero temperature the
state with a gap in the core is energetically favorable compared with the
normal metallic state up to quite high doping.  It will be interesting to
extend these measurements to even more highly overdoped samples to see when
the gap in the vortex core finally fills in.  At the same time, it will be
interesting to extend the tunneling into the vortex core in overdoped samples
to higher temperatures, to see if the gap will fill in at some temperature
below $T_c$.

\section{Projected trial wavefunctions and other numerical results} 

In the original RVB article, \Ref{A8796} 
proposed a projected trial wavefunction as a description of the RVB state.
\be
\Psi = P_G|\psi_0\rangle
\label{Eq.12}
\en
where $P_G = \prod_{\v i}(1-n_{\v i\up}n_{\v i\down})$ is called the Gutzwiller
projection operator.  It has the effect of suppressing all amplitudes in
$|\psi_0\>$ with double occupation, thereby enforcing the constant of the $t$-$J$
model exactly.  The unprojected wavefunction contains variational parameters
and its choice is guided by mean-field theory (see section XIII).  The full
motivation for the choice of $|\psi_0\>$ becomes clear only after the discussion
of mean-field theory, but we discuss the projected wavefunction first because
the results are concrete and the concepts are simple. The projection operator
is too complicated to be treated analytically, but properties of the trial
wavefunction can be evaluated using Monte Carlo sampling.

\subsection{The half-filled case} 

We shall first discuss the half-filled case,
where the problem reduces to the Heisenberg model.  While the original
proposal was for $|\psi_0\>$ to be the $s$-wave BCS wavefunction, it was soon
found that the $d$-wave BCS state is a better trial wavefunction, \ie
consider
\begin{align}
H_d & = 
-\sum_{\langle \v i\v j \rangle ,\sigma}  \left(
\chi_{\v i\v j} f_{\v i\sigma}^\dagger f_{\v i\sigma} + c.c.  \right)-
\sum_{\v i,\sigma} 
\mu f_{\v i\sigma}^\dagger f_{\v i\sigma} +
\nonumber\\
 & +\sum_{\langle \v i\v j \rangle} 
\left[\Delta_{\v i\v j} \left( f_{\v i\up}^\dagger f_{\v j\down}^\dagger -
f_{\v i\down}^\dagger f_{\v j\up}^\dagger \right) + c.c.\right]
\label{Eq.13}
\end{align}
where $\chi_{\v i\v j} = \chi_0$ for nearest neighbors, and $\Delta_{\v i\v j} = \Delta_0$
for $\v j =\v i + \hat{x}$ and $-\Delta_0$ for $j = i + \hat{y}$.  The eigenvalues
are the well known BCS spectrum
\be
E_{\v k} = \sqrt{(\epsilon_{\v{k}}-\mu)^2 + \Delta_{\v{k}}^2}
\label{Eq.14} 
\en
where 
\be
\epsilon_{\v{k}} = -2 \chi_0 \left(
\cos k_x + \cos k_y
\right)
\label{Eq.15}
\en
\be
\Delta_{\v{k}} = 2 \Delta_0 \left(
\cos k_x - \cos k_y
\right)
\label{Eq.16}
\en
At half filling, $\mu = 0$ and $|\psi_0\>$ is the usual BCS wavefunction 
$|\psi_0\>
= \prod_{\v{k}} \left( u_{\v{k}} + v_{\v{k}} f_{\v{k}\up}^\dagger
f_{-\v{k}\down}^\dagger \right) |0\rangle.  $

\begin{figure}[t]
\centerline{
\includegraphics[width=3in]{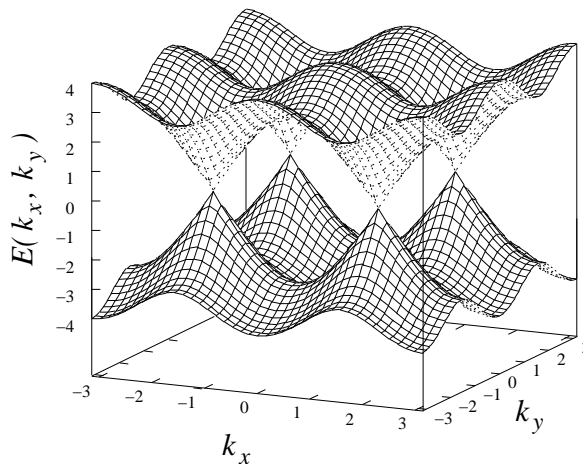}
}
\caption{
The energy dispersion of the $\pi$ flux phase.  Note the massless Dirac spectrum at the nodal
points at $\left( \pm {\pi \over 2}, \pm{\pi \over 2} \right)$.}
\label{flux}
\end{figure}

A variety of mean-field wavefunctions were soon discovered which give
identical energy and dispersion.  Notable among these is the staggered flux
state 
\cite{AM8874}.  In this state the hopping $\chi_{\v i\v j}$ is
complex, $\chi_{\v i\v j} = \chi_0 \exp \left( i (-1)^{i_x+j_y} \Phi_0 \right) $,
and the phase is arranged in such a way that it describes free fermion hopping
on a lattice with a fictitious flux $\pm 4 \Phi_0$ threading alternative
plaquettes.  Remarkably, the eigenvalues of this problem are identical to that
of the $d$-wave superconductor given by eq.~(\ref{Eq.14}), with
\be
\tan \Phi_0 = {\Delta_0 \over \chi_0}  .
\label{Eq.17}
\en
The case $\Phi_0 = \pi/4$, called the $\pi$ flux phase, is special in that it
does not break the lattice translation symmetry.  As we can see from
eq.~(\ref{Eq.17}),
the corresponding $d$-wave problem has a very large energy gap and its
dispersion is shown in Fig.~\ref{flux}.  The key feature is that the energy
gap vanishes at the nodal points located at $\left( \pm{\pi\over 2},
\pm{\pi\over 2} \right)$.  Around the nodal points the dispersion rises
linearly, forming a cone which resembles the massless Dirac spectrum.  For the
$\pi$ flux state the dispersion around the node is isotropic.  For $\Phi_0$
less than $\pi /4$ the gap is smaller and the Dirac cone becomes progressively
anisotropic.    The anisotropy can be characterized by two velocities, $v_F$
in the direction towards $(\pi,\pi)$ and $v_\Delta$ in the direction towards
the maximum gap at $(0,\pi)$.

The reason various mean-field theories have the same energy was explained by
\Ref{AZH8845} and \Ref{DFM8826}  as being due to a
certain $SU(2)$ symmetry.  We defer a full discussion of the $SU(2)$ symmetry to
section X but we only mention here that it corresponds to the following
particle-hole transformation
\be
f_{\v i\up}^\dagger &\rightarrow& \alpha_{\v i} f_{\v i\up}^\dagger + \beta_{\v i} f_{\v i\down} \label{Eq.18} \\ \nonumber
f_{\v i\down} &\rightarrow& -\beta_{\v i}^\ast f_{\v i\up}^\dagger + \alpha_{\v i}^\ast f_{\v i\down}  .
\en
Note that the spin quantum number is conserved.  It describes the physical
idea that adding a spin-up fermion or removing a spin-down fermion are the
same state after projection to the subspace of singly occupied fermions.  It
is then not a surprise to learn that the Gutzwiller projection of the $d$-wave
superconductor and that of the staggered flux state gives the same trial
wavefunction, up to a trivial overall phase factor, provided $\mu = 0$ and
eq.~(\ref{Eq.17}) is satisfied.  A simple proof of this is given by 
\Ref{ZGR8836}.  The energy of this state is quite good.  The best estimate for the
ground state energy of the square lattice Heisenberg antiferromagnet which is
a N\'{e}el ordered state is $\langle S_{\v i} \cdot S_{\v j} \rangle = -0.3346$~J
(\cit{TC8937}; \cit{R9292}).  The projected $\pi$ flux state
\cite{G8831,YO9615} gives $-0.319$J, which is excellent
considering that there is no variational parameter.  
 
We note that the projected $d$-wave state has power law decay for the
spin-spin correlation function.  The equal time spin-spin correlater decays as
$r^{-\alpha}$ where $\alpha$ has been estimated to be 1.5 \cite{I00,PRT0453}.
This projection has considerably enhanced the spin correlation
compared with the exponent of 4 for the unprojected state.  One might expect a
better trial wavefunction by introducing a sublattice magnetization in the
mean-field Hamiltonian.  A projection of this state gives an energy which is
marginally better than the projected flux state, $-0.3206$J.  It also has a
sublattice magnetization of 84\% which is too classical.  The best trial
wavefunction is one which combines staggered flux and sublattice magnetization
before projection 
\cite{G8831,G8853,LF8809}.  
It gives an energy of $-0.332$~J and a sublattice magnetization of about 70\%,
both in excellent agreement with the best estimates.

\begin{figure}[t]
\centerline{
\includegraphics[width=3in]{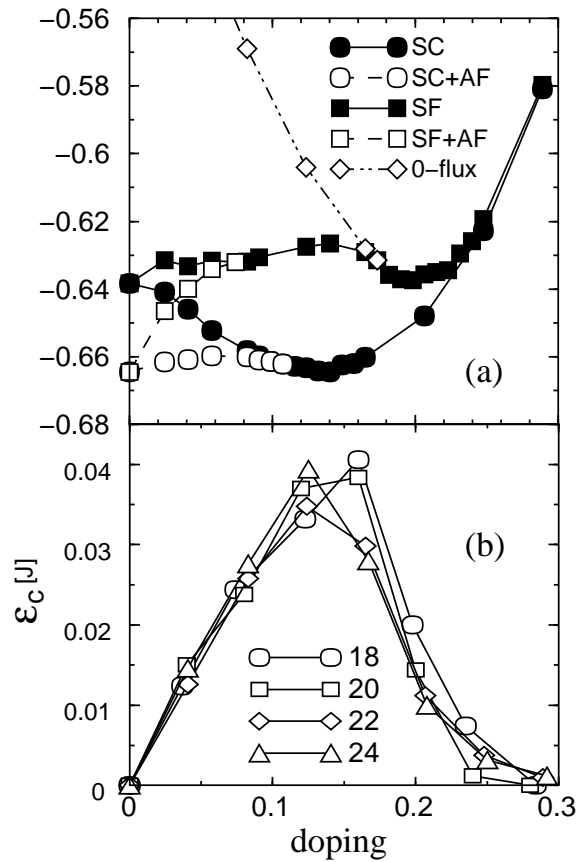}
}
\caption{
(a) Comparison of the energy of various projected trial wavefunctions. From 
\Ref{I0365}.
(b)~The condensation energy estimated from the difference of the projected $d$-wave superconductor and
the projected staggered flux state. From 
\Ref{IL0301}.}
\label{Ivanov}
\end{figure} 

\subsection{Doped case} 

In the presence of a hole, the projected wavefunction
eq.~(\ref{Eq.12}) has been studied for a variety of mean-field states $\psi_0$.  Here
$P_G$ stands for a double projection: the amplitudes with double occupied
sites are projected out and only amplitudes with the desired number of holes
are kept.  The ratio $\Delta_0/\chi_0$, $\mu/\chi_0$ and $h_s/\chi_0$, where
$h_s$ is the field conjugate to the sublattice magnetization, are the
variational parameters.  It was found that the best state is a projected
$d$-wave superconductor and the sublattice magnetization is nonzero for $x  <
x_c$, where $x_c = 0.11$ for $t/J = 3$. 
\cite{YO9615}  The
energetics of various state are shown in Fig.~(\ref{Ivanov}(a)).  It is
interesting to note that  the projected staggered-flux state always lies above
the projected $d$-wave superconductor, but the energy difference is small and
vanishes as $x$ goes to zero, as expected.  The staggered-flux state also
prefers antiferromagnetic order for small $x$, and the critical $x_c^{SF}$ is
now 0.08, considerably less than that for the projected $d$-wave
superconductor.  The energy difference between the projected flux state and
projected $d$ superconductor (with antiferromagnetic order) is shown in
Fig.~(\ref{Ivanov}(b)).  As we can see from Fig.~(\ref{Ivanov}(a)), inclusion
of AF will only give a small enhancement of the energy difference for small $x
\leq 0.05$.  The projected staggered flux state is the lowest energy
non-superconducting state that has been constructed so far.  For $x > 0.18$,
the flux $\Phi_0$ vanishes and this state connects smoothly to the projected
Fermi sea, which one ordinarily thinks of as the normal state.  It is then
tempting to think of the projected staggered flux state as the ``normal''
state in the underdoped region $(x < 0.18)$ and interpret the energy
difference shown in Fig.~(\ref{Ivanov}(b)) as the condensation energy.  Such a
state may serve as the ``competing'' state that we have argued must live
inside the vortex core.  The fact that the energy difference vanishes at $x =
0$ guarantees that it is small for small $x$.

\Ref{I0365} pointed out that the concave nature of the energy curves shown
in Fig.~\ref{Ivanov}(a) for small $x$ indicate that the system is prone to
phase separation.  Such a phase separation may be suppressed by long-range
Coulomb interaction and the energy curves are indeed sensitive to
nearest-neighbor repulsion.  Thus we believe that Fig.~\ref{Ivanov}(a) still
provides a useful comparison of different trial wavefunctions.

\subsection{Properties of projected wavefunctions} It is interesting to put
aside the question of energetics and study the nature of the projected
$d$-wave superconductor.  A thorough study by 
\Ref{PRT0102}, \citeyear{PRT0404}
showed that it correctly captures many of the properties
of the cuprate superconductors.  For example, the superfluid density vanishes
linearly in $x$ for small $x$.  This is to be expected since the projection
operator is designed to yield an insulator at half-filling.  The momentum
distribution exhibits a jump near the noninteracting Fermi surface. The size
of the jump is interpreted as the quasiparticle weight $z$ according to Fermi
liquid theory and again goes to zero smoothly as $x \rightarrow 0$.  Using the
sum rule and assuming Ferm liquid behavior for the nodal quasiparticles, the
Fermi velocity is estimated and found to be insensitive to doping, in
agreement with photoemission experiments.

A distinctive feature of the projected staggered flux state is that it breaks
translational symmetry and orbital currents circulate the plaquettes in a
staggered fashion as soon as $x \neq 0$.  Motivated by the $SU(2)$ symmetry
which predicts a close relationship between the projected $d$-wave
superconductor and the projected staggered flux states, 
\Ref{ILW0053}
examined whether there are signs of the orbital current in the
projected $d$-wave superconductor.  Since this state does not break
translation or time-reversal symmetry, there is no static current.  However,
the current-current correlation
\be
G_{\v j} = \langle
j(\alpha) j(\beta)
\rangle
\label{Eq.19}
\en
where $j(\alpha)$ is the current on the $\alpha$ bond, shows a power law-type
decay and its magnitude is much larger than the naive expectation that it
should scale as $x^2$.  Note that before projection the $d$-wave
superconductor shows no hint of the staggered current correlation.  The
correlation that emerges is entirely a consequence of the projection.  We
believe the emergence of orbital current fluctuations provides strong support
for the importance of $SU(2)$ symmetry near half filling.  Orbital current
fluctuations of similar magnitude were found in the exact ground sate
wavefunction of the $t$-$J$ model on a small lattice, two holes on 32 sites.
(\cit{L0012}; \cit{LS0371}) showed that the orbital current correlation
has the same power law decay as the hole-chirality correlation,
\be
G_{\chi_h} = \langle
\chi_h(i) \chi_h(j) 
\rangle \nonumber
\en
 where $\chi_h$ is defined on a plaquette $i$ as $n_h(4) \v{S}_1 \cdot
(\v{S}_2 \times \v{S}_3)$ where 1 to 4 labels the sites around the plaquette
and $n_h(\v i) = 1 - c_{\v i\sigma}^\dagger c_{\v i\sigma}$ is the hole density
operator.  This is in agreement with the notion  that a hole moving around the
plaque experiences a Berry's phase due to the non-colinearity of the spin
quantization axis of the instantaneous spin configurations.  For $S = {1\over
2}$ the Berry's phase is given by ${1\over 2} \phi$ where $\phi$ is the solid
angle subtended by the instantaneous spin orientations $\v{S}_1$, $\v{S}_2$
and $\v{S}_3$. 
(\cit{WWZcsp}; \cit{Fra91}) This solid angle
is related to the spin chirality $\v{S}_1 \cdot (\v{S}_2 \times \v{S}_3)$.
This phase drives the hole in a clockwise or anti-clockwise direction
depending on its sign, just as a magnetic flux through the center of the
plaquette would.  Thus the flux $\Phi_0$ of the staggered flux state has its
physical origin in the coupling between the hole kinetic energy and the spin
chirality.

It is important to emphasize that the projected $d$-wave state possesses long
range superconducting pairing order, while at the same time exhibiting power
law correlation in antiferromagnetic order and staggered orbital current.  On
the other hand, projection of a staggered flux phase at finite doping will
possess long range orbital current order, but short range pairing and
antiferromagnetic order.  A useful analogy is to think of these projected
states as a person with a variety of personalities.  He may be courteous and
friendly at one time, and aggressive and even belligerent at another, depending
on his environment.  Thus different versions of projected states shown in
Fig.~\ref{Ivanov}(a) all have the same kinds of fluctuations; it is just
that one kind of order may dominate over the others.  Then it is easy to
imagine that the system may shift from one state to another in different
environments.  For instance, in section XII.C we will argue that the pairing
state will switch to a projected staggered flux state inside the vortex core.
Note that this is a different picture from the traditional Landau picture of
competing states as advocated by 
\Ref{CLM0203}, for instance.
These authors suggested on phenomenological grounds that the pseudogap region
is characterized by staggered orbital current order, which they call
$d$-density waves (DDW).  The symmetry of this order is indistinguishable from
the doped staggered flux phase 
\cite{HMA9166,L0249}.  According
to Landau theory, the competition between DDW and superconducting order will
result in either a first order transition or a region of co-existing phase at
low temperatures.  This view of competing order is very different from the one
proposed here, where a single quantum state possesses a variety of fluctuating
orders.

\subsection{Improvement of projected wavefunctions, effect of $t^\prime$, and
the Gutzwiller approximation} 

The projected wavefunction is the starting point
for various schemes to further improve the trial wavefunction.  Indeed, the
variational energy can be lowered and 
\Ref{SMB0202}
 provide
strong evidence that  a $d$-wave superconducting state may be the ground state
of the $t$-$J$ model.  On the other hand, other workers 
\cite{HR9373,SCL9894}
 found that the superconducting tendency decreases
with the improvement of the trial wavefunctions.  Studies based  on  other
methods such as DMRG 
\cite{WS9953} found that next-nearest
neighbor hopping $t^\prime$ with $t^\prime/t > 0$ is needed to stabilize the
$d$-wave superconductor.  Otherwise the holes are segregated into strip-like
structures.  All these computational schemes suffer from some form of
approximation and cannot give definitive answers.  What is clear is that the
$d$-wave superconductor is a highly competitive candidate for the ground state
of the $t$-$J$ model.

Recently 
\Ref{SLE0402} have examined the pairing correlation in
projected wavefunctions including the effect of $t^\prime$.  They find that
for moderate doping ($x \gtrsim 0.1$) $t^\prime/t$ with a negative sign
greatly enhances the pairing correlation.  The effect increases with
increasing $t^\prime$ and is maximal around $t^\prime/t \approx -0.4$.  Their
result contradicts expectations based on earlier DMRG work 
\cite{WS9953} which found a suppression of superconductivity with negative
$t^\prime/t$.  However, Shih {\em et al.} pointed out that the earlier work
was limited to very low doping and is really not in disagreement with their
finding for $x \gtrsim 0.1$.  This result should be confirmed by improving the
wavefunction but the pair correlation with $t^\prime$ is so robust that the
controversy surrounding the $t^\prime = 0$ case may well be avoided.  It
should be noted that a negative $t^\prime$ is what band theory predicts.
Furthermore, 
\Ref{PDS0103} have noted a correlation of $T_c$
with $| t^\prime |$ and shown that the Hg and Tl compounds which have the
highest $T_C$ have $t^\prime/t$ in the range $-0.3$ to $-0.4$.  Thus the role
of $t^\prime$ may well explain the variation of $T_c$ among different families
of cuprates.

The Gutzwiller projection is a rather cumbersome machinery to implement and a
simple approximate scheme has been proposed, called the Gutzwiller
approximation
(\cit{ZGR8836}; \cit{H9079}).  The essential step is to construct an effective Hamiltonian
\be
H_\text{eff} = -g_t t\sum_{\langle \v i\v j \rangle \sigma} c_{\v i\sigma}^\dagger c_{\v i\sigma} + g_J \sum_{\langle \v i\v j
\rangle} \v{S}_{\v i} \cdot \v{S}_{\v j}
\label{Eq.20}
\en
and treat this in the Hartree-Fock-BCS approximation. The projection operator
in the original $t$-$J$ model is eliminated in favor of the reduction factors
$g_t = 2x/(1 + x)$ and $g_J = 4/(1 + x)^2$, which are estimated by assuming
statistical independence of the population of the sites
\cite{V8499}.
The important point is that $g_t = 2x/(1 + x)$ reduces the kinetic energy to
zero in the $x \rightarrow 0$ limit, in an attempt to capture the physics of
the approach to the Mott insulator.  The Gutzwiller approximation bears a
strong resemblance to the slave-boson mean-field theory and is just as easy to
handle analytically.  It has the advantage that the energetics compare well
with the Monte Carlo projection results.  The Gutzwiller approximation has
been applied to more complicated problems such as impurity and vortex
structure 
\cite{TTO0005,TOT0309} with good results.

%
%
%
%
%
%


\section{The single hole problem}

The motion of a single hole doped into the antiferromagnet is a most
fundamental issue to start with.  The $t$-$J$ type model is again the
canonical Hamiltonian to study this problem.  The key physics of the problem
is the competition between the antiferromagnetic (AF) correlation/long range
ordering and the kinetic energy of the hole. The motion of the single hole
distort the AF ordering when it hops between different sublattices.
\Ref{SS8867} studied this distortion in a semiclassical way,
and found the new coupling between the spin current of the hole and the
magnetization current of the background. This coupling leads to the long
range dipolar distortion of the staggered magnetization and the minimum of the
hole dispersion at $k=(\pi/2,\pi/2)$.  This position of the energy minimum is
interpreted as follows.  Even if we start with the pure $t$-$J$ model, the
direct hopping between nearest neighbor sites is suppressed, while the second
order processes in $t$ leads to the effective hopping between the sites
belonging to the same sublattice.  This effective $t'$ and $t''$ has the
negative sign and hence lower the energy of $k=(\pi/2,\pi/2)$ compared with
$k=(\pi,0),(0,\pi)$.

The dynamics of the single hole, \ie the spectral function of the Green's
function is also studied by analytic method.  When the spin excitation is
approximated by the magnon (spin wave), the Hamiltonian for the signle hole is
given by \cite{KLR8980}
\be
H = { t \over N} \sum_{\v k ,q} M_{\v k,\v q} [ h^\dagger_{\v k} h_{\v k - \v q}
\alpha_{\v q}
+ h.c.] + \sum_{q} \Omega_{\v q} \alpha_{\v q}^\dagger \alpha_{\v q}
\label{Eq.21}
\en
where 
\be
\Omega_{\v q} =  2J \sqrt{ 1 - \gamma_{\v q}^2}
\label{Eq.22}
\en
with $\gamma_{\v q} = ( \cos q_x + \cos q_y )/2$ and
\be
M(\v k,\v q) = 4(u_{\v q} \gamma_{\v k-\v q} + v_{\v q} \gamma_{\v k})
\label{Eq.23}
\en
with $u_{\v k} = \sqrt{(1 + \nu_{\v k})/(2 \nu_{\v k})}$, $v_{\v k} = - \text{sign}(\gamma_{\v k})
\sqrt{(1 - \nu_{\v k})/(2 \nu_{\v k})}$, and $\nu_{\v k} = \sqrt{1 -\gamma_{\v k}^2}$.  This
Hamiltonian dictates that the magnon is emitted or absorbed every time the
hole hops.  The most widely accepted method to study this model is the
self-consistent Born approximation (SCBA) initiated by Kane, Lee and
Read where the Feynman diagrams with the crossing magnon
propagators are neglected. This leads to the self-consistent equation for the
hole propagator:
\be
G(\v k,\omega) = [ \omega - \sum_{\v q} g(\v k,\v q)^2 G(\v k-\v q,\omega-\Omega_{\v q}) ]^{-1}.
\label{Eq.24}
\en
The result is that there are two components of the spectral function $A(\v
k,\omega) = - (1/\pi) \Im G^R(\v k,\omega)$: One is the coherent sharp peak
corresponding to the quasi-particle and the other is the incoherent
background. The former has the lowest energy at $k=(\pi/2,\pi/2)$ at the
energy $\sim -t$ and disperses of the order of $J$, while the latter does not
depend on the momentum $k$ so much and extends over the energy of the order of
$t$.  Intuitively the hole has to wait for the spins to flip to hop, which
takes a time of the order of $J^{-1}$.  Therefore the bandwidth is reduced
from $\sim t$ to $\sim J$.  This mass enhancement leads to the reduced weight
$z \sim J/t$ for the quasi-particle peak.  Later, more detailed studies have
been done in SCBA\cite{LM9225}.  The conclusions obtained are the followings:
(i) At $k=(\pi/2,\pi/2)$, there appear two additional two peaks at $E_{2,3}$
in addition to the ground state delta-functional peak at $E_1$.

These energies are given for $J/t<0.4$ by
\be
E_n/t = - b + a_n (J/t)^{2/3}
\label{Eq.25}
\en
where $a_1=2.16,a_2=5.46,a_3=7.81$, and $b=3.28$.  (ii) The spectral weight
$z$ at $k=(\pi/2,\pi/2)$ scales as $z = 0.65(J/t)^{2/3}$.

These can be understood as the "string" excitation of the hole moving in the
linear confining potential due to the AF background.  It has also been
interpreted in terms of the confining interaction between spinon and holon
\cite{L9726}.  Exact diagonalization studies have reached consistent results
with SCBA.  Experimentally angle-resolved-photoemission spectroscopy
(ARPES)(\cit{WSM9564}; \cit{Ro9867}) in undoped cuprates has revealed the spectral function of
the single doped hole.  The energy dispersion of the hole looks like that of
the $\pi$-flux state shifted by the Mott gap to the low energy 
\cite{L9726}.  However, in real materials the second ($t'$) and third ($t"$)
nearest neighbor hoppings are important.  The calculated energy dispersion is
found to be sensitive to $t^\prime$ and $t^{\prime\prime}$.  For $t^\prime =
t^{\prime\prime} = 0$, the dispersion is flat between $(\pi/2, \pi/2)$ and
$(0,\pi)$ and does not agree with the data.  It turns out that the data is
well fitted by $J/t=0.3$, $t'/t=-0.3$, $t"/t=0.2$.  On the other hand, ARPES
in slightly electron-doped Ne$_{2-x}$Ce$_x$CuO$_2$ showed that the electron is
doped into the point $k=(\pi,0)$ and $(0,\pi)$ \cite{Ao0103}. This difference
will be discussed below.

The variational wavefunction approach to the antiferromagnet and single hole
problem has been pursued by several authors \cite{LHN0301,LLH0301}.  A good
ground state variational wavefunction (vwf) at half-filling is 
\be
| \Psi_0\> = P_G
\biggl[ \sum_{\v k} ( A_{\v k} a_{\v k \uparrow}^\dagger a_{-\v k \downarrow}^\dagger
+ B_{\v k} b_{\v k \uparrow}^\dagger b_{-\v k \downarrow}^\dagger \biggr]^{N/2}
| 0\>
\label{Eq.26}
\en
with $N$ being the number of atoms.  The operators $a_{\v k \sigma}^\dagger$,
$b_{\v k \sigma}^\dagger$ are those for the upper and lower bands split by SDW
with the energy $\pm \xi_{\v k}$, respectively, and $A_{\v k} = (E_{\v k} + \xi_{\v k})/\Delta_{\v k}$,
$B_{\v k} =  (-E_{\v k} + \xi_{\v k})/\Delta_{\v k}$ with $E_{\v k} = \sqrt{ \xi_{\v k}^2+ \Delta_{\v k}^2}$ and
$\Delta_{\v k} = (3/8)J \Delta( \cos k_x - \cos k_y)$.  The picture here is that in
addition to the SDW, the RVB singlet formation represented by $\Delta$ is
taken into account.  As mentioned in the last section, this vwf gives much
better energy compared to  that with $\Delta=0$. Hence the ground state is far
from the classical N\"{e}el state and includes strong quantum fluctuations.
Next the vwf in the case of single doped hole with momentum $q$ and $S^z =
1/2$ is
\be
| \Psi_{\v q}\> = P_G c^\dagger_{q \uparrow}
\biggl[ \sum_{\v k (\ne q)} ( A_{\v k} a_{\v k \uparrow}^\dagger 
a_{-\v k \downarrow}^\dagger
+ B_{\v k} b_{\v k \uparrow}^\dagger b_{-\v k \downarrow}^\dagger \biggr]^{N/2-1}
| 0\>
\label{Eq.27}
\en

This vwf does not contain the information of $t'$,$t''$ except the very small
dependence of $A_{\v k}$, and  $B_{\v k}$.  The robustness of this vwf is the
consequence of the large quantum fluctuation already present in the
half-filled case, so that the hole motion is possible even without disturbing
the spin liquid-like state.  Although the vwf does not depends on the
parameters $t'$,$t''$, the energy dispersion $E(\v k)$ is given by the
expectation value as
\begin{equation}
E(\v k) = \< \Psi_{\v k} | H_{t-J} + H_{t'-t''}| \Psi_{\v k} \>,
\label{Eq.28}
\end{equation}
and depends on these parameters.  This expression gives a reasonable agreement
with the experiments both in undoped material \cite{Ro9867} and electron-doped
material \cite{Ao0103}.  Here an important question is the relation
between the hole- and electron-doped cases.  There is a particle-hole symmetry
operation which relates the $t$-$t'$-$t''$-$J$ model for a hole to that for an
electron.  The conclusion is that the sign change of $t'$, and $t''$ together
with the shift in the momentum by $(\pi,\pi)$ gives the mapping between the
two cases.  Using this transformation, one can explain the difference between
hole- and electron-doped cases in terms of the common vwf eq.~(\ref{Eq.27}).
The former has the minimum at $k=(\pi/2,\pi/2)$ while the latter at
$k=(\pi,0),(0,\pi)$.

Exact diagonalization study \cite{T009} has shown that the electronic state is
very different between $k = (\pi/2,\pi/2)$ and $k =(\pi,0)$ for the
appropriate values of $t'$ and $t''$ for hole doped case.
The spectral weight becomes very small at $(\pi,0)$ and the hole is surrounded
by anti-parallel spins sitting on the same sublattice.  Both these features
are captured by a trial wavefunction which differs from eq.~(\ref{Eq.27}) in
that the momentum $\v{q}$ of the broken pair is different from the momentum
of the inserted electron.  This can also be interpreted as the decay of the
quasiparticle state via the emission of a spin wave \cite{LLH0301}.  There are
thus two types of wf's with qualitatively different nature, {\em i.e.}, one
describes the quasi-particle state and another  which is highly incoherent and
may be realized as a spin-charge separated state.

One important discrepancy between experiment and theory is the line-shape of
the spectral function. Namely the experiments show broad peak with the width
of the order of $\sim 0.3$eV in contrast to the delta-functional peak expected
for the ground state at $k=(\pi/2,\pi/2)$. One may attribute this large width
to the disorder effect in the sample. However the ARPES in the overdoped region
shows even sharper peak at the Fermi energy even though the doping introduces
further disorder.  Therefore the disorder effect is unlikely to explain this
discrepancy.  Recently the electron-phonon coupling to the single hole in
$t$-$J$ model has been studied using quantum Monte Carlo simulation
\cite{MN0402}.  It is found that the small polaron 
formation
in the presence of strong
correlation 
reduces the dispersion and the weight of the zero-phonon line, while the center of
mass of the spectral weight for the originally "quasi-particle" peak remain
the same as the pure $t$-$J$ model, even though the shape is broadened.
Therefore the polaron effect is a promising scenario to explain the spectral
shape.

Recently, 
\Ref{SRL0402} pointed out that the polaron picture also
explains a long standing puzzle regarding the location of the chemical
potential with doping.  Naive expectation based on doping a Hubbard model
predicts that the chemical potential should lie at the top of the valence
band, whereas experimentally in Na-doped Ca$_2$CuO$_2$Cl$_2$ it was found that
the chemical potential appears to lie somewhere in mid-gap, \ie with a small
but finite density of holes, the chemical potential is several tenths of eV
higher than the energy of the peak position of the one-hole spectrum.  This is
naturally explained if the one-hole spectrum has been shifted down by polaron
effects, so that the top of the valence band should be at the zero-phonon
line, rather than the center of mass of the one-hole spectrum.

\section{Slave boson formulation of $t$-$J$ model and mean field theory}

As has been discussed in II, it is widely believed that the low energy physics
of high-Tc cuprates is described in terms of $t$-$J$ type model, which is
given by \cite{LN9221}
\be
H= \sum_{\<\v i\v j\>} J\left(
{\v{S}}_{\v i}\cdot {\v{S}}_{\v j}-{1\over
4}n_{\v i} n_{\v j} \right)
-\sum_{\v i\v j}t_{\v i\v j}
\left(c_{\v i\sigma}^\dagger
c_{\v j\sigma}+{\rm H.c.}\right).
\label{t-J}
\en
where $t_{\v i\v j}=t$, $t'$, $t''$ for the nearest, second nearest and 3rd nearest
neighbor pairs, respectively. The effect of the strong Coulomb repulsion is
represented by the fact that the electron operators $c^\dagger_{\v i\sigma}$ and
$c_{\v i\sigma}$ are the projected ones, where the double occupation is
forbidden.  This is written as the inequality
\be
\sum_{\sigma} c^\dagger_{\v i\sigma} c_{\v i \sigma} \le 1,
\label{Eq.30}
\en
which is very difficult to handle.  A powerful method to treat this constraint
is so called the slave-boson method 
\cite{B7675,C8435}.  In most
general form, the electron operator is represented as
\be
c^\dagger_{\v i\sigma} = f_{\v i\sigma}^\dagger b_{\v i} + 
\epsilon_{\sigma \sigma'} f_{\v i \sigma'} d_{\v i}^\dagger
\label{Eq.31}
\en
where $\epsilon_{\uparrow \downarrow} = 
- \epsilon_{\downarrow \uparrow} = 1$ is the antisymmetric tensor.
  $f_{\v i\sigma}^\dagger$, $f_{\v i \sigma}$ are the fermion operators, while
$b_{\v i}$, $d^\dagger_{\v i}$ are the slave-boson operators.  This representation
together with the constraint 
\be
f_{\v i\uparrow}^\dagger f_{\v i\uparrow}
+ f_{\v i\downarrow}^\dagger f_{\v i\downarrow} + b^\dagger_{\v i} b_{\v i}
+ d^\dagger_{\v i} d_{\v i} = 1
\label{Eq.32}
\en
reproduces all the algebra of the electron (fermion) operators.  From eqs.
(\ref{Eq.31}) and (\ref{Eq.32}), the physical meaning of these operators is
clear. Namely, there are 4 states per site and $b^\dagger$, $b$ corresponds to
the vacant state, $d^\dagger$,$d$ to double occupancy, and $f^\dagger_\sigma$,
$f_{\sigma}$ to the single electron with spin $\sigma$.  With this formalism
it is easy to exclude the double occupancy just by deleting $d^\dagger$, $d$
from the above equations (\ref{Eq.31}) and (\ref{Eq.32}).  Then the projected
electron operator is written as
\be
c_{\v i\sigma}^\dagger = f_{\v i\sigma}^\dagger b_{\v i}
\label{Eq.33}
\en
with the condition
\be
f_{\v i\uparrow}^\dagger f_{\v i\uparrow}
+ f_{\v i\downarrow}^\dagger f_{\v i\downarrow} + b^\dagger_{\v i} b_{\v i} = 1.
\label{Eq.34}
\en
This constraint can be enforced with a Lagrangian multiplier $\lambda_{\v i}$.
Note that unlike eq. (\ref{Eq.31}), eq. (\ref{Eq.33}) is not an operator
identity and the R.H.S. does not satisfy the fermion commutation relation.
Rather, the requirement is that both sides have the correct matrix elements in
the reduced Hilbert space with no doubly occupied states.  For example, the
Heisenberg exchange term is written in terms of $f^\dagger_{\v i\sigma}$,
$f_{\v i\sigma}$ only \cite{BZA8773}
\be
{\v{S}}_{\v i}\cdot {\v{S}}_{\v j} &=& -{1\over 4} f_{\v i\sigma}^\dagger f_{\v j\sigma}
f_{\v j\beta}^\dagger f_{\v i\beta}  \nonumber  \label{Eq.35} \\
&-& {1\over 4} \left(
f_{\v i\up}^\dagger f_{\v j\down}^\dagger - f_{\v i\down}^\dagger f_{\v j\up}^\dagger
\right) \left(
f_{\v j\down} f_{\v i\up} - f_{\v j\up} f_{\v i\down}
\right)  \nonumber \\ 
&+& {1\over 4} \left( f_{\v i\alpha}^\dagger f_{\v i\alpha}  \right) .
\en
We write
\be
n_{\v i}n_{\v j} = (1 - b_{\v i}^\dagger b_{\v i} ) (1 - b_{\v j}^\dagger b_{\v j} )  .
\label{Eq.36}
\en
Then ${\v{S}}_{\v i}\cdot {\v{S}}_{\v j} - {1\over 4}n_{\v i}n_{\v j}$ can be written in terms of
the first two terms of eq. (\ref{Eq.35})  plus quadratic terms, provided we
ignore the nearest-neighbor hole-hole interaction ${1\over 4}b_{\v i}^\dagger
b_{\v i}b_{\v j}^\dagger b_{\v j} $.  We then decouple the exchange term in both the
particle-hole and particle-particle channels via the Hubbard-Stratonovich (HS)
transformation.  

Then the partition function is written in the 
form
\be
Z = \int{D f D f^\dagger Db D\lambda D\chi D\Delta} \exp \left(
-\int^\beta _0 d\tau L_1
\right)
\label{Eq.38}
\en
where
\be
\label{Eq.39}
L_1 &=& \tilde{J} \sum_{\<\v i\v j\>} \left(
|\chi_{\v i\v j}|^2 + | \Delta_{\v i\v j} |^2 \right)
+\sum_{\v i\sigma}f_{\v i\sigma}^\dagger (\partial_\tau - i\lambda_{\v i})
f_{\v i\sigma} 
\nonumber\\
&-& \tilde{J} 
\left[
 \sum_{\<\v i\v j \>} \chi_{\v i\v j}^\ast \left( \sum_{\sigma} f_{\v i\sigma}^\dagger
f_{\v j\sigma} \right) + c.c. \right]  
\\
&+& \tilde{J} \left[ \sum_{\<\v i\v j\>} \Delta_{\v i\v j} \left(
f^\dagger_{\v i\up}f^\dagger_{\v j\down} - f^\dagger_{\v i\down} f^\dagger_{\v j\up} \right) + c.c. \right] \nonumber \\
&+&\sum_{\v i} b_{\v i}^\ast(\partial_\tau - i\lambda_{\v i} + \mu_B ) b_{\v
i} 
- \sum_{\v i\v j}{t}_{\v i\v j}b_{\v i}b_{\v j}^\ast f_{\v i\sigma}^\dagger f_{\v j\sigma} ,
\nonumber 
\en
with $\chi_{\v i\v j}$ representing fermion hopping and $\Delta_{\v i\v j}$
representing fermion pairing corresponding to the two ways of representing the
exchange interaction in terms of the fermion operators. From eqs.
(\ref{Eq.35}) and (\ref{Eq.39})  it is concluded that $\tilde{J}=J/4$, but in
practice the choice of $\tilde{J}_{\v i\v j}$ is not so trivial, namely one
would like to study the saddle point approximation (SPA) and the Gaussian
fluctuation around it, and requires SPA to reproduce the mean field theory.
The latter requirement is satisfied when only one HS variable is relevant, but
not for the multicomponent HS variables 
\cite{NegO87,UL9234}.  In the latter case, it is better to chose the
parameters in the Lagrangian to reproduce the mean field theory. In the
present case, $\tilde{J} = 3J/8$ reproduces the mean field self-consistent
equation which is obtained by the Feynman variational principle \cite{BL0102}.

We note that $L_1$ in \Eq{Eq.39} is invariant under a local $U(1)$
transformation
\begin{align}
\label{U1gaugetrans}
 f_{\v i} &\to e^{i\th_{\v i}} f_{\v i} \nonumber\\
 b_{\v i} &\to e^{i\th_{\v i}} b_{\v i} \nonumber\\
 \chi_{\v i\v j} &\to e^{-i\th_{\v i}} \chi_{\v i\v j} e^{i\th_{\v j}} \nonumber\\
 \Del_{\v i\v j} &\to e^{i\th_{\v i}} \Del_{\v i\v j} e^{i\th_{\v j}} \nonumber\\
 \la_{\v i} &\to  \la_{\v i} +\prt_\tau \th_{\v i}
\end{align}
which is called $U(1)$ gauge transformation.
Due to such a $U(1)$ gauge invariance, the phase fluctuations of $\chi_{\v i\v j}$ and $\la_{\v i}$
have a dynamics of $U(1)$ gauge field (see section IX).

Now we describe the various mean field theory corresponding to the saddle
point solution to the functional integral.  The mean field conditions are
\begin{align}
\label{Eq.39a}
\chi_{\v i\v j} &= \sum_\sigma \langle
f^\dagger_{\v i\sigma} f_{\v j\sigma} \rangle \\
\label{Eq.40} 
\Delta_{\v i\v j} &= \langle  f_{\v i\up}f_{\v j\down} - f_{\v i\down}f_{\v i\up}  \rangle
\end{align}

Let us first consider the $t$-$J$ model in the undoped case,
\ie the half-filled case.  There are no bosons in this case, and the theory
is purely that of fermions.  The original one, \ie uniform RVB state,
proposed by Baskaran-Zou-Anderson \cite{BZA8773} is given by
\be
\chi_{\v i\v j} = \chi = \text{real}
\label{Eq.41}
\en
for all the bond and $\Delta_{\v i\v j}=0$.  The fermion spectrum is that of the
tight binding model
\be
H_\text{uRVB} = - \sum_{\v k \sigma} 2 {\tilde J}
\chi ( \cos k_x + \cos k_y )
f^\dagger_{\v k \sigma} f_{\v k \sigma}, 
\label{Eq.42}
\en
with the saddle point value to the Lagrange multiplier $\lambda_{\v i}=0$.  The so
called ``spinon Fermi surface'' is large, \ie it is given by the condition
$k_x \pm k_y = \pm \pi$ with a diverging density of states (van Hove
singularity) at the Fermi energy. 
Soon after, many authors 
found lower energy states than the uniform RVB state. One can easily
understand that lower energy states 
exist because the Fermi surface is perfectly nested with the nesting
wavevector ${\vec Q} = (\pi,\pi)$ and the various instabilities with $\vec Q$
are expected. Of particular importance are the $d$-wave state [see (Eq.
\ref{Eq.13})] and the staggered flux state [see (Eq. \ref{Eq.17})] which give
identical energy dispersion.  This was explained as being due to a local
$SU(2)$ symmetry when the spin problem is formulated in terms of fermions
\cite{AZH8845,DFM8826}.  We write
\be
\Phi_{\v i\up} =
\bpm f_{\v i\up} \\
f_{\v i\down}^\dagger \epm \,\, , \,\,\,\,
\Phi_{\v i\down} =
\bpm f_{\v i\down} \\
-f_{\v i\up}^\dagger \epm \,\, ,
\label{Eq.37}
\en
Then eq.~(\ref{Eq.39}) can be written in the more compact form
\be
L_1 &=& {\tilde{J}\over 2} \sum_{\<\v i\v j\>}\Tr [U_{\v i\v j}^\dagger U_{\v i\v j}] +
{\tilde{J}\over 2} \sum_{\<\v i\v j\>,\sigma} \left (
\Phi_{\v i\sigma}^\dagger U_{\v i\v j} \Phi_{\v j\sigma} + c.c.
\right) \nonumber \\
&+&\sum_{\v i\sigma}f_{\v i\sigma}^\dagger (\partial_\tau - i\lambda_{\v i})
f_{\v i\sigma} \nonumber \\
&+&\sum_{\v i} b_{\v i}^\ast(\partial_\tau - i\lambda_{\v i} + \mu_B ) b_{\v i} \nonumber \\
&-& \sum_{\v i\v j}{t}_{\v i\v j}b_{\v i}b_{\v j}^\ast f_{\v i\sigma}^\dagger f_{\v j\sigma} ,
\label{Eq.44}
\en
where
\be
U_{\v i\v j}=
\bpm -\chi_{\v i\v j}^\ast&\Delta_{\v i\v j}\\
\Delta_{\v i\v j}^\ast&\chi_{\v i\v j} \epm.
\label{Eq.45} 
\en
At half filling $b = \mu_B =0$ and the mean field solution corresponds to
$\lambda_{\v i} = 0$.  The Lagrangian is invariant under
\be
\Phi_{\v i\si} &\rightarrow&  W_{\v i}\Phi_{\v i\si} \\
U_{\v i\v j} &\rightarrow& W_{\v i} U_{\v i\v j} W_{\v j}^\dagger
\en
where $W_{\v i}$ is an $SU(2)$ matrix [see eq.~(\ref{Eq.18})].  We reserve a fuller discussion of the $SU(2)$ gauge symmetry to Section X, but here we just give a simple example.
In terms of the link variable $U_{\v i\v j}$, the $\pi$-flux and $d$-RVB states are
represented as
\be  
U_{\v i\v j}^{\pi\text{-flux}} = -\chi ( \tau^3 - i(-1) ^{i_x+j_y} ),
\label{Eq.49}
\en
and
\be
U_{\v i,i+\mu}^d = -\chi ( \tau^3 + \eta_\mu \tau^1 ),
\label{Eq.50}
\en
respectively.  These two are related by
\be
U_{\v i\v j}^{SF} = W_{\v i}^\dagger U_{\v i\v j}^d W_{\v j}
\label{Eq.51}
\en
where
\be
W_{\v j} = \exp \left[
i(-1)^{j_x+j_y} {\pi\over 4} \tau^1 \right].
\label{Eq.52}
\en
Therefore the $SU(2)$ transformation of the fermion variable
\be
\Phi'_{\v i} = W_{\v i} \Phi_{\v i}
\label{Eq.53}
\en
relates the $\pi$-flux and d-RVB states.  Here some remarks are in order.
First it should be noted that we are discussing the Mott insulating state and
its spin dynamics. The charge transport is completely suppressed by the
constraint eq. (\ref{Eq.34}). This will be discussed in sec. X where the mean
field theory is elaborated into gauge theory.  Secondly, it is now established
that the ground state of the two-dimensional antiferromagnetic Heisenberg
model shows the antiferromagnetic long range ordering (AFLRO).  This
corresponds to the third (and most naive) way of decoupling the exchange
interaction, \ie
\be
{\v{S}}_{\v i}\cdot {\v{S}}_{\v j} = 
{1\over 4} f_{\v i\alpha}^\dagger \sigma^{\mu}_{\alpha \beta} f_{\v i \beta}
f_{\v j \gamma}^\dagger \sigma^{\mu}_{\gamma \delta} f_{\v j\delta}
\label{Eq.54}
\en
However even with the AFLRO, the singlet formation represented by $\chi_{\v i\v j}$
and $\Delta_{\v i\v j}$ dominates and AFLRO occurs on top of it. This view has been
stressed by Hsu \cite{H9079,HMA9166} generalizing the $\pi$-flux state to
include the AFRLO, and is in accord with the energetics of the projected
wavefunctions, as discussed in section VI.A.

\begin{figure}[t]
\centerline{
\includegraphics[width=2.5in]{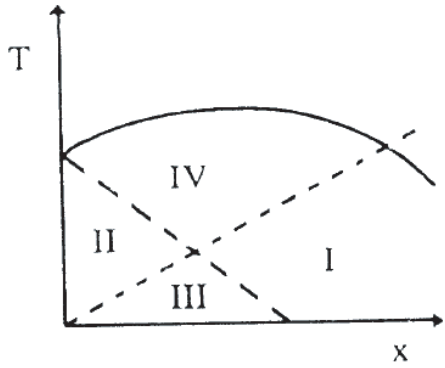}
}
\vspace{0.5cm}
\caption{ 
Schematic phase diagram of the $U(1)$ mean field theory.  The solid line
denotes the onset of the uniform RVB state $(\chi \neq 0)$.  The dashed line
denotes the onset of fermion pairing $(\Delta \neq 0)$ and the dotted line
denotes mean field Bose condensation $(b \neq 0)$.  The four regions are
(I)~Fermi liquid $\chi \neq 0$, $b \neq 0$; (II)~spin gap $\chi \neq 0$,
$\Delta \neq 0$; (III)~$d$-wave superconductor $\chi \neq 0$, $\Delta \neq 0$,
$b \neq 0$; and (IV)~ strange metal $\chi \neq 0$ \cite{LN9221}.
}
\label{U1}
\end{figure}

Now we turn to the doped case, \ie $x \ne 0$.  Then the behavior of the
bosons are crucial for the charge dynamics. At the mean field theory, the
bosons are free and condensed at $T_{BE}$.  In three-dimensional system,
$T_{BE}$ is finite while $T_{BE} = 0$ for purely two-dimensional system.
Theories assume weak three dimensional hopping between layers, and obtain the
finite $T_{BE}$ roughly proportional to the boson density $x$
\cite{KL8842,SHF8868}.  This materializes the original idea by Anderson
\cite{A8796} that the preformed spin superconductivity (RVB) turns into the
real superconductivity via the Bose condensation of holons.  
\Ref{KL8842}
and 
\Ref{SHF8868} found the d-wave superconductivity in the
slave-boson mean field theory presented above, and the schematic phase diagram
is given in Fig.~\ref{U1}.  There are 5 phases classified by the order
parameters $\chi$, $\Delta$, and $b=\<b_{\v i}\>$ for the Bose condensation.  In
the incoherent state at high temperature, all the order parameters are zero.
In the uniform RVB state (IV in Fig.~\ref{U1}), only $\chi$ is finite. In the
spin gap state (II), $\Delta$ and $\chi$ are nonzero while $b=0$. This
corresponds to the spin singlet ``superconductivity'' with the incoherent
charge motion, and can be viewed as the precursor phase of the
superconductivity. 
This state has been interpreted as the pseudogap phase \cite{F9287}.
We note that at the mean field level, the $SU(2)$ symmetry is broken by the
nonzero $\mu_B$ in eq.~(\ref{Eq.44}) and the $d$-wave pairing state is chosen
because it has lower energy than the staggered flux state.  We shall return to
this point in Section X.  In the Fermi liquid state (I), both $\chi$ and $b$
are nonzero while $\Delta=0$. This state is similar to the slave-boson
description of heavy fermion state.  Lastly when all the order parameter is
nonzero, we obtain the $d$-wave superconducting state (III). This mean field
theory, in spite of its simplicity, captures rather well the experimental
features as described in sections III and IV.

Before closing this section, we mention the slave fermion method and its mean
field theory (\cit{AA8816}; \cit{Y8916}; \cit{CRK9019}).  One can exchange the
statistics of fermion and boson in eqs.  (\ref{Eq.31}) and (\ref{Eq.33}). Then
the bosons has the spin index, \ie $b_{\v i \sigma}$ while the fermion
becomes spinless, \ie $f_{\v i}$. This boson is called Schwinger boson, and
is suitable to describe the AFLRO state. The large $N$-limit of Schwinger
boson theory gives the AFLRO state for $S=1/2$.  The holes are represented by
the spinless fermion forming a small hole pockets around $k=(\pi/2,\pi/2)$.
The size of the hole pocket is twice as large as the usual doped SDW state due
to the absence of the spin index. Therefore the slave fermion method violates
the Luttinger theorem.  Finally we mention that by introducing a phase-string
in the slave fermion approach, one obtains a phase-string formulation of high
$T_c$ superconductivity (\cit{WST0067}; \cit{W0361}).  In such an approach both
spin-1/2 neutral particles and spin-0 charged particles are bosons with a
non-trivial mutual statistics between them.

\section{$U(1)$ gauge theory of the uRVB state}

The mean field theory only enforces the constraint on the average.
Furthermore, the fermions and bosons introduce redundancy in representing the
original electron, which results in an extra gauge degree of freedom.  The
fermions and bosons are not gauge invariant and should not be thought of as
physical particles.  To include these effects we need to consider fluctuations
around the mean field saddle points, which immediately become gauge theories,
as first pointed out by 
\Ref{BA8880}.  Here, we review the
early work on the $U(1)$ gauge theory, which treats gauge fluctuations on the
Gaussian level 
(\cit{IL8988}; \cit{NL9050}; \cit{LN9221}; \cit{IK9048}).  The theory can be
worked out in some detail, leading to a nontrivial recipe for obtaining
physical response functions in terms of the fermion and boson ones, called the
Ioffe-Larkin composition rule.  It highlights the importance of calculating
gauge invariant quantities and the fact that the fermion and bosons only enter
as useful intermediate steps.  The Gaussian $U(1)$ gauge theory was mainly
designed for the high temperature limit of the optimally doped cuprate, \ie
the so-called strange metal phase in Fig.~\ref{U1}.  We will describe its
failure in the underdoped region, which leads to the $SU(2)$ formulation of
the next two sections.  The Gaussian theory also misses the confinement
physics which is important for the ground state.

\subsection{Effective gauge action and non-Fermi-liquid behavior}


As has been discussed in section III, the phenomenology of the optimally doped
Mott insulator is required to describe the two seemingly contradicting
features, \ie the doped insulator with small hole carrier concentration and
the electrons forming the large Fermi surface. The former is supported various
transport and optical properties, representatively the Drude weight
proportional to $x$, while the latter by the angle resolved photoemission
spectra (ARPES) in the normal state of optimal doped samples. In the
conventional single-particle picture, the reduction of the 1st Brillouin zone
due to the antiferromagnetic long range ordering (AFLRO) distinguishes these
two. Namely small hole pockets with area $x$ are formed in the reduced
1st BZ in the AFLRO state, while the large metallic Fermi surface of area
$1-x$ appears otherwise. The challenge for the theory of the optimally doped
case is that aspects of the doped insulator appear in some experiments even
with the large Fermi surface. Also it is noted that the ARPES shows that there
is no sharp peak corresponding to the quasi-particle in the normal state,
especially at the anti-nodal region near ${\v k} = (\pi,0)$. The fermi
surface is defined by a rather broad peak dispersing near the Fermi energy.
These strongly suggests that the normal state of high temperature
superconductors is not described in terms of the usual Landau Fermi liquid
picture. 

A promising theoretical framework to describe this dilemma is the slave-boson
formalism introduced above. It has the two species of particles, \ie
fermions and bosons, due to the strong correlation, and the electron is
``fractionalized'' into these two particles.  However, one must not take
naively this conclusion, because the fermions and bosons cannot be regarded as
``physical'' particles in that they are not gauge invariant as explained
below.  Furthermore, they are not noninteracting particles; they are strongly
coupled to the gauge field. This arises from the fact that the conservation of
the gauge charge $Q_{\v i} = \sum_\sigma f^\dagger_{\v i \sigma} f_{\v i \sigma} +
b^\dagger_{\v i} b_{\v i}$ can be derived by the Noether theorem starting from the local
$U(1)$ gauge transformation
\be
f_{\v i \sigma} \to e^{\v i \varphi_{\v i}} f_{\v i \sigma}
\nonumber  \\
b_{\v i} \to e^{i \varphi_{\v i}} b_{\v i \sigma}.
\label{Eq.55}
\en
Therefore the constraint  eq.~(\ref{Eq.34})  is equivalent to a local gauge
symmetry.  The Green's functions for fermions and bosons $G_F(\v i,\v j;\tau) = -
\<T_\tau f_{\v i \sigma}(\tau) f^\dagger_{\v j \sigma} \>$ and $G_B(\v i,\v j;\tau) = -
\<T_\tau b_{\v i}(\tau) b^\dagger_{\v j} \>$ transforms as
\be
G_F(\v i,\v j;\tau) \to e^{i (\varphi_{\v i} - \varphi_{\v j})} G_F(\v i,\v j;\tau)
\nonumber \\
G_B(\v i,\v j;\tau) \to e^{i (\varphi_{\v i} - \varphi_{\v j})} G_B(\v i,\v j;\tau).
\label{Eq.56}
\en
Therefore these fermions and bosons are not gauge invariant and should be
regarded as only the particles which are useful in the intermediate step of
the theory to calculate the physical (gauge invariant) quantities as will be
done in the next section. 

At the mean field level, the constraint was replaced by the averaged one
$\<Q_{\v i}\> = 1$. This average is controlled by the saddle point value of the
Lagrange multiplier field $\<\lambda_{\v i}\> = \lambda$. Originally $\lambda_{\v i}$ is
the functional integral variable and is a function of (imaginary) time. When
this integration is done exactly, the constraint is imposed. Therefore we have
to go beyond the mean field theory and take into account the fluctuation
around it.  In other words, the local gauge symmetry is restored by the gauge
fields which transform as
\be
a_{\v i\v j} \to a_{\v i\v j} + \varphi_{\v i} - \varphi_{\v j}
\nonumber \\
a_0 (\v i)  \to  a_0(\v i) + { {\partial \varphi_{\v i}(\tau)} \over {\partial \tau}},
\label{Eq.57}
\en
corresponding to eq.~(\ref{Eq.55}).  The fields satisfying this condition are
already in the Lagrangian eq.~(\ref{Eq.39}). Namely the phase of the HS
variable $\chi_{\v i\v j}$ and the fluctuation part of the Lagrange multiplier
$a_0(\v i) = \lambda_{\v i}$ are these fields. 

Let us study this $U(1)$ gauge theory for the uRVB state in the phase diagram
Fig.~\ref{U1}. This state is expected to describe the normal state of the
optimally doped cuprates, where the $SU(2)$ particle-hole symmetry 
described by eq.~(\ref{Eq.37}) is not so important.   Here we neglect
$\Delta$-field, and consider $\chi$ and $\lambda$ field. There are amplitude
and phase fluctuations of $\chi$-field, but the former one is massive and does
not play important roles in the low energy limit.  Therefore the relevant
Lagrangian to start with is
\be
L_1 &=& \sum_{ i, \sigma} f^*_{\v i \sigma} 
\biggl( {\partial \over {\partial \tau}} - \mu_F + i a_0({\v r}_{\v i})
\biggr) f_{\v i \sigma} 
\nonumber \\
 &+& \sum_{ i} b^*_{\v i} 
\biggl( {\partial \over {\partial \tau}} - \mu_B + i a_0({\v r}_{\v i})
\biggr) b_{\v i} 
\nonumber \\
&-& {\tilde J}\chi \sum_{\<\v i\v j\> \sigma} 
( e^{i a_{\v i\v j}} f^*_{\v i \sigma} f_{\v j \sigma} + h.c.) \nonumber \\
&-& t \eta \sum_{\<\v i\v j\>} 
( e^{i a_{\v i\v j}} b^*_{\v i} b_{\v j} + h.c.)
\label{Eq.58}
\en
where $\eta$ is the saddle point value of another HS variable to decouple the
hopping term.  We can take $\eta = \chi$ using eq.~(\ref{Eq.39a}). Here the
lattice structure and the periodicity with respect to $a_{\v i\v j} \to a_{\v
i\v j} + 2 \pi$ are evident, and the problem is that of the lattice gauge
theory coupled to the fermions and bosons. It is also noted here that there is
no dynamics of the gauge field at this starting Lagrangian. Namely the
coupling constant of the gauge field is infinity, and the system is in the
strong coupling limit.  This is because the gauge field represents the
constraint; by integrating over the gauge field we obtain the original problem
with the constraint.  This raises the issue of confinement as will be
discussed in section  IX.D and XI.F.  Here we exchange the order of the
integration between the gauge field ($a_{\v i\v j},a_0$) and the matter fields
(fermions and bosons). Namely the matter fields are integrated over first, and
we obtain the effective action for the gauge field.
\be
e^{- S_{\rm eff.} (a)} = 
\int D f^* Df D b^* Db e^{ - \int_0^\beta L_1 }
\label{Eq.59}
\en
However this integration can not be done exactly, and the approximation is
introduced here. The most standard one is the Gaussian approximation or RPA,
where the effective action is obtained by perturbation theory up to the
quadratic order in $a$.  For this purpose we introduce here the continuum
approximation to the Lagrangian $L_1$ in eq.~(\ref{Eq.58}).
\be
L &=& \int d^2 {\v r} 
\biggl[  \sum_\sigma f^*_{ \sigma}({\v r}) 
\biggl( {\partial \over {\partial \tau}} - \mu_F + i a_0({\v r})
\biggr) f_{ \sigma}({\v r}) 
\nonumber \\
 &+& b^*({\v r}) 
\biggl( {\partial \over {\partial \tau}} - \mu_B + i a_0({\v r})
\biggr) b({\v r}) 
\nonumber \\
 &-& {1 \over {2 m_F}} 
\sum_{\sigma, j = x,y} 
 f^*_{ \sigma}({\v r}) 
\biggl( {\partial \over {\partial x_j} } + i a_j
\biggr)^2 f_{ \sigma}({\v r}) 
\nonumber \\
 &-& {1 \over {2 m_B}} 
\sum_{j = x,y} 
b^*({\v r}) 
\biggl( {\partial \over {\partial x_j} } + i a_j
\biggr)^2 b({\v r}) 
\biggr],
\label{Eq.60}
\en
where the vector field ${\v a}$ is introduced by $a_{\v i\v j} = ({\v r}_{\v i} - {\v
r}_{\v j}) \cdot {\v a}[ ({\v r}_{\v i} + {\v r}_{\v j})/2]$.  Note $1/m_F \approx J$ and
$1/m_B \approx t$.  The coupling between the matter fields and gauge field is
given by
\be
L_{\rm int.} = \int d^2 {\v r} (j^F_\mu + j^B_\mu) a_\mu
\label{Eq.61}
\en
where $j^F_\mu$ ($j^B_\mu$) is the fermion (boson) current density. 

Note that integration over $a_0$ recovers the constraint eq.~(\ref{Eq.34}) and
integration over the vector potential $\v{a}$ yields the constraint
\be
\v{j}_F + \v{j}_B = 0 ,
\label{Eq.61A}
\en
\ie the fermion and boson can move only by exchanging places.  Thus the
Gaussian approximation apparently enforces the local constraint exactly
\cite{L0094}.  We must caution that this is true only in the continuum limit,
and an important lattice effect related to the $2\pi$ periodicity of the phase
variable has been ignored.  These latter effects lead to instantons and
confinement, as will be discussed later in section IX.D.  Thus it is not
surprising that the ``exact'' treatment of D.H. Lee yields the same
Ioffe-Larkin composition rule which is derived based on the Gaussian theory
(see section IX.C).

We now proceed to reverse the order of integration.  We integrate out the
fermion and boson fields to obtain an effective action for $a_\mu$.  We then
consider the coupling of the fermions and bosons to the gauge fluctuations
which are controlled by the effective action.  To avoid double counting, it
may be useful to consider this procedure in the renormalization group sense,
\ie we integrate out the high energy fermion and boson fields to produce an
effective action of the gauge field which in turn modifies the low energy
matter field. This way we convert the initial problem of infinite coupling to
one of finite coupling.  The coupling is of order unity but may be formally
organized as a $1/N$ expansion by artificially introducing $N$ species of
fermions.  Alternatively, we can think of this as an RPA approximation, \ie
a sum of fermion and boson bubbles.  The effective action for $a_\mu$ is given
by  the following 
\be
S_{\rm eff.}^\text{RPA} (a) = 
(\Pi^F_{\mu \nu} (q) + \Pi^B_{\mu \nu} (q) ) a_\mu(q) a_\nu(-q)
\label{Eq.62}
\en
where $q= ({\v q},\omega_n)$ is a three dimensional vector.  The
current-current correlation function $\Pi^F_{\mu \nu}(q)$ ($\Pi^B_{\mu
\nu}(q)$) of the fermions (bosons) is given  by
\be
\Pi^\alpha_{\mu\nu}(q) = \< j_\mu^\alpha(q) j_\nu^\alpha(-q)\>
\label{Eq.63}
\en
with $\alpha=F,B$.  Taking the transverse gauge by imposing the gauge fixing
condition
\be
\nabla \cdot {\v a} = 0
\label{Eq.64}
\en
the scalar ($\mu=0$) and vector parts of the gauge field dynamics are
decoupled. The scalar part $\Pi^\alpha_{00}(q)$ corresponds to the
density-density response function and does not show any singular behavior in
the low energy/momentum limit. On the other hand, the transverse
current-current response function shows singular behavior for small $\v q$ and
$\om$.  Explicitly the fermion correlation function is given by
\be
\Pi^F_T(q) = i \omega \sigma^T_{F1}({\v q},\omega) - \chi_F 
{\v q}^2
\label{Eq.65}
\en
where $\chi_F = 1/(24 \pi m_F)$ is the fermion Landau diamagnetic
susceptibililty.  The first term describes the dissipation and the static limit
of $\sigma^T_{F1}$ (real part of the fermion conductivity) for $\omega <
\gamma_{\v q}$ is $\sigma^T_{F1}({\v q},\omega) = \rho_F/(m_F \gamma_{\v
q})$ where $\rho_F$ is the fermion density and
\be
\gamma_{\v q} &=& \tau_{\rm tr}^{-1} \ \ \ {\rm for}\
|{\v q}| < (v_F  \tau_{\rm tr})^{-1}
\nonumber \\
 &=& v_F |{\v q}|/2  \ \ \ {\rm for}\ 
|{\v q}| > (v_F  \tau_{\rm tr})^{-1}
\label{Eq.66}
\en
where $\tau_{\rm tr}$ is the transport lifetime due to the scatterings by the
disorder and/or the gauge field.  A similar expression is obtained for the
bosonic contribution as
\be
\Pi^B_T(q) = i \omega \sigma^T_{B1}({\v q},\omega) - \chi_B 
{\v q}^2
\label{Eq.67}
\en
where $\chi_B = n(0)/(48 \pi m_B)$ where $n(\epsilon) $ is the Bose occupation
factor.  $\chi_B$ diverges at the Bose condensation temperature $T_{BE}^{(0)}
= 2 \pi x/ m_B$ when we assume a weak 3D transfer of the bosons. Assuming that
the temperature is higher than $T_{BE}^{(0)}$, the boson conductivity is
estimated as
\be
\sigma^T_{B1}
 \cong x^{1/2}/|{\v q}|
\label{Eq.68}
\en
for $|{\v q}|> \ell_B^{-1}$, where $\ell_B$ is the mean free path of the
bosons. It can be seen from eqs. (\ref{Eq.66}) and (\ref{Eq.68}),
$\sigma^T_{B1} \ll \sigma^T_{F1}$.

Summarizing, the propagator of the transverse gauge field is given by
\be
\<a_\alpha(q) a_{\beta}(-q)\> = ( \delta_{\alpha \beta}
- q_\alpha q_\beta / |{\v q}|^2 ) D_T(q)
\label{Eq.69}
\en
\be
D_T(q) = [\Pi^F_T(q) + \Pi^B_T(q) ]^{-1}
\cong [ i \omega \sigma ({\v q}) - \chi_d {\v q}^2 ]^{-1}.
\label{Eq.70}
\en
Here  
\be
\sigma({\v q}) &\cong& k_0/|{\v q}| \ \ \ {\rm for}\ |{\v q}|\ell>1
\nonumber \\
 &\cong& k_0 \ell \ \ \ {\rm for}\ |{\v q}|\ell<1
\label{Eq.71}
\en
where $\ell$ is the fermion mean free path and $k_0$ is of the order $k_F$ of
the fermions.

This gauge field is coupled to the fermions and bosons and leads to their
inelastic scatterings.  By estimating the lowest order self-energies of the
fermion and boson propagators, it is found that these are diverging at any
finite temperature.  It is because of the singular behavior of $D_T(q)$ for
small $|{\v q}|$ and $\omega$.  This kind of singularity was first noted by
\Ref{R8902} for the problem of electrons coupled to a transverse
electromagnetic field, even though related effects such as non-Fermi liquid
corrections for the specific heat have been noted earlier 
\cite{HNP7349}.  However this does not cause any trouble since the propagators
of fermions and bosons are not the gauge invariant quantity and hence is not
physical as discussed above.  As the representative of gauge invariant
quantities, we consider the conductivity of fermions and bosons.  (Note that
these are not still ``physical'' because one must combine these to obtain the
physical conductivity as discussed in the next section.) For example the
integral for the (inverse of) transport life-time $\tau_{\rm tr}$ contains the
factor $1 - \cos \theta$ where $\theta$ is the angle between the initial and
final momentum for the scattering. This factor scales with $|{\v q}|^2$ for
small ${\v q}$, and gets rid of the divergence.  The explicit estimate gives
\be
{ 1 \over {\tau^F_{\rm tr}} } &\cong& \xi_{\v k}^{4/3} \ \ \ {\rm for} \ \ 
\xi_{\v k} > kT
\nonumber \\
 &\cong& T^{4/3} \ \ \ {\rm for} \ \ 
\xi_{\v k} < kT
\label{Eq. 72}
\en
for the fermions while
\be
{ 1 \over {\tau^B_{\rm tr}} } &\cong& 
{{ k T} \over { m_B \chi_d} }
\label{Eq.73}
\en
for bosons.  These results are interpreted as the scattering by the
fluctuating gauge flux 
whose propagator is given by the loop representing the particle-hole
propagator for the two-particle current-current correlation function.
 
Now some words on the physical meaning of the gauge field are in order. For
simplicity let us consider the three sites, and that the electron is moving
around these.  The quantum mechanical amplitude for this process is
\be
P_{123} = \<\chi_{12} \chi_{23} \chi_{34}\>
=\< f^\dagger_{1 \alpha} f_{2 \alpha}  
 f^\dagger_{2 \beta} f_{3 \beta}  
 f^\dagger_{3 \gamma} f_{1 \gamma}\>.  
\label{Eq.74}
\en
One can prove that
\be
(P_{123} - P_{132})/(4i)  =  {\v S}_1 \cdot
( {\v S}_2 \times {\v S}_3)
\label{Eq.75}
\en
and the righthand side of the above equation corresponds to the solid angle
subtended by the three vectors ${\v S}_1,{\v S}_2, {\v S}_3$, and is called
spin chirality \cite{WWZcsp}.  Therefore the gauge field fluctuation is
regarded as that of the spin chirality. Recently it is discussed that the spin
chirality will produce the anomalous Hall effect in some ferromagnets such as
manganites and pyrochlore oxides, where the non-coplanar spin configurations
are realized by thermal excitation of the Skymion or the strong spin anisotropy
in the ground state 
(\cit{YKM9937}; \cit{TOY0173}).  This
phenomenon can be interpreted as the static limit of the gauge field, while
the gauge field discussed here has both quantum and thermal fluctuations.

\subsection{Ioffe-Larkin composition rule}

In order to discuss the physical properties of the total system, we have to
combine the information obtained for fermions and bosons.  This has been first
discussed by 
\Ref{IL8988}.  Let us start with the physical
conductivity $\sigma$, which is given by
\be
\sigma^{-1} = \sigma_F^{-1} + \sigma_B^{-1}
\label{Eq.76}
\en
in terms of the conductivities of fermions ($\sigma_F$) and bosons
($\sigma_B$).  This formula corresponds to the sequential circuit (not
parallel) of the two resistance, and is intuitively understood from the fact
that both fermions and bosons have to move subject to the constraint.  This
formula can be derived in terms of the shift of the gauge field $\v{a}$, and
resultant backflow effect.  In the presence of the external electric field
${\v E}$, the gauge field $\v{a}$ and hence the internal electric field ${\v
e}$ is induced.  Let us assume that the external electric field ${\v E}$ is
coupled to the fermions. Then the effective electric field 
seen by the fermions 
is
\be
{\v e}_F = {\v E} + {\v e}
\label{Eq.77}
\en
while that for the boson is
\be
{\v e}_B = {\v e}.
\label{Eq.78}
\en
The fermion current ${\v j}_F$ and boson current ${\v j}_B$ are induced,
respectively as
\be
{\v j}_F = \sigma_F {\v e}_F, \ \ \ {\v j}_B = \sigma_F {\v e}_B.
\label{Eq.79}
\en
The constraint ${\v j}_F + {\v j}_B = {\v 0}$ given by eq.~(\ref{Eq.61A})
leads to the relation
\be
{\v e} = - { {\sigma_F} \over { \sigma_F + \sigma_B} } {\v E} .
\label{Eq.81}
\en
The physical current ${\v j}$ given by
\be
{\v j} = {\v j}_F = - {\v j}_B = 
{ {\sigma_F \sigma_B } \over { \sigma_F + \sigma_B} } {\v E}
\label{Eq.82}
\en
leading to the expression for the physical conductivity $\sigma$ in
eq.~(\ref{Eq.76}).  It is also noted here that the same result 
is obtained 
if instead we couple the e.m. field to bosons. 
In this case the internal electric field ${\v e}$ is different, but ${\v
e}_F$ and ${\v e}_B$ remain unchanged.  Therefore it is not a physical
question which particle is charged, \ie fermion or boson. This is related to
the fact that both fermions and bosons are not physical particles as
repeatedly stated.  Note that $\sigma_F \gg \sigma_B$ in the uRVB state, we
conclude that $\sigma \cong \sigma_B = x \tau_{\rm tr}^B/m_B$ which is
inversely proportional to the temperature $T$. Furthermore the Drude weight
of the optical conductivity is determined by $x/m_B$ as is observed
experimentally. It remains true that the superfluidity density $\rho_S$ in the
superconducting state is given by the missing oscillator strength below the
gap, this also means that $\rho_s \propto x$.

A more formal way of deriving the physical electromagnetic response follows.
We can generalize the discussion of the effective action $S_{\rm eff.}(\v{a})$
for the gauge field to include the external e.m. field $A_\mu$. Let us couple
$A_\mu$ again to the fermions. Then the effective action becomes instead of
eq.~(\ref{Eq.62})
\be
S_{\rm eff.}^\text{RPA} (a,A) &=& 
\Pi^F_{\mu \nu} (q) (a_\mu(q)+ A_\mu(q))( a_\nu(-q) + A_\nu(-q))
\nonumber \\
 &+& \Pi^B_{\mu \nu} (q)  a_\mu(q) a_\nu(-q).
\label{Eq.83}
\en
Then after integrating over the gauge field $a_\mu$, we end up with the
effective action for $A_\mu$ only as
\be
S_{\rm eff.}^\text{RPA} (A) = 
\Pi_{\mu \nu} (q) A_\mu(q)A_\nu(-q)
\label{Eq.84}
\en
with the physical e.m. response function 
\be
\Pi_\alpha(q)^{-1} = (\Pi_\alpha^F(q))^{-1}+(\Pi_\alpha^B(q))^{-1}
\label{Eq.85}
\en
where $\alpha = 0$ or $T$ stands for the longitudinal and the transverse
parts.  Then the physical diamagnetic susceptibility ${\tilde \chi}$ is given
by ${\tilde \chi}^{-1} = \chi_F^{-1}+ \chi_B^{-1}$.  Again in the
superconducting state, $\Pi_T^F \propto \rho^F_s$ and $\Pi_T^B \propto
\rho^B_s$, where $\rho^F_s$ and $\rho_s^B$ are superfluidity density of the
fermion pairing and boson condensation.  This leads to the composition rule
for $\rho_s$ as $\rho_s^{-1} = (\rho^F_s)^{-1} + (\rho_s^B)^{-1} \cong
(\rho^B_s)^{-1} \propto x^{-1}$ with $\rho^F_s \gg \rho_s^B $, reproducing the
same result as suggested from the Drude weight.  
On the other hand the temperature dependence of $\rho_s^F$ is of the form
$\rho_s^F(T)=\rho_s^F(0)(1-aT)$ where $a$ is given by the nodal fermion
dispersion, while the temperature dependence of $\rho_s^B$ is expected to be 
higher power in $T$ and negligible. The Ioffe-Larkin composition rule then
predicts that
\begin{align}
\label{rhosT}
 \rho_s(T) &\approx \rho_s^B(1-\frac{\rho_s^B}{\rho_s^F}) \nonumber\\
&\approx \rho_s^B(0) - \frac{(\rho_s^B(0))^2}{\rho_s^F(0)} aT .
\end{align} 
Since $\rho_s^B(0)\sim x$, this predicts that the temperature dependence of
the superfluid density is proportional to $x^2$.
Comparison with \Eq{Eq.5} implies that $\al\sim x$ in the slave-boson theory.
As shown in Fig. \ref{Lemberger}, this prediction does not agree with
experiment and is probably an indication of the breakdown of gaussian
fluctuations which underlines the Ioffe-Larkin rule.

We conclude this section by remarking that the Ioffe-Larkin rule can be extended to
various other physical quantities.  For example the Hall constant $R_H$ is
given by
\be
R_H = { {R_H^F \chi_B + R_H^F \chi_F} \over
{\chi_B + \chi_F} }
\label{Eq.86}
\en
while the thermopower $S = S_B + S_F$ and the electronic thermal conductivity
$\kappa = \kappa_B + \kappa_F$ are sum of the bosonic and fermionic
contributions.

Compared with the two-particle correlation functions discussed above, the single
particle Green's function is more complicated.  At the mean field level, the
electron Green's function is given by the product of those of fermions and
boson in the $({\v r},\tau)$ space. Therefore in the momentum-frequency
space, it is given by the convolution. The spectral function is composed of
the two contributions, one is the quasi-particle peak with the weight $\sim x$
while the other is the incoherent background. Even the former one is broadened
due to the momentum distribution of the noncondensed bosons, \ie there is no
quasi-particle peak in the strict sense.  This absence of the delta-functional
peak occurs also in the $SU(2)$ theory in sec. XI indicating that the fermions 
are not free and hence can not be regarded as the quasi-particle.  On the
other hand, the dispersion of this ``quasi-particle'' peak is determined by
that of fermions, and hence its locus of zero energy constitutes the large
Fermi surface enclosing the area $1-x$.  However this simple calculation does
not reproduce some of the novel features in the ARPES experiments such as the
``Fermi arc'' in underdoped samples, which will be discussed later in section
XI.

Combined with the discussion on the transport properties and the electron
Green's function, the present uniform RVB state in the $U(1)$ formulation
offers an explanation on the dichotomy between the doped Mott insulator and
the metal with large Fermi surface.  In particular, the conclusion that the
conductivity is dominated by the boson conductivity $\sigma \approx \sigma_B
\approx x \tau_{tr}^B/m_B \approx xt T$ explains the linear $T$ resistivity
which has been taken as a sign of non-Fermi liquid behavior from the beginning
of high $T_c$ research.  However, we must caution that this conclusion was
reached for $T > T_{BE}^{(0)}$ while in the experiment the linear $T$ behavior
persists to much lower temperature near optimal doping.  It is possible that
gauge fluctuations suppress the effective Bose condensation.  
\Ref{LKL9601}
attempted  to include the effect of strong gauge fluctuations
on the boson conductivity by assuming a quasi-static gauge fluctuation and
treating the problem by quantum Monte Carlo.  The picture is that the boson
tends to make self-retracing paths to cancel out the effect of the gauge
field \cite{NL9133}.  They indeed find that the boson conductivity remains linear in $T$
down to much lower temperature than $T_{BE}^{(0)}$.

\subsection{Ginzburg-Landau theory and vortex structure}
 
Up to now, we have focused on the uRVB state where the pairing amplitude
$\Delta$ of the fermions is zero. In this subsection we review the
phenomenological Ginzburg-Landau theory to treat this pairing field.  The free
energy for a single CuO$_2$ layer is given by
\be
F = F_F[\psi, {\v a}, {\v A}] + 
F_B[\phi, {\v a}]
+ F_{\rm gauge} [ {\v a}]
\label{Eq.87}
\en
with
\be
F_F[\psi, {\v a},{\v A}]
&=& {{H_{cF}^2} \over {8 \pi}}
\int d^2 r 
\biggl[ 2 \xi_F^2 |( \nabla - 2 i {\v a} - i { {2e} \over c}
{\v A}) \psi |^2
\nonumber \\
&+& 2 \text{sign}(T-T_D^{(0)}) |\psi|^2 + |\psi|^4 \biggr],
\label{Eq.88}
\en
\be F_B[\psi, {\v a}] &=& {{H_{cB}^2} \over {8 \pi}} \int d^2 r \biggl[ 2
\xi_B^2 |( \nabla - i {\v a}) \phi |^2 \nonumber \\ &+& 2
\text{sign}(T-T_{BE}^{(0)}) |\phi|^2 + |\phi|^4 \biggr], \label{Eq.89} \en and
\be
F_{\rm gauge}[{\v a}]
= \int d^2 r 
\biggl[ \chi_F [ \nabla \times ({\v a}+(e/c){\v A})]^2
+ \chi_B (\nabla \times {\v a})^2
\biggr]
\label{Eq.90}
\en
where ${\v A}$ is the e.m. vector potential, $c$ is the velocity of light,
and $\hbar$ is put to be unity. In the above equations, the optimal value of
the order parameter is scaled to be unity, and hence the correlation lengths
$\xi_B,\xi_F$ and the thermodynamic critical fields $H_{cF},H_{cB}$ are
temperature dependent both for fermion pairing and Bose condensation. It is
noted that the penetration length of the fermion pairing (boson condensation)
$\lambda_F$ ($\lambda_B$) is related to $H_{cF}$ ($H_{cB}$) as $H_{cF} =
\phi_0/(2 \sqrt{2} \pi \xi_F \lambda_F)$ ($H_{cB} = \phi_0/( \sqrt{2} \pi
\xi_B \lambda_B)$).  We take the lattice constant as the unit of length.  Then
$\xi_F(0) \sim J/\Delta$, $\xi_B \sim x^{-1/2}$, and the condensation energy
per unit area is given by $H_{cF}(0)^2/(8 \pi) \sim \Delta^2/J$, and
$H_{cB}(0)^2/(8 \pi) \sim t x^2$.

Now we consider the consequences derived from this GL free energy. One is on
the interplay between the Berezinskii-Kosterlitz-Thouless (BKT) transitions
for the fermion pairing and boson condensation. We consider the type II limit,
and neglect ${\v A}$ for the moment. As is well known, the binding-unbinding
of the topological vortex excitations leads to the novel phase transition (BKT
transition) in 2D.  This is due to the logarithmic divergence of the vortex
energy with respect to the sample size.  This energy is competing with the
entropy term which is also logarithmically diverging. Above some critical
temperature the entropy dominates, and the free vortex excitations are
liberated resulting in the exponential decay of the order parameter. However
this logarithmic divergence is cut-off when the order parameter is coupled to
the massless gauge field ${\v a}$. Namely the gauge field screens the vortex
current, and $| (\nabla - i{\v a}) \phi|$ and $|(\nabla - 2 i{\v a})\psi|$
decays exponentially beyond some penetration lengths. This means that the BKT
transition for the fermion pairing and boson condensation disappear when the
gauge field ${\v a}$ is massless. In other words, these two order parameters
are coupled through the gauge field, and the BKT transition occurs only
simultaneously where the gauge field becomes massive due to the Higgs
mechanism.  Therefore the phase transition lines for fermion pairing and boson
condensation in the phase diagram Fig.~\ref{U1}  become the crossover
lines and only the superconducting transition remains to be the real BKT
transition. 

Now we turn to the vortex structures in the superconducting state. The most
intriguing issue here is the quantization of the magnetic flux. Because the
boson has charge $e$ while the fermion pairing $-2e$, the question is whether
the $hc/e$ vortex may be more stable than the conventional $hc/2e$ vortex. To
study this issue, we compare the energy cost of the two types of vortex
structure, \ie (i) type A: the fermion pairing order parameter $\psi$
vanished at the core, with its phase winding around it. The boson condensation
does not vanish and the vortex core state is the Fermi liquid. The flux
quantization is $hc/2e$. (ii) type B: the Bose condensation is destroyed at the
core and the fermion pairing remains finite. Then the vortex core state is the
spin gap state. The flux quantization is $hc/e$ in this case.  The energy of
each vortex is estimated as follows. First the Ioffe-Larkin composition rule
results in the penetration length $\lambda$ of the magnetic field as 
\be
\lambda^2 = \lambda_F^2 + \lambda_B^2,
\label{Eq.91}
\en
which is equivalent to $\rho_s^{-1} = (\rho^F_s)^{-1} + (\rho_s^B)^{-1}$
derived in the previous subsection.  The contribution from the region where
the distance from the core is larger than $\xi_F,\xi_B$ is estimated similarly
to the usual case.
\be
E_0 = \biggl[ { {\phi_0} \over { 4 \pi \lambda} }\biggr]^2
\ln \biggl[ { {\lambda} \over { {\rm max}(\xi_F, \xi_B)} } \biggr]
\label{Eq.92}
\en
for the type A, and $4E_0$ for the type B because the quantized flux is
doubled in the latter case.  The core energy $E_c$ is given by the
condensation energy per area times the area of the core.  For type A vortex
\be
E_c^{(A)} \approx H_{cF}^2 \xi_F^2 \approx J
\label{Eq.93}
\en
while for type B vortex
\be
E_c^{(B)} \approx H_{cB}^2 \xi_B^2 \approx tx .
\label{Eq.94}
\en
Then the vortex energies are estimated to be $E^{(A)} \approx E_0 + E_c^{(A)}$
and $E^{(B)} = 4E_0 + E_c^{(B)}$, respectively.  Note that $E_0$ is
proportional to $\lambda^{-2}$ which is dominated by  $\lambda_B^{-2} = x$ and
hence $E_0$, $E_c^{(B)}$ are proportional to $x$ while $E_c^{(A)}$ is a
constant of order $J$.  The latter energy is in agreement with the estimate of
the vortex in the BCS theory discussed in Section V.B and is the dominant
energy for sufficiently small $x$.  We come to the conclusion that type B
vortex (with $hc/e$ flux quantization) will be more stable in the underdoped
region. This conclusion was reached by 
\Ref{S9289}
and by
\Ref{NL9266} and appears to be a general feature of the $U(1)$ gauge
theory.  Unfortunately, the experimental search for stable $hc/e$ vortices
have so far come up negative 
\cite{WBG0102}.   In section XII.C we
will describe how this problem is fixed by the $SU(2)$ gauge theory, which is
designed to be more accurate for small doping.

\subsection{Confinement-deconfinement problem}

Despite the qualitative success of the mean field and $U(1)$ gauge field theory,
there are several difficulties with this picture. One is that the gauge
fluctuations are strong and one can not have a well controlled small expansion
parameter, except rather formal ones such as the large $N$ expansion.  This
issue is closely related to the confinement problem in lattice gauge theory,
and will be discussed  below and also in section X.H and XI.F.

The coupling constant of the gauge field is defined as the inverse of the
coefficient of $f_{\mu \nu}^2$ in the Lagrangian. It is well-known that the
strong coupling gauge field leads to confinement.  In the confining phase,
only the gauge singlet particles appear in the physical spectrum, which
corresponds for example to the physical electron and magnon in the present
context. Below we give a brief introduction to this issue.

Up to now the discussion is at the Gaussian fluctuation level where the
effective action for the gauge field has been truncated at the quadratic order
in the continuum approximation.  However we are starting from the
infinite-coupling limit, and even if the finite coupling is produced by
integrating over the matter field, the strong coupling effect must be
considered seriously. In the original problem the gauge field is defined on
the lattice and the periodicity with respect to $a_{\v i\v j} \to a_{\v i\v j} + 2 \pi$
must be taken into account. Namely the relevant model is that of the {\em
compact} lattice gauge theory.  Let us first consider the most fundamental
model without the matter field;
\be
S_{\rm gauge}  =  - { 1 \over g} \sum_{\rm plaquette}
( 1 - \cos f_{\mu \nu})
\label{Eq.95}
\en
where 
\be
f_{\mu \nu} =
  a_{\v i, \v i+\v \mu} 
+ a_{\v i+\v \mu, \v i+ \v \mu + \v \nu}
- a_{\v i+ \v \nu, \v i + \v \mu + \v \nu} 
- a_{\v i, \v i +\v \nu}  
\label{Eq.96}
\en
is the flux penetrating through the plaquette in the (d+1)-dimensional space,
and $\mu,\nu=x,y,\cdots$.
Now $S_{\rm gauge}$ is a periodic function of $f_{\mu\nu}$ with period $2\pi$
and one can consider tunneling between different potential minima.  This leads
to the ``Bloch state'' of $f_{\mu \nu}$ when the potential barrier height
$1/g$ is low enough, while it is ``localized'' near one minimum when $1/g$ is
high.  The former corresponds to the quantum disordered $f_{\mu \nu}$, and
leads to the linear confining force as shown below (confining state).  On the
other hand, in the latter case, one can neglect the compact nature of the
gauge field, and the analysis in previous sections are justified (deconfining
state).  For this confinement-deconfinement transition, one can define the
following order parameter, \ie the Wilson loop:
\be 
W( C )= \< \exp [ i q \oint_C d {x}_\mu a_\mu (x) ]\>  
\label{Eq.97}
\en
where the loop $C$ consists of the paths of length $T$ along the time
direction and those of length $R$ along the spatial direction. It is related
to the gauge potential $V(R)$ between the two static gauge charges $\pm q$
with opposite sign put at the distance $R$ as
\be
W( C ) = \exp[ - V(R)T ].
\label{Eq.98}
\en
There are two types of behavior of $W(C)$, \ie (i) area law: $W( C ) \sim
e^{- \alpha RT }$, and (ii) perimeter law: $W( C ) \sim e^{- \beta (R+T) }$,
where $\alpha,\beta$ are constants.
In the first case (i), the potential $V(R)$ is increasing linearly 
in $R$, and hence the two gauge charges can never be free. Therefore it
corresponds to confinement, while the other case (ii) to 
deconfinement.

It is known that the compact QED (pure gauge model) 
in (2+1)D is always confining however small the coupling 
constant is \cite{Pol87}.
The argument is based on the instanton configuration,
which is enabled by the compactness of the gauge field.
This instanton is the source of the flux with the field distribution 
\be 
{\v b}({\v x}) = { { {\v x}} \over { 2|{\v x}|^3} }.
\label{Eq.99}
\en
where ${\v x} = ({\v r},\tau)$ is the (2+1)D coordinates in the imaginary time
formalism, and ${\v b}({\v x}) = ( e_y({\v x}), - e_x({\v x}), b({\v x}))$ is
the combination of the ``electric field'' $e_\alpha({\v x})$ and ``magnetic
field'' $b({\v x})$.  This corresponds to the tunneling phenomenon of the flux
because the total flux slightly above (future) or below (past) of the
instanton differs by $2 \pi$. The anti-instanton corresponds to the sink of
the flux.  This instanton/anti-instanton corresponds to the singular
configuration in the continuous approximation, but is allowed in the compact
model on a lattice.  Therefore (anti)instantons take into account the compact
nature of the original model in the continuum approximation.  It is also clear
from eq. (\ref{Eq.99}) that the (anti)instanton behaves as the (negative)
positive magnetic charge. Then it is evident that when we plug in the
(anti)instanton configurations into the action
\be
S= \int d^3 {\v x}{ 1 \over {2g}}  [{\v b}({\v x})]^2
\label{Eq.100}
\en
( $g$ is the coupling constant), we obtain the Coulomb $1/|{\v
x}|$-interaction between the (anti)instantons as
\be
S_{\rm inst} =  \sum_{i<j} {{q_i q_j} \over {| {\v x}_i - {\v x}_j |} }.
\label{101}
\en
where $q_i$ is the magnetic charge, which is $\sqrt{g}/2$ for instanton and $
-\sqrt{g}/2$ for anti-instanton.

Now it is well-known that the Coulomb gas in 3D is always in the screening
phase, namely the long range Coulomb interaction is screened to be the short
range one due to the cloud of the opposite charges surrounding the charge.
Therefore the creation energy of the (anti)instanton is finite and the free
magnetic charges are liberated. This free magnetic charges disorder the gauge
field and  makes the Wilson loop show the area law, \ie confinement. 

The discussion up to now is for the pure gauge model without matter field.
With matter field the confinement issue becomes very subtle since the Wilson
loop does not work as the order parameter any more. Furthermore the
confinement disappears above some transition temperature even in the pure
gauge model. In the presence of matter field, the confinement-deconfinement
transition at finite temperature is replaced by the gradual crossover to the
plasma phase in the high temperature limit (\cit{P7877}; \cit{S7910}; \cit{S861}). 
Therefore we can expect that the
slave-boson theory without  confinement describes the physics of the
intermediate energy scale even though the ground state is the confining state.
Indeed, within the $U(1)$ gauge theory, the ground states are either
antiferromagnetic, superconductor or Fermi-liquid and are all confining.
Nevertheless, the pseudogap region which exists only at finite temperatures
may be considered ``deconfined'' and describable by fermions and bosons
coupled to noncompact gauge fields.  We emphasize once again that in this
scenario the fermions and bosons are not to be considered free physical
objects.  Their interaction with gauge fields are important and physical
gauge-invariant quantities are governed by the Ioffe-Larkin rule within the
Gaussian approximation.

It is of great interest to ask the question of whether a deconfined ground
state is possible in a $U(1)$ gauge theory in the presence of matter field. This
issue was first addressed in  a seminal paper by 
\Ref{FS7982}
who considered a boson field coupled to a compact $U(1)$ gauge field.  The
following bosonic action is added to $S_{\rm gauge}$ :
\be
S_B = t \sum_{\v i} \cos \left(
\Delta_\mu \theta(\v r_{\v i}) - q a_\mu (\v r_{\v i})
\right)
\label{Eq.102}
\en
Here the Bose field is represented by phase fluctuation only, $\Delta_\mu$ is
the lattice derivative and $a_\mu(\v r_{\v i}) = a_{\v i,\v i+\v \mu}$ 
is the gauge field on
the link $\v i, \v i+\v \mu$ and $q$ is an integer.  It is interesting to consider the
phase diagram in the $t, g$ plane.  Along the $t=0$ line, we have pure gauge
theory which is always confining in $2+1$ dimension.  For $g \ll 1$, gauge
fluctuations are weak and $S_B$ reduces to the XY model weakly coupled to a
$U(1)$ gauge field, which exhibits an ordered phase called the Higgs phase at
zero temperature.  Note that in the Higgs phase, the gauge field is gapped by
the Anderson-Higgs mechanism. On the other hand,  it is also gapped in the
confinement phase due to the screening of magnetic charges described earlier.
There is no easy way to distinguish between these two phases and the central
result of Fradkin and Shenker is that for $q = 1$ the Higgs phase and the
confinement phases are smoothly connected to each other.  Indeed, it was
argued by 
\Ref{NL0066} that for the 1+2D case the entire $t$-$g$ plane is covered by
the Higgs-confinement phase, with the exception of the line $g = 0$, which
contains the XY transition.

The situation is dramatically different for $q = 2$, \ie if the boson field
corresponds to a pairing field.  Then it is possible to distinguish between
the Higgs phase and the confinement phase by asking whether two $q = \pm 1$
have a linear confinement potential between them or not.  In this case there
is a phase boundary between the confined and the Higgs phase, and the Higgs
phase (the pairing phase) is deconfined.  One way of understanding this
deconfinement is that the paired phase has a residual $Z_2$ gauge symmetry,
\ie the pairing order parameter is invariant under a sign change of the
underlying $q = 1$ fields which make up the pair.  Furthermore, it is known
that the $Z_2$ gauge theory has a confinement-deconfinement transition in $2 +
1$ dimensions.  Thus the conclusion is that a compact $U(1)$ gauge theory
coupled to a pair field can have a deconfined phase.  This is indeed the route
to a deconfined ground state proposed by \Ref{RS9173} and \Ref{Wsrvb}.  
In the context
of the $U(1)$ gauge theory, the fermion pair field $\Delta$ plays the role of
the $q = 2$ boson field in eq.~(\ref{Eq.102}). In such a phase,
the spinon and holons are deconfined, leading to the phenomenon of spin and
charge fractionalization.  A third elementary excitation in this theory is the
$Z_2$ vortex, which is gapped.

\Ref{SF0050} pointed out that the square root of $\Delta$ carried unit gauge
charge and one can combine this with the fermion to form a gauge invariant
spinon and with the boson to form a gauge invariant ``chargon''.  The spinon
and chargon only carry $Z_2$ gauge charges and can be considered almost free,
They propose an experiment to look for the gapped $Z_2$ vortex but the results
have so far been negative.  The connection between the $U(1)$ slave-boson
theory and their $Z_2$ gauge theory was clarified by 
\Ref{SF0119}.  

There is yet another route to a deconfined ground state, and that is a
coupling of a compact $U(1)$ gauge field to gapless fermions.  
\Ref{N9310}
suggested that dissipation due to gapless excitations lead to  deconfinement.
The special case of coupling to gapless Dirac fermions is of special interest.
This route (called the $U(1)$ spin liquid) appears naturally in the $SU(2)$
formulation and will be discussed in detail in section XI.F.  The hope
expressed in section XII is to use the proximity to this deconfined state to
understand the pseudogap state.  This is a more attractive scenario compared
with the reliance purely on finite temperature to see deconfinement effects as
described earlier in this section.

In the literature there have been some confusing discussions of the role of
confinement in the gauge theory approach to strong correlation.  In
particular, 
\Ref{N0078}, \citeyear{N0193} has claimed that slave particles are
always confined in $U(1)$ gauge theories.  His argument is based on the fact
that since these gauge fields are introduced to enforce constraint, they
do not have restoring force and the coupling constant is infinite.  What he
overlooks is the possibility that partially integrating out the matter fields
will generate restoring forces, which brings the problem to one of strong but
finite coupling, and then sweeping conclusions can no longer be made.
Comments by 
\Ref{IM0142}, 
\Ref{IMO0116}, and by 
\Ref{O0301} have clarified the issues and
in our opinion adequately answered Nayak's objections.  For example, 
\Ref{IM0142} pointed out that $3+1$ dimensional $SU(3)$ gauge theory
coupled to $N$ fermions is in the deconfined phase even at infinite coupling
for $N > 7$.  Another counter example is found by
\Ref{Wqoslpub}, \Ref{RWspin} and 
\Ref{HSF0451}
who showed that the $U(1)$ gauge theory coupled to massless Dirac fermions is
in a gapless phase (or the deconfined phase) for sufficiently large $N$ (see
section XI.F).  There is also numerical evidence from Monte Carlo studies that
the $SU(N)$ Hubbard-Heisenberg model at $N=4$ exhibits a gapless spin liquid
phase, \ie a Mott insulator with power law spin correlation without breaking
of lattice translation symmetry \cite{A0474}.  This spin liquid state is
strongly suggestive of the stability of a deconfined phase with $U(1)$ gauge
field coupled to Dirac fermions.

\subsection{Limitations of the $U(1)$ gauge theory} 

The $U(1)$ gauge theory, which only includes Gaussian fluctuations about mean
field theory, suffers from several limitations which are especially serious in
the underdoped regime.  Apart from the confinement issue discussed in the last
section, we first mention a difficulty with the linear $T$ coefficient of the
superfluid density.  As long as the gauge fluctuation is treated as Gaussian,
the Ioffe-Larkin law holds and one predicts that the superfluid density
$\rho_s(T)$ behaves as $\rho_s(T) \approx ax - bx^2T$.  The $ax$ term agrees
with experiment while the $-bx^2T$ term does not (\cit{LW9711}; \cit{IM0109})
as already explained in section V.A.  This failure is traced to the fact that
in the Gaussian approximation, the current carried by the quasiparticles in
the superconducting state is proportional to $x v_F$.  We believe this failure
is a sign that nonperturbative effects again become important and confinement
takes place, so that the low energy quasiparticles near the nodes behave like
BCS quasiparticles which carry the full current $v_F$.  This is certainly
beyond the Gaussian fluctuation treatment described here.

A second difficulty is that experimentally it is known from neutron scattering
that spin correlations at $(\pi,\pi)$ are enhanced in the underdoped regime.
This happens at the same time while a spin gap is forming in the pseudogap
regime.  The $U(1)$ mean field theory explains the existence of the spin gap
as due to fermion pairing.  However, this reduces the fermion density of
states and it is not clear how one can get an enhancement of the spin
correlation unless one introduces phenomenologically RPA interactions
\cite{BL0102}.  The problem is more serious because the gauge field is gapped
in the fermion paired state and one cannot use gauge fluctuation to enhance
the spin correlation.  The gapping of the gauge field also tends to suppress
fermion pairing self-consistently 
\cite{UL9453}.  We shall see that both these difficulties are resolved by the
$SU(2)$ formulation.

A third difficulty has to do with the structure of the vortex core in the
underdoped limit.\cite{WLsu2}  As mentioned in section IX.C, the $U(1)$ gauge
theory predicts the stability of $hc/e$ vortices, which has not been observed.
This is a serious issue especially because the STM experiments show that the
pseudogap remains in the vortex core. Therefore it should be type B in the
$U(1)$ theory, which carries $hc/e$ flux.  On the other hand, the $hc/2e$ vortex
is not ``cheap'' because the pairing amplitude vanishes and one has to pay the
pairing energy at the core.  These difficulties arise because in the $U(1)$
theory, the fermions becomes ``strong'' superconductor at low temperature in
the underdoped region. However this contradicts with the fact that at
half-filling the d-wave RVB state is equivalent to the $\pi$-flux state, which
is not ``superconducting''.  In short, the $U(1)$ theory misses the important
low lying fluctuation related to the $SU(2)$ particle-hole symmetry at
half-filling.  By incorporating this symmetry to the gauge field even at
finite doping, we will be lead to the $SU(2)$ gauge theory of high Tc
superconductors, which we will next discuss.

\section{$SU(2)$ slave-boson representation for spin liquids}

In this section we are going to develop $SU(2)$ slave-boson theory for spin
liquids and underdoped high $T_c$ superconductors.  The $SU(2)$ slave-boson
theory is equivalent to the $U(1)$ slave-boson theory discussed in the last
section. However, the $SU(2)$ formalism makes more symmetries of the
slave-boson theory explicit. This makes it easier to see the low energy
collective modes in the $SU(2)$ formalism, which in turn allows us to resolve
some difficulties of the $U(1)$ slave-boson theory.\footnote{We would like to
point out that those difficulties are not because the $U(1)$ slave-boson
theory is incorrect. The difficulties are results of incorrect treatment of
the $U(1)$ slave-boson theory, for example, overlooking some low energy soft
modes.} To develop the $SU(2)$ slave-boson theory, let us first describe
another way to understand the $U(1)$ gauge fluctuations in the slave-boson
theory. In this section we will concentrate on undoped case where the model is
just a pure spin system.  Even though the theory involves only fermionic
representation of the spin in the underdoped case, we continue to refer to
the theory as slave-boson theory in anticipation of the doped case.  We
generalize the $SU(2)$  salve-boson theory to doped model in the next section.

\subsection{Where does the gauge structure come from?}

According to the $U(1)$ slave-boson mean-field theory, the fluctuations around
the mean-field ground state are described by gauge fields and fermion fields.
Remember that the original model is just a interacting spin model which is a
purely bosonic model. How can a purely bosonic model contain excitations
described by gauge fields and fermion fields?  Should we believe the result?

Let us examine how the results are obtained. We first split the bosonic spin
operator into a product of two fermionic operators $\v S_{\v i}=\frac12
f^\dag_{\v i}\v \si f_{\v i}$.  We then introduce a gauge field to glue the
fermions back into a bosonic spin.  From this point of view it appears that
the gauge bosons and the fermions are fake and their appearance is just a
mathematical artifact. The appearance of the fermion field and gauge field in
a purely bosonic model seems only indicates that the slave-boson theory is
incorrect.

However, we should not discard the slave-boson mean-field theory too quickly.
It is actually capable of producing pictures that agree with the common sense:
the excitations in a bosonic spin system are bosonic excitations corresponding
to spin flips, provided that the gauge field is in a confining phase.
In the confining phase of the $U(1)$ gauge theory, the fermions interact with
each other through a linear potential and can never appear as quasiparticles
at low energies. The gauge bosons have a large energy gap in the confining
phase and are absent from the low energy spectrum.  The only low energy
excitations are bound state of two fermions which carry spin-1 and are bosons.
So the mean-field theory plus the gauge fluctuations, may not be very useful,
but is not wrong. 

On the other hand, the slave-boson mean-field theory (plus gauge fluctuations)
is also capable of producing pictures that defy the common senses, if the
gauge field is in a deconfined phase. In this case the fermions and gauge
bosons 
may appear as well defined quasiparticles. The question is do we believe the
picture of deconfined phase? Do we believe the possibility of emergent gauge
bosons and fermions from a purely bosonic model?  Clearly, the slave-boson
construction outlined above is far too formal to convince most people to
believe such drastic results.  However, recently, it was realized that some
models \cite{K032,LWsta,Wqoem} can be solved by the slave-boson theory exactly
\cite{Wqoexct}.  Those models are in deconfined phases and confirm the
striking results of emergence of gauge bosons and fermions
from the slave-boson theory.  


To have an intuitive picture of the correlated ground state
which leads to emergent gauge bosons and fermions, let us try to understand
how a mean-field ansatz $\chi_{\v i\v j}$ is connected to a physical spin wave
function.
We know that the ground
state, $|\Psi_\text{mean}^{(\chi_{\v i \v j})}\>$, of the mean-field Hamiltonian
\begin{align}
\label{HmeanU1}
H_\text{mean}=\tilde J \sum (\chi_{\v i\v j}f^\dag_{\v i}f_{\v j}+h.c.) 
+\sum a_0(f^\dag_{\v i}f_{\v i}-1),
\end{align}
is not a valid wavefunction for
the spin system, since it may not have one fermion per site.  To connect to
physical spin wavefunction, we need to include fluctuations of $a_0$ to
enforce the one-fermion-per-site constraint. With this understanding, we may
obtain a valid wave-function of the spin system $\Psi_\text{spin}(\{\al_{\v
i}\})$ by projecting the mean-field state to the subspace of
one-fermion-per-site:
\begin{equation}
\Psi_\text{spin}^{(\chi_{\v i \v j})}(\{\al_{\v i}\}) 
= \<0_f|\prod_{\v i} f_{\al_{\v i}\v i} |
\Psi_\text{mean}^{(\chi_{\v i \v j})}\> .
\label{PsiPsichi}
\end{equation}
where $|0_f\>$ is the state with no $f$-fermions: $f_{\al\v i}|0_f\>=0$.
Eq. (\ref{PsiPsichi}) connects the mean-field ansatz to 
physical spin wavefunction.
It allows us to understand the physical meaning of the mean-field ansatz and
mean-field fluctuations.

For example, the projection \Eq{PsiPsichi} give the gauge transformation
\Eq{U1gaugetrans} a physical meaning.  Usually, for different choices of
$\chi_{\v i\v j}$, the ground states of $H_\text{mean}$ \Eq{HmeanU1} correspond to
different mean-field wavefunctions $|\Psi_\text{mean}^{(\chi_{\v i \v j})}\>$.
After projection they lead to different physical spin wavefunctions
$\Psi_\text{spin}^{(\t \chi_{\v i \v j})}(\{\al_{\v i}\})$. Thus we can regard
$\chi_{\v i\v j}$ as labels that label different physical spin states.
However, two mean-field ansatz $\chi_{\v i \v j}$ and $\t \chi_{\v i \v j}$
related by a gauge transformation
\begin{equation}
\label{tchichiga}
 \t \chi_{\v i \v j} = e^{i\th_{\v i}}  \chi_{\v i \v j}e^{-i\th_{\v j}}
\end{equation}
give rise to the same physical spin state after the projection
\begin{equation}
\label{tchichi}
 \Psi_\text{spin}^{(\t \chi_{\v i \v j})}(\{\al_{\v i}\})
= e^{i\sum_{\v i} \th_{\v i}}
\Psi_\text{spin}^{(\chi_{\v i \v j})}(\{\al_{\v i}\})
\end{equation}
Thus $\chi_{\v i \v j}$ is not a one-to-one label, but a many-to-one label.
This property is important for us to understand the unusual dynamical
properties of $\chi_{\v i\v j}$ fluctuations.  Using many labels to label the
same physical state also make our theory a gauge theory.

Let us consider how the many-to-one property or the gauge structure of
$\chi_{\v i\v j}$ affect its dynamical properties.  If $\chi_{\v i\v j}$ was
an one-to-one label of physical states, then $\chi_{\v i\v j}$ would be like
the condensed boson amplitude $\<\phi(\v x,t)\>$ in boson superfluid or the
condensed spin moment $\<\v S_{\v i}(t)\>$ in SDW state. The fluctuations of
$\chi_{\v i\v j}$ would correspond to a bosonic mode similar to sound mode or
spin-wave mode.\footnote{More precisely, the sound mode and spin-wave mode are
so called scaler bosons. The fluctuations of local order parameters always
give rise to scaler bosons.} However, $\chi_{\v i\v j}$ does not behave like
local order parameters, such as $\<\phi(\v x,t)\>$ and $\<\v S_{\v i}(t)\>$,
which label physical states without redundancy.  $\chi_{\v i\v j}$ is a
many-to-one label as discussed above.  The many-to-one label creates a
interesting situation when we consider the fluctuations of $\chi_{\v i \v j}$
-- some fluctuations of $\chi_{\v i \v j}$ do not change the physical state
and are unphysical. Those fluctuations are called the pure gauge fluctuations.
The effective theory for $\chi_{\v i\v j}$ must be gauge invariant: for
example, the energy for the ansatz $\chi_{\v i\v j}$ satisfies
\begin{equation*}
 E(\chi_{\v i\v j})=E(e^{i\th_{\v i}}  \chi_{\v i \v j}e^{-i\th_{\v j}}).
\end{equation*}
If we consider the phase fluctuations of $\chi_{\v i\v j}=\bar\chi_{\v i\v
j}e^{ia_{\v i\v j}}$, then the energy for the fluctuations $a_{\v i\v j}$
satisfies
\begin{equation*}
 E(a_{\v i\v j})=E(a_{\v i \v j}+\th_{\v i}-\th_{\v j}).
\end{equation*}
This gauge invariant property of the energy (or more precisely, the action)
drastically change the dynamical properties of the fluctuations.  It is this
property that makes fluctuations of $a_{\v i \v j}$ behave like gauge bosons,
which are very different from sound mode and spin-wave mode.\footnote{In the
continuum limit, the gauge bosons are vector bosons -- bosons described by
vector fields.}

If we believe that gauge bosons and fermions do appear as low energy
excitations in the deconfined phase, then a natural question will be what do
those excitations looks like?  The slave-boson construction \Eq{PsiPsichi}
allows us to construct an explicit physical spin wavefunction that
corresponds to a gauge fluctuation $a_{\v i\v j}$
\begin{equation*}
\Psi_\text{spin}^{(a_{\v i\v j})}
= \<0_f|\prod_{\v i} f_{\al_{\v i}\v i} 
|\Psi_\text{mean}^{(\bar \chi_{\v i \v j}e^{ia_{\v i\v j}})}\> .
\end{equation*}
We would like to mention that the gauge fluctuations affect the average
\begin{equation*}
 P_{123} = \<\chi_{12} \chi_{23} \chi_{31}\>=
  \<\bar\chi_{12} \bar\chi_{23} \bar\chi_{31}\> e^{i(a_{12}+a_{23}+a_{31})}
\end{equation*}
Thus the $U(1)$ gauge fluctuations $a_{\v i\v j}$, or more precisely the flux
of $U(1)$ gauge fluctuations $a_{12}+a_{23}+a_{31}$, correspond to the
fluctuations of the spin chirality ${\v S}_1 \cdot ( {\v S}_2 \times {\v
S}_3)=\frac{P_{123} - P_{132}}{4i}$ as pointed out in the last section.

Similarly, the slave-boson construction also allows us to construct a physical
spin wavefunction that corresponds to a pair of the fermion excitations.  We
start with the mean-field ground state with a pair of particle-hole
excitations.  After the projection \Eq{PsiPsichi}, we obtain the physical spin
wavefunctions that contain a pair of fermions:
\begin{equation*}
\Psi_\text{spin}^\text{ferm}(\v i_1,\la_1; \v i_2,\la_2)
= \<0| (\prod_{\v i} f_{\al_{\v i}\v i}) 
f^\dag_{\la_1\v i_1} f_{\la_2\v i_2} 
|\Psi_\text{mean}^{(\bar \chi_{\v i \v j})}\> .
\end{equation*}
We see that the gauge fluctuation $a_{\v i\v j}$ and fermion excitation do
have a physical ``shape'' given by the spin wavefunction
$\Psi_\text{spin}^{(a_{\v i\v j})}$ and $\Psi_\text{spin}^\text{ferm}$,
although the shape is too complicated to picture. 

Certainly, the two types of excitations, the gauge fluctuations and the
fermion excitations, interact with each other. The form of the interaction is
determined by the fact that the fermions carry unit charge of the $U(1)$ gauge
field. The low energy effective theory is given by \Eq{Eq.39} with $\Delta_{\v
i\v j} = 0$ and $b_{\v i} = 0$.



\subsection{What determines the gauge group?}

We have mentioned that the collective fluctuations around the a slave-boson
mean-field ground state are described by $U(1)$ gauge field. Here we would
like to ask why the gauge group is $U(1)$?  The reason for the gauge group to
be $U(1)$ is that the fermion Hamiltonian and the mean-field Hamiltonian are
invariant under the local $U(1)$ transformation
\begin{equation*}
\label{lclU1}
 f_{\v i} \to e^{i\th_{\v i}} f_{\v i},\ \ \ \ \ \ \
 \chi_{\v i \v j} \to e^{i\th_{\v i}}  \chi_{\v i \v j}e^{-i\th_{\v j}}
\end{equation*}
The reason that the fermion Hamiltonian is invariant under the local $U(1)$
transformation is that the fermion Hamiltonian is a function of spin operator
$\v S_{\v i}$ and the spin operator $\v S_{\v i}=\frac12 f_{\v i}^\dag \v \si
f_{\v i}$ is invariant under the local $U(1)$ transformation.  So the gauge
group is simply the group formed all the transformations between $f_{\up\v i}$
and $f_{\down\v i}$ that leave the physical spin operator invariant.

\subsection{From $U(1)$ to $SU(2)$}

This deeper understanding of gauge transformation allows us to realize that
$U(1)$ is only part of the gauge group. The full gauge group is actually
$SU(2)$. To see the gauge group to be $SU(2)$ let us introduce
\begin{equation*}
 \psi_{1\v i}=f_{\up\v i},\ \ \ \ \ \
 \psi_{2\v i}=f^\dag_{\down\v i}
\end{equation*}
We find
\begin{align*}
 S^+_{\v i}=& f_{\v i}^\dag  \si^+ f_{\v i} = 
\frac12 (
\psi_{1\v i}^\dag \psi_{2\v i}^\dag -\psi_{2\v i}^\dag \psi_{1\v i}^\dag )
\nonumber\\
 S^z_{\v i} = & \frac12 f_{\v i}^\dag  \si^z f_{\v i} = 
\frac12 (\psi_{1\v i}^\dag \psi_{1\v i} 
+ \psi_{2\v i}^\dag \psi_{2\v i} -1)
\end{align*}
Now it is clear that $\v S_{\v i}$ and any Hamiltonian expressed in terms of
$\v S_{\v i}$ are invariant under local $SU(2)$ gauge transformation:
\begin{equation*}
\bpm \psi_{1\v i}\\ \psi_{2\v i} \epm
\to
W_{\v i}\bpm \psi_{1\v i}\\ \psi_{2\v i} \epm,
\ \ \ \ \ \ W_{\v i} \in SU(2)
\end{equation*}
The local $SU(2)$ invariance of the spin Hamiltonian implies that the
mean-field Hamiltonian not only should have the $U(1)$ gauge invariance, it
should also have the $SU(2)$ gauge invariance.

To write down the mean-field theory with explicit $SU(2)$ gauge invariance, we
start with the mean-field ansatz that includes pairing correlation:
\begin{align}
\label{sl2a.4}
\chi_{\v i \v j}\del_{\al\bt}&= 2 \< f_{\v i\al}^{\dag} f_{\v j \bt} \>,
& \chi_{\v i \v j} &= \chi_{\v j \v i}^* .
\nonumber \\
\Del_{\v i \v j} \eps_{\al\bt}&=  2 \< f_{\v i \alpha}~f_{\v j \bt} \> ,
& \Del_{\v i \v j} &=\Del_{\v j \v i} ,
\end{align}
After replacing fermion bi-linears with $\chi_{\v i\v j}$ and $\Del_{\v i\v
j}$ in \Eq{Eq.35}, we obtain the following mean-field Hamiltonian with pairing
\begin{align*}
H_\text{mean}= &
\ \ \sum_{\<\v i\v j\>} -\frac{3}{8} J_{\v i\v j} 
\left[(\chi_{\v j \v i}f^\dag_{\v i\al}f_{\v j\al}  - \Del_{\v i \v j}
~f^\dag_{ i \alpha} f^\dag_{ j \bt}\eps_{\alpha \bt}) + h.c
\right. \nonumber\\
& \left.
 - |\chi_{\v i \v j}|^2  -|\Del_{\v i \v j}|^2 \right]
\end{align*}

However, the above mean-field Hamiltonian is incomplete.  We know that the
physical Hilbert space is formed by states with one $f$-fermion per site. Such
states correspond to states with even $\psi$-fermion per site.  The states
with even $\psi$-fermion per site are $SU(2)$ singlet one every site. The
operators $\psi_{\v i}^\dag \v \tau \psi_{\v i}$ that generate local $SU(2)$
transformations vanishes within the physical Hilbert space, where $\v
\tau=(\tau^1,\tau^2, \tau^3)$ are the Pauli matrices.  In the mean-field
theory, we replace the constraint $\psi_{\v i}^\dag \v \tau \psi_{\v i}=0$ by
its average
\begin{equation*}
\< \psi_{\v i}^\dag \v \tau \psi_{\v i} \>=0 .
\end{equation*}
The averaged constraint can be enforced by including Lagrangian multiplier
$\sum_{\v i} a^l_0(\v i)\psi_{\v i}^\dag \tau^l \psi_{\v i}$ in the mean-field
Hamiltonian. This way we obtain the mean-field Hamiltonian of $SU(2)$
slave-boson theory \cite{AZH8845,DFM8826}:
\begin{align}
\label{sl2.6su2}
&H_\text{mean} \\
=&
\ \ \sum_{\<\v i\v j\>} -\frac{3}{8} J_{\v i\v j} 
\Big[(\chi_{\v j \v i}f^\dag_{\v i\al}f_{\v j\al}  - \Del_{\v i \v j}
~f^\dag_{ i \alpha} f^\dag_{ j \bt}\eps_{\alpha \bt}) + h.c
\nonumber\\
&\ \ \ \ \ \ \ \ \ \ \ \ \ \ \ \ \
 - |\chi_{\v i \v j}|^2  -|\Del_{\v i \v j}|^2 \Big]
\nonumber \\
& 
\ \  +\sum_{\v i} \left[ a_0^3(f_{\v i\al}^\dag f_{\v i\al} -1)+[(a_0^1+i a_0^2)
 f_{\v i\al} f_{\v i\bt}\eps_{\al\bt}+h.c.]\right]
\nonumber 
\end{align}
So the mean-field ansatz that describes a $SU(2)$ slave-boson mean-field state
is really given by $\chi_{\v i\v j}$, $\Del_{\v i\v j}$, and $\v a_0$.  We
note that $\chi_{\v i\v j}$, $\Del_{\v i\v j}$, and $\v a_0$ are invariant
under spin rotation. Thus the mean-field ground state of $H_\text{mean}$ is a spin
singlet. Such a state describes a spin liquid state.

The $SU(2)$ mean-field Hamiltonian \Eq{sl2.6su2} is invariant under local
$SU(2)$ gauge transformation. To see such an invariance explicitly, we need to
rewrite \Eq{sl2.6su2} in terms of $\psi$:
\begin{align}
\label{sl2a.8}
H_\text{mean} =& \sum_{\<\v i \v j\>} \frac{3}{8}
 J_{\v i\v j} \left[ 
\frac12 \Tr (U_{\v i \v j}^\dag~U_{\v i \v j}) +(\psi^\dag_ i
U_{\v i \v j}
\psi_ j +~h.c.)\right]  \nonumber\\
&
+\sum_{\v i}  a_0^l \psi_{\v i}^\dag \tau^l \psi_{\v i} 
\end{align}
where
\begin{equation}
U_{\v i \v j }= \bpm - \chi_{\v i \v j}^\ast &\Del_{\v i \v j} \\
               \Del_{\v i \v j}^\ast & \chi_{\v i \v j}
                                        \epm
=U_{\v j \v i}^\dag
\label{sl2a.7}
\end{equation}
Note that $\det (U) < 0$, so that $U_{\v j \v k}$ is not a member of $SU(2)$,
but $iU_{\v j \v k}$ is a member up to a normalization constant.  From
\Eq{sl2a.8} we now can see clearly that the mean-field Hamiltonian is
invariant under a local $SU(2)$ transformation $W_{\v i}$:
\begin{align}
\psi_{\v i} \to & W_{\v i}\psi_{\v i}  \nonumber \\
U_{\v i \v j} \to &W_{\v i}~U_{\v i \v j}~W^\dag_{\v j}    
\label{sl2a.9}
\end{align}

We note that in contrast to $\Phi_{i\up}$ and $\Phi_{i\down}$ introduced in
eq.~(\ref{Eq.37}), the doublet $\psi_{\v i}$ does not carry a spin index.  Thus the
redundancy in the $\Phi_{i\sigma}$ representation is avoided, which accounts
for a factor of 2 difference in front of the bilinear $\psi_{\v i}$ term in
\Eq{sl2a.8} vs eq.~(\ref{Eq.44}).  However, the spin-rotation symmetry is not
explicit in our formalism and it is hard to tell if \Eq{sl2a.8} describes a
spin-rotation invariant state or not. In fact, for a general $U_{\v i \v j}$
satisfying $U_{\v i \v j}=U_{\v j \v i}^\dag$, \Eq{sl2a.8} may not describe a
spin-rotation invariant state.  But, if $U_{\v i \v j}$ has a form
\begin{align}
\label{UuW}
U_{\v i \v j} =& \chi^\mu_{\v i\v j}\tau^\mu,\ \ \ \ \ \ \mu=0,1,2,3,\nonumber\\
\chi^0_{\v i\v j}=& \text{imaginary}, \ \ \ \ \ \ 
\chi^l_{\v i\v j} = \text{real},\ \ \ \ \ \ l=1,2,3,
\end{align}
then \Eq{sl2a.8} will describe a spin-rotation invariant state.  This is
because the above $U_{\v i \v j}$ has the form of \Eq{sl2a.7}.  In this case
\Eq{sl2a.8} can be rewritten as \Eq{sl2.6su2} where the spin-rotation
invariance is explicit.  In \Eq{UuW}, $\tau^0$ is the identity matrix.

Now the mean-field ansatz can be more compactly represented by $(U_{\v i\v j},
\v a_0(\v i))$. Again the mean-field ansatz $(U_{\v i\v j}, \v a_0(\v i))$ can
be viewed as a many-to-one label of physical spin states. The physical spin
state labeled by $(U_{\v i\v j}, \v a(\v i))$ is given by
\begin{equation*}
 |\Psi^{(U_{\v i\v j},\v a_0(\v i))}_\text{spin}\>
=\cP |\Psi^{(U_{\v i\v j},\v a_0(\v i))}_\text{mean}\>
\end{equation*}
where $|\Psi^{(U_{\v i\v j},\v a_0(\v i))}_\text{mean}\>$ is the ground state of
the mean-field Hamiltonian \Eq{sl2a.8} and $\cP$ is the projection that
project into the subspace with even numbers of $\psi$-fermion per site.  From
the relation between the $f$-fermion and the $\psi$-fermion, we note that the
state with zero $\psi$-fermion correspond to the spin-down state and the state
with two $\psi$-fermions correspond to the spin-up state.  Since the states
with even numbers of $\psi$-fermion per site are $SU(2)$ singlet on every
site, we find that two mean-field ansatz $(U_{\v i\v j}, \v a_0(\v i))$ and
$(\t U_{\v i\v j}, \t{\v a}(\v i))$ related by a local $SU(2)$ gauge
transformation
\begin{equation*}
\t U_{\v i \v j} = W_{\v i}~U_{\v i \v j}~W^\dag_{\v j},\ \ \ \ \ \ \ \ 
\t {\v a}_0(\v i)\cdot \v\tau  
= W_{\v i}~\v a_0(\v i)\cdot \v\tau ~W^\dag_{\v i} .
\end{equation*}
label the same physical spin state
\begin{equation*}
\cP |\Psi^{(U_{\v i\v j},\v a_0(\v i))}_\text{mean}\>
=\cP |\Psi^{(\t U_{\v i\v j},\t{\v a}_0(\v i))}_\text{mean}\>
\end{equation*}
This relation represents the physical meaning of the $SU(2)$ gauge structure.

Just like the $U(1)$ slave-boson theory, the fluctuations of the mean-field
ansatz correspond to collective excitations.  In particular, the ``phase''
fluctuations of $U_{\v i\v j}$ represent the potential gapless excitations.
However, unlike the $U(1)$ slave-boson theory, the ``phase'' of $U_{\v i\v j}$
is described by a two by two hermitian matrix $a^l_{\v i\v j}\tau^l$,
$l=1,2,3$, on each link.  If $(\bar U_{\v i\v j},\bar{\v a}(\v i))$ is the
ansatz that describe the mean-field ground state, then the potential gapless
fluctuations are described by
\begin{equation*}
 U_{\v i\v j}= \bar U_{\v i\v j}e^{ia^l_{\v i\v j}\tau^l},\ \ \ \ \
\v a_0(\v i)= \bar{\v a}_0(\v i)+ \del \v a_0(\v i).
\end{equation*}
Since $(U_{\v i\v j},\v a_0(\v i))$ is a many-to-one labeling, the
fluctuations $(\v a_{\v i\v j}, \del \v a_0(\v i))$ correspond to $SU(2)$
gauge fluctuations rather than usual bosonic collective modes such as phonon
modes and spin waves.

\subsection{A few mean-field ansatz for symmetric spin liquids}

After a general discussion of the $SU(2)$ slave-boson theory, let us discuss a
few mean-field ansatz that have spin rotation, translation $T_{x,y}$, and
parity $P_{x,y,xy}$ symmetries.  We will call such a spin state symmetric spin
liquid.  Here $T_x$ and $T_y$ are translation in $x$- and $y$-directions, and
$P_x$, $P_y$ and $P_{xy}$ are parity transformations $(x, y)\to (-x,y)$, $(x,
y)\to (x,-y)$, and $(x, y)\to (y,x)$ respectively.  We note that  $P_{x,y,xy}$
parity symmetries imply the $90^\circ$ rotation symmetry.  


We will concentrate on three simple mean-field ansatz that describe symmetric
spin liquids:\\

\noindent
$\pi$-flux liquid ($\pi$fL) state\footnote{
This state was called $\pi$-flux ($\pi$F) state in literature.}
\cite{AM8874}
\begin{align}
\label{piF}
U_{\v i, \v i + \v x} & = - i(-)^{i_y} \chi ,
\nonumber\\
U_{\v i, \v i + \v y} & = - i\chi ,
\end{align}
staggered flux liquid (sfL)
\footnote{In
\Ref{WLsu2} and \Ref{LNNWsu2}, this phase was called simply the staggered
flux (sF) state.  In this paper we reserve sF to denote the $U(1)$ mean field
state which explicitly breaks translational symmetry and which exhibits
staggered orbital currents, as originally described by 
\Ref{HMA9166}.  This latter state is also called $d$-density wave (ddw),
following 
\Ref{CLM0203}.}
state \cite{AM8874}
\begin{align}
\label{sF}
U_{\v i, \v i + \v x} & = - \tau^3 \chi - i (-)^{\v i} \Delta ,
\nonumber\\
U_{\v i, \v i + \v y} & = - \tau^3 \chi + i (-)^{\v i} \Delta ,
\end{align}
$Z_2$-gapped state \cite{Wsrvb}
\begin{align}
U_{\v i,\v i+\v x }=&  U_{ \v i,\v i+\v y }= -\chi \tau^3 \nonumber \\
U_{\v i,\v i+ \v x + \v y} =&  \eta\tau^1 +\la\tau^2 \nonumber \\
U_{\v i,\v i-\v x + \v y} =&  \eta\tau^1 -\la\tau^2 \nonumber\\
a_0^{2,3} =&  0, \ \ \ a_0^1 \neq 0
\label{Z2gA}
\end{align}
where $(-)^{\v i}\equiv (-)^{i_x+i_y}$.  Note that the $Z_2$ mean-field state
has pairing along the diagonal bond.

At first sight, those mean-field ansatz appear not to have all the symmetries.
For example, the $Z_2$-gapped ansatz are not invariant under the $P_x$ and
$P_y$ parity transformations and the $\pi$fL and sfL ansatz are not invariant
under translation in the $y$-direction. However, those ansatz do describe spin
states that have all the symmetries. This is because the mean-field ansatz are
many-to-one labels of the physical spin state, the non-invariance of the
ansatz does not imply the non-invariance of the corresponding physical spin
state after the projection.  We only require the mean-field ansatz to be
invariant up to a $SU(2)$ gauge transformation in order for the projected
physical spin state to have a symmetry. For example, a $P_{xy}$  parity
transformation changes the sfL ansatz to
\begin{align*}
U_{\v i, \v i + \v x} & = - \tau^3 \chi + i (-)^{\v i} \Delta ,
\nonumber\\
U_{\v i, \v i + \v y} & = - \tau^3 \chi - i (-)^{\v i} \Delta ,
\end{align*}
The reflected ansatz can be transformed into the original ansatz via a $SU(2)$
gauge transformation $W_{\v i}=(-)^{\v i}i\tau^1$.  Therefore, after the
projection, the sfL ansatz describes a $P_{xy}$ parity symmetric spin state.
Using the similar consideration, one can show that the above three ansatz
are invariant under translation $T_{x,y}$ and parity $P_{x,y,xy}$ symmetry
transformations followed by corresponding $SU(2)$ gauge transformations
$G_{T_x,T_y}$ and $G_{P_x,P_y,P_{xy}}$, respectively. Thus the three ansatz all
describe symmetric spin liquids.  In the following, we list the corresponding
gauge transformations $G_{T_x,T_y}$ and $G_{P_x,P_y,P_{xy}}$ for the above
three ansatz:\\

\noindent
$\pi$fL state
\begin{align}
\label{GssF}
G_{T_x}(\v i) =& (-)^{i_x}G_{T_y}(\v i) = \tau^0,  
&
G_{P_{xy}}(\v i) =&  (-)^{i_xi_y} \tau^0, 
\nonumber\\
(-)^{i_x}G_{P_x}(\v i) =& (-)^{i_y}G_{P_y}(\v i) = \tau^0, 
& 
G_0(\v i) =& e^{i\th^l\tau^l}
\end{align}
sfL state
\begin{align}
\label{GspiF}
G_{T_x}(\v i) =& G_{T_y}(\v i) = i(-)^{\v i}\tau^1,  
&
G_{P_{xy}}(\v i) =&  i(-)^{\v i}\tau^1, 
\nonumber\\
G_{P_x}(\v i) =& G_{P_y}(\v i) = \tau^0, 
& 
G_0(\v i) =& e^{i\th\tau^3}
\end{align}
$Z_2$-gapped state
\begin{align}
\label{GsZ2gA}
G_{T_x}(\v i) =& G_{T_y}(\v i) = i\tau^0,  
&
G_{P_{xy}}(\v i) =&  \tau^0, 
\nonumber\\
G_{P_x}(\v i) =& G_{P_y}(\v i) = (-)^{\v i}\tau^1, 
& 
G_0(\v i) =& -\tau^0
\end{align}
In the above we also list the pure gauge transformation $G_0(\v i)$ that leave
the ansatz invariant.

\subsection{Physical properties of the symmetric spin liquids at mean-field
level}

To understand the physical properties of the above three symmetric spin
liquids, let us first ignore the mean-field fluctuations of $U_{\v i\v j}$ and
consider the excitations at mean-field level.

At mean-field level, the excitations are spin-1/2 fermions $\psi$ (or $f$).
Their spectrum is determined by the mean-field Hamiltonian \Eq{sl2a.8} (or
\Eq{sl2.6su2}).  In the $\pi$fL state, the fermions has a dispersion
\begin{equation*}
 E_{\v k}=\pm \frac34 J|\chi| \sqrt{\sin^2 k_x + \sin^2 k_y}
\end{equation*}
In the sfL state, the dispersion is given by
\begin{equation*}
 E_{\v k}=\pm \frac34 J\sqrt{\chi^2 (\cos k_x+\cos k_y)^2 
 + \Del^2 (\cos k_x - \cos k_y)^2 }
\end{equation*}
In the $Z_2$-gapped state, we have
\begin{align*}
E_{\v k}=&\pm \sqrt{\eps_{\v k}^2+\Del^2_{1\v k}+\Del^2_{2\v k}}
\nonumber\\
\eps_{\v k} = & -\frac34 J \chi (\cos k_x + \cos k_y) 
\nonumber\\
\Delta_{1\v k} = & \frac34  \eta J' \cos (k_x + k_y) +a^1_0
\nonumber\\
\Delta_{2\v k} = & \frac34 \la J' \cos (-k_x + k_y)
\end{align*}
where $J$ is the nearest-neighbor spin coupling and $J'$ is the
next-nearest-neighbor spin coupling.  We find that the $\pi$fL state and the
sfL
state, at the mean-field level, have gapless spin-1/2 fermion excitations,
while the $Z_2$-gapped state has gapped spin-1/2 fermion excitations.


Should we trust the mean-field results from the slave-boson theory?  The
answer is that it depends on the importance of the gauge fluctuations.  Unlike
usual mean-field theory, the fluctuations in the slave-boson theory include
gauge fluctuations which can generate confining interactions between the
fermions. In this case the gauge interactions represent relevant perturbations
and the mean-field state is said to be unstable.  The mean-field results from
an unstable mean-field ansatz cannot be trusted and cannot be applied to real
physical spin state. In particular, the spin-1/2 fermionic excitations in the
mean-field theory in this case will not appear in the physical spectrum of
real spin state.

If the dynamics of the gauge fluctuations is such that the gauge interaction
is short ranged, then  the gauge interactions represent irrelevant
perturbations and can be ignored.  In this case the mean-field state is said
to be stable and the mean-field results can be applied to the real physical
spin liquids.  In particular, the corresponding physical spin state contain
fractionalized spin-1/2 fermionic excitations.

\subsection{Classical dynamics of the $SU(2)$ gauge fluctuations}

We have seen that the key to understand the physical properties of a spin
liquid described by a mean-field ansatz $(U_{\v i\v j}, a^l_0)$ is to
understand the dynamics of the $SU(2)$ gauge fluctuations.  To gain some
intuitive understanding, let us treat the mean-field ansatz $(U_{\v i\v j}, \v
a_0(\v i))$ as classical fields and study classical dynamics of their
fluctuations.  The  dynamics of the fluctuations is determined by the
effective Lagrangian $L_\text{eff}(U_{\v i\v j}(t), \v a_0(\v i)(t))$.  To obtained
the effective Lagrangian, we start with the Lagrangian representation of the
mean-field Hamiltonian
\begin{align*}
 L(\psi_{\v i}, U_{\v i\v j}, \v a_0)
=&\sum_{\v i} i\psi^\dag_{\v i}\prt_t\psi_{\v i} -H_\text{mean}
\end{align*}
where $H_\text{mean}$ is given in \Eq{sl2a.8}. The effective Lagrangian $L_\text{eff}$
is then obtained by integrating out $\psi$:
\begin{equation*}
 e^{i\int dt L_\text{eff}(U_{\v i\v j},\v a_0)}
=\int \cD\psi\cD\psi^\dag e^{i\int dt L(\psi,U_{\v i\v j},\v a_0)}
\end{equation*}
We note that $L$ describes a system of fermions $\psi_{\v i}$ and $SU(2)$
gauge fluctuations $U_{\v i\v j}$.  Thus the effective Lagrangian is invariant
under the $SU(2)$ gauge transformation
\begin{align}
\label{Egauge}
& L_\text{eff}(\t U_{\v i\v j}, \t{\v a}_0)=L_\text{eff}(U_{\v i\v j}, \v a_0),
\nonumber\\
& 
\t U_{\v i\v j}= W_{\v i}(U_{\v i\v j})W_{\v j},\ \ \ \ 
\t a^l_0(\v i)\tau^l= W_{\v i}a^l_0(\v i)\tau^lW_{\v i},
\nonumber\\
& W_{\v i}\in SU(2)
\end{align}
The classical equation of motion obtained from $L_\text{eff}(U_{\v i\v j},\v
a_0)$ determines the classical dynamics of the fluctuations.

To see if the collective fluctuations are gapless, we would like to examine if
the frequencies of the collective fluctuations are bound from below. We know
that the time independent saddle point of $L_\text{eff}(U_{\v i\v j},\v
a_0)$, $(\bar U_{\v i\v j},\bar {\v a}_0)$, corresponds to mean-field ground
state ansatz, and $-L_\text{eff}(\bar U_{\v i\v j},\bar {\v a}_0)$ is the
mean-field ground state energy. If we expand $-L_\text{eff}(\bar U_{\v i\v
j}e^{ia^l_{\v i\v j}\tau^l},\bar {\v a}_0)$ to the second order in the
fluctuation $a_{\v i\v j}$, then the presence or the absence of the mass term
$a_{\v i\v j}^2$ will determine if the collective $SU(2)$ gauge fluctuations
have an energy gap or not.


To understand how the mean-field ansatz $\bar U_{\v i\v j}$ affect the
dynamics of the gauge fluctuations, it is convenient to introduce the loop
variable of the mean-field solution \index{flux!$SU(2)$}
\begin{equation}
\label{sl2a.15}
P(C_{\v i})= (i\bar{U}_{\v i\v j}) (i\bar{U}_{\v j\v k})...  (i\bar{U}_{\v k\v i})
\end{equation}
Following the comment after \Eq{sl2a.7} $P(C_{\v i})$ belongs to $SU(2)$ and we can
write $P(C_{\v i})$ as $P(C_{\v i})=e^{i\Phi(C_{\v i})}$, where $\Phi$ is the
$SU(2)$ flux through the loop $C_{\v i}$: $\v i\to \v j\to \v k\to ..\to \v
l\to \v i$ with base point $\v i$.
The $SU(2)$ flux correspond to gauge field strength in the continuum limit.
Compare with the $U(1)$ flux, the $SU(2)$ flux has two new features.  First
the flux $\Phi$ is a two-by-two traceless Hermitian matrix.  If we expand
$\Phi$ as $\Phi=\Phi^l\tau^l$, $l=1,2,3$ we can say that the flux is
represented by a vector $\Phi^l$ in the $\tau^l$ space.  Second, the flux is
not gauge invariant.  Under the gauge transformations, $\Phi(C_{\v i})$
transforms as
\begin{equation}
\label{PWPW}
 \Phi(C_{\v i}) \to W_{\v i} \Phi(C_{\v i}) W_{\v i}
\end{equation}
Such a transformation rotates the direction of the vector $\Phi^l$.
Since the direction of the $SU(2)$ flux for loops with different base point
can be rotated independently by the local $SU(2)$ gauge transformations, it is
meaningless to directly compare the directions of $SU(2)$ flux for different
base points.  However, it is quite meaningful to compare the directions of
$SU(2)$ flux for loops with the same base point. We can divide different
$SU(2)$ flux configurations into three classes based  on the $SU(2)$ flux
through loops with the \emph{same} base point: (a) trivial $SU(2)$ flux where
all $P(C)\propto \tau^0$, (b) collinear $SU(2)$ flux where all the $SU(2)$
fluxes point in the same direction, and (c) non-collinear $SU(2)$ flux where
$SU(2)$ flux for loops with the same base point are in different directions.
\index{SU2 flux@$SU(2)$ flux!trivial} \index{SU2 flux@$SU(2)$ flux!collinear}
\index{SU2 flux@$SU(2)$ flux!non-collinear} We will show below that different
$SU(2)$ flux can lead to different dynamics for the gauge field \cite{Wsrvb}.

\subsubsection{Trivial $SU(2)$ flux}

First let us consider an ansatz $\bar U_{\v i\v j}$ with trivial $SU(2)$ flux
$\Phi(C)=0$ for all the loops (such as the $\pi$fL ansatz in \Eq{piF}). We will
call the state described by such an ansatz the $SU(2)$ state. 
We can perform a $SU(2)$ gauge transformations to transform the ansatz into a
form where all $\bar U_{\v i\v j} \propto \tau^0$.
In this case, the gauge invariance of the effective Lagrangian implies that
\begin{equation} 
L_\text{eff}(\bar U_{\v i\v j}e^{ia_{\v i\v j}^l\tau^l} )=
L_\text{eff}(\bar U_{\v i\v j}e^{i\th^l_{\v i}\tau^l} e^{ia_{\v i\v j}^l\tau^l}
e^{-i\th^l_{\v j}\tau^l} ).
\end{equation}
Under gauge transformation $e^{i\th^1_{\v i}\tau^1}$, $a_{\v i\v j}^1$
transform as $a_{\v i\v j}^1=a_{\v i\v j}^1+\th^1_{\v i}-\th^1_{\v j}$.  The
mass term $(a^1_{\v i\v j})^2$ is not invariant under such a transformation
and is thus not allowed.  Similarly, we can show that none of the mass terms
$(a^1_{\v i\v j})^2$, $(a^2_{\v i\v j})^2$, and $(a^3_{\v i\v j})^2$ are
allowed in the expansion of $L_\text{eff}$.  Thus the $SU(2)$ gauge fluctuations
are gapless and appear at low energies.  

We note that all the pure gauge transformations $G_0(\v i)$ that leave the
ansatz invariant form a group. We will call such a group invariant gauge group
(IGG).  For the ansatz $\bar U_{\v i\v j} \propto \tau^0$,  the IGG is an
$SU(2)$ group formed by uniform $SU(2)$ gauge transformation $G_0(\v
i)=e^{i\th^l\tau^l}$. 
We recall from the last paragraph that the (classical) gapless gauge
fluctuations 
is also $SU(2)$.  Such a relation between the IGG and the gauge group of the
gapless classical gauge fluctuations is general and applies to all the ansatz
\cite{Wqoslpub}.  

To understand the dynamics of the gapless gauge fluctuations beyond the
classical level, we need to treat two cases separately. In the first case, the
fermions have a finite energy gap. Those fermions will generate the following
low energy effective Lagrangian for the gauge fluctuations
\begin{equation*}
 \cL = \frac{g}{8\pi} \Tr f_{\mu\nu}f^{\mu\nu}
\end{equation*}
where $f_{\mu\nu}$ is a 2 by 2 matrix representing the field strength of the
$SU(2)$ gauge field $a_{\v i\v j}$ in the continuum limit.  At classical
level, such an effective Lagrangian leads to $\sim g\log (r)$ interaction
between $SU(2)$ charges in two spatial dimensions.  So the gauge interaction
at classical level is not confining (\ie not described by a linear potential).  However, if we go beyond the classical
level (\ie beyond the quadratic approximation) and include the interactions
between gauge fluctuations, the picture is changed completely.  In 1+2D, the
interactions between gauge fluctuations change the $g\log (r)$ interaction to
a linear confining interaction, regardless the value of the coupling constant
$g$. So the $SU(2)$ mean-field states with gapped fermions  are not stable.
The mean-field results from such ansatz cannot be trusted.

In the second case, the fermions are gapless and have a linear dispersion.  In
the continuum limit, those fermions correspond to massless Dirac fermions.
Those fermions will generate a non-local low energy effective Lagrangian for
the gauge fluctuations, which roughly has a form, $ \cL = \frac{g}{8\pi} \Tr
f_{\mu\nu}\frac{1}{\sqrt{-\prt^2}}f^{\mu\nu}$.  Due to the screening of
massless fermions the interaction potential between $SU(2)$ charges becomes
$\sim g/r$ at classical level.  Such an interaction represents a marginal
perturbation.  It is a quite complicated matter to determine if the $SU(2)$
states with gapless Dirac fermions are stable or not beyond the quadratic
approximation.

\subsubsection{Collinear $SU(2)$ flux}

Second, let us assume the $SU(2)$ flux is collinear.  This means the $SU(2)$
flux for different loops with the same base point all point in the same
direction. 
However, the $SU(2)$ flux for loops with different base points may still point
in different directions (even for the collinear $SU(2)$ flux).  Using the
local $SU(2)$ gauge transformation we can always rotate the $SU(2)$ flux for
different base points into the same direction, and we can pick this direction
to be $\tau^3$ direction.  In this case all the $SU(2)$ flux have a form
$P(C)\propto \chi^0(C)+ i\chi^3(C)\tau^3$.  We can choose a gauge such that
the mean-field ansatz  have a form $\bar U_{\v i\v j}=ie^{i\phi_{\v i\v j}
\tau^3}$.  
The gauge invariance of the energy implies that
\begin{equation} 
L_\text{eff}(\bar U_{\v i\v j}e^{ia_{\v i\v j}^l\tau^l} )=
L_\text{eff}(\bar U_{\v i\v j}e^{i\th_{\v i}\tau^3} e^{ia_{\v i\v j}^l\tau^l}
e^{-i\th_{\v j}\tau^3} ).
\end{equation}
When $a^{1,2}_{\v i\v j}=0$, The above reduces to
\begin{equation} 
L_\text{eff}(\bar U_{\v i\v j}e^{ia_{\v i\v j}^3\tau^3} )=
L_\text{eff}(\bar U_{\v i\v j}e^{i(a_{\v i\v j}^3+\th_{\v i}-\th_{\v j})\tau^3}
).
\label{sl2a.16}
\end{equation}
We find that the mass term $(a^3_{\v i\v j})^2$ is incompatible with
\Eq{sl2a.16}.  Therefore at least the gauge field $a^3_{\v i\v j}$ is 
gapless.  How
about $a^1_{\v i\v j}$ and $a^2_{\v i\v j}$ gauge fields?  Let $P_A(\v i)$ be
the $SU(2)$ flux through a loop with base point $\v i$.  If we assume all the
gauge invariant terms that can appear in the effective Lagrangian do appear,
then $L_\text{eff}(U_{\v i\v j})$ will contain the following term
\begin{equation} 
L_\text{eff}=a\Tr[P_A(\v i)iU_{\v i,\v i+\v x}P_A(\v i+\v x)iU_{\v i+\v x, \v i}]+...
\label{sl3a.6c}
\end{equation}
If we write $iU_{\v i, \v i+\v x}$ as $\chi e^{i\phi_{\v i\v j}\tau^3}e^{i
a_x^l\tau^l}$, using the fact $U_{\v i, \v i+\v x}=U^\dag_{\v i+\v x,\v i}$
(see \Eq{sl2a.7}), and expand to $(a_x^l)^2$ order, \Eq{sl3a.6c} becomes
\begin{equation} 
L_\text{eff}=-\frac{1}{2} a\chi^2 \Tr ([P_A, a_x^l\tau^l]^2) +...
\label{sl3a.6d}
\end{equation}
We see from \Eq{sl3a.6d} that the mass term for $a^1_{\v i\v j}$ and $a^2_{\v
i\v j}$ are generated if $P_A\propto \tau^3$.  

To summarize, we find that if the $SU(2)$ flux is collinear, then the ansatz
is invariant only under a $U(1)$ rotation $e^{i\th \v n\cdot\tau}$ where $\v
n$ is the direction of the $SU(2)$ flux.  Thus the IGG=$U(1)$.  The collinear
$SU(2)$ flux also break the $SU(2)$ gauge structure down to a $U(1)$ gauge
structure, \ie the low lying gauge fluctuations are described by a $U(1)$
gauge field.  Again we see that the IGG of the ansatz is the gauge group of
the (classical) gapless gauge fluctuations.
We will call the states with collinear $SU(2)$ flux the $U(1)$ states.  The
sfL ansatz in \Eq{sF} is an example of collinear $SU(2)$ flux.

For the $U(1)$ states with gapped fermions, the fermions will generate the
following effective Lagrangian for the gauge fluctuations
\begin{equation*}
 \cL = \frac{g}{8\pi} (\v e^2 - b^2)
\end{equation*}
where $\v e$ is the ``electric'' field and $b$ is the ``magnetic'' field of
the $U(1)$ gauge field.  Again at classical level, the effective Lagrangian
leads to $\sim g\log (r)$ interaction between $U(1)$ charges and the gauge
interaction at classical level is not confining.  If we go beyond the
classical level and include the interactions between gauge fluctuations
induced by the space-time monopoles,  the $g\log (r)$ interaction will be
changed to a linear confining interaction, regardless the value of the
coupling constant $g$ \cite{P7729}. So the $U(1)$ mean-field states with
gapped fermions  are not stable.  

If the fermions in the $U(1)$ state are gapless and are described by massless
Dirac fermions (such as those in the sfL state), those fermions will generate a non-local low energy effective
Lagrangian, which, at quadratic level, has a form
\begin{equation}
\label{LaNL}
 \cL = \frac{g}{8\pi} f_{\mu\nu}\frac{1}{\sqrt{-\prt^2}}f^{\mu\nu}
\end{equation}
Again the screening of massless fermions change the $g\log (r)$ interaction to
$g/r$ interactions between $U(1)$ charge, at lease at classical level.  Such
an interaction represents a marginal perturbation.  Beyond the classical
level, we will show in subsections \ref{asl} and \ref{screen} that, when there
are many Dirac fermions, the $U(1)$ gauge interactions with Dirac fermions are
exact marginal perturbations.  So the $U(1)$ states with enough gapless Dirac
fermions are not unstable.  The mean-field theory can give us a good starting
point to study the properties of the corresponding physical spin state (see
subsection \ref{asl}).

\subsubsection{Non-collinear $SU(2)$ flux}

Third, we consider the situation where the $SU(2)$ flux is non-collinear.  In
the above, we have shown that an $SU(2)$ flux $P_A$ can induce a mass term of
form $\Tr([P_A, a_x^l\tau^l]^2)$. For a non-collinear $SU(2)$ flux
configuration, we can have in \Eq{sl3a.6c} another $SU(2)$ flux, 
$P_B$, pointing in a different direction from $P_A$. The mass term will 
contain in addition to \Eq{sl3a.6d} a term $\Tr([P_B, a_x^l\tau^l]^2)$. In
this case, the mass terms for all the $SU(2)$ gauge fields $(a^1_{\v i\v
j})^2$, $(a^2_{\v i\v j})^2$, and $(a^3_{\v i\v j})^2$ will be generated.  All
$SU(2)$ gauge bosons will gain an energy gap.  

We note that ansatz $U_{\v i\v j}$ is always invariant under the global $Z_2$
gauge transformation $-\tau^0$. So the IGG always contains a $Z_2$ subgroup
and the $Z_2$ gauge structure is unbroken at low energies.  The global $Z_2$
gauge transformation is the only invariance for the non-collinear ansatz. Thus
IGG=$Z_2$ 
and the low energy effective theory is a $Z_2$ gauge theory.  
We can show that the low energy properties of non-collinear states, such as
the existence of $Z_2$ vortex and ground state degeneracy, are indeed
identical to those of a $Z_2$ gauge theory. So we will call the state with
non-collinear $SU(2)$ flux a $Z_2$ state.

In a $Z_2$ state, all the gauge fluctuations are gapped. Those fluctuations
can only mediate short range interactions between fermions. The low energy
fermions interact weakly and behave like free fermions. Therefore, including
mean-field fluctuations does not qualitatively change the properties of the
mean-field state. The gauge interactions are irrelevant and the $Z_2$
mean-field state is stable at low energies.

A stable mean-field spin liquid state implies the existence of a real physical
spin liquid. The physical properties of the stable mean-field state apply to
the physical spin liquid. If we believe these two statements, then we can
study the properties of a physical spin liquid by studying its corresponding
stable mean-field state. Since the fermions are not confined in mean-field
$Z_2$ states, the physical spin liquid derived from a mean-field $Z_2$ state
contain neutral spin-1/2 fermions as its excitation. 

The $Z_2$-gapped ansatz in \Eq{Z2gA} is an example where the $SU(2)$ flux is
non-collinear. To see this, let us consider the $SU(2)$ flux through two
triangular loops $(\v i,\v i+\v y,\v i-\v x)$ and $(\v i,\v i+\v x,\v i+\v y)$
with the same base point $\v i$:
\begin{align*}
U_{\v i,\v i+\v y} ~U_{\v i+\v y,\v i-\v x} ~U_{\v i-\v x,\v i} = 
 -\chi^2 ( \eta\tau^1 +\la\tau^2 ),
\nonumber\\ 
U_{\v i,\v i+\v x} ~U_{\v i+\v x,\v i+\v y} ~U_{\v i+\v y,\v i} = 
 -\chi^2 ( \eta\tau^1 -\la\tau^2 ).
\end{align*}
We see that when $\eta$ and $\la$ are non-zero, the $SU(2)$ flux is not
collinear. Therefore, after projection, the $Z_2$-gapped ansatz give rise to a
real physical spin liquid which contains fractionalized spin-1/2 neutral
fermionic excitations \cite{Wsrvb}. The spin liquid also contains a $Z_2$
vortex excitation.  The bound state of a spin-1/2 fermionic excitation and a
$Z_2$ vortex give us a spin-1/2 bosonic excitation
\cite{RC8933,Wsrvb}.

\subsection{The relation between different versions of slave-boson theory}

We have discussed two version of the slave-boson theories, the $U(1)$
slave-boson theory and the $SU(2)$ slave-boson theory.  In \Ref{SF0050}, a
third slave-boson theory -- $Z_2$ slave-boson theory -- was also proposed.
Here we would like to point out that all the three version of the slave-boson
theory are \emph{equivalent} description of the same spin-1/2 Heisenberg model
on square lattice, if we treat the $SU(2)$, $U(1)$ or $Z_2$ gauge fluctuations
exactly.

To understand the relation between the three version of the slave-boson
theory, we would like to point out that the $SU(2)$, $U(1)$ or $Z_2$ gauge
structures are introduced to project the fermion Hilbert space (which has four
states per site) to the smaller spin-1/2 Hilbert space (which has two states
per site).  In the $SU(2)$ slave-boson theory, we regard the two fermions
$\psi_{1\v i}$ and $\psi_{2\v i}$ as an $SU(2)$ doublet.  Among the four
fermion-states on each site, $|0\>$, $\psi^\dag_{1\v i}|0\>$, $\psi^\dag_{2\v
i}|0\>$, and $\psi^\dag_{1\v i}\psi^\dag_{2\v i}|0\>$, only the $SU(2)$
invariant state correspond to the physical spin state.  There are only two
$SU(2)$ invariant states on each site: $|0\>$ and $\psi^\dag_{1\v
i}\psi^\dag_{2\v i}|0\>$ which correspond to the spin-up and the spin-down
states.  So the spin-1/2 Hilbert space is obtained from the fermion Hilbert
space by projecting onto the local $SU(2)$ singlet subspace.

In the $U(1)$ slave-boson theory, we regard $\psi_{1\v i}$ as a charge $+1$
fermion and $\psi_{2\v i}$ as a charge $-1$ fermion.  The spin-1/2 Hilbert
space is obtained from the fermion Hilbert space by projecting onto the local
charge neutral subspace.  Among the four fermion-states on each site, only two
states $|0\>$ and $\psi^\dag_{1\v i}\psi^\dag_{2\v i}|0\>$ are charge neutral.

In the $Z_2$ slave-boson theory, we regard $\psi_{a\v i}$ as a fermion that
carries a unit $Z_2$-charge.  The spin-1/2 Hilbert space is obtained from the
fermion Hilbert space by projecting onto the local $Z_2$-charge neutral
subspace.  Again the two states $|0\>$ and $\psi^\dag_{1\v i}\psi^\dag_{2\v
i}|0\>$ are the only $Z_2$-charge neutral states.

In the last subsection we discussed  $Z_2$, $U(1)$, and $SU(2)$ spin liquid
states.  
These must not be confused with $Z_2$, $U(1)$, and $SU(2)$ slave-boson
theories.  We would like to stress that $Z_2$, $U(1)$, and $SU(2)$ in the
$Z_2$, $U(1)$, and $SU(2)$ spin liquid states  are gauge groups that appear in
the low energy effective theories of those spin liquids. We will call those
gauge group low energy gauge group.  They should not be confused with the
$Z_2$, $U(1)$, and $SU(2)$ gauge groups in the $Z_2$, $U(1)$, and $SU(2)$
slave-boson theories.  
We will call the latter high energy gauge groups.
The high energy gauge groups have nothing to do with the low energy gauge
groups.
A high energy $Z_2$ gauge theory (or a $Z_2$ slave-boson approach) can have a
low energy effective theory that contains $SU(2)$, $U(1)$ or $Z_2$ gauge
fluctuations.  Even the Heisenberg model, which has no gauge structure at
lattice scale, can have a low energy effective theory that contains $SU(2)$,
$U(1)$ or $Z_2$ gauge fluctuations.  The spin liquids studied in this paper
all contain some kind of low energy gauge fluctuations. Despite their
different low energy gauge groups, all those spin liquids can be constructed
from any one of $SU(2)$, $U(1)$, or $Z_2$ slave-boson approaches.  After all,
all those slave-boson approaches describe the same Heisenberg model and are
equivalent.

The high energy gauge group is related to the way in which we write down the
Hamiltonian. We can write Hamiltonian of the Heisenberg model in many
different ways which can contain arbitrary high energy gauge group of our
choice. We just need to split the spin into two, four, six, or some other even
numbers of fermions.  While the low energy gauge group is a property of ground
state of the spin model. It has nothing to do with how are we going to write
down the Hamiltonian.  
Thus we should not regard $Z_2$ spin liquids as the spin liquids constructed
using $Z_2$ slave-boson approach.  A $Z_2$ spin liquid can be constructed and
was first constructed within the $U(1)$ or $SU(2)$ slave-boson/slave-fermion
approaches.  
	      However, when we study a particular spin liquid state, a certain
version of the slave-boson theory may be more convenient than other versions.
Although a spin liquid can be described by all the versions of the slave-boson
theory, sometimes a particular version may have the weakest fluctuations. 

\subsection{The emergence of gauge bosons and fermions in condensed matter
systems}

In the early days, it was believed that a pure boson system can never generate
gauge bosons and fermions. Rather, the gauge bosons and the fermions were
regarded as fundamental.   The spin liquids discussed in this paper suggest
that gauge bosons (or gauge structures) and fermions are not fundamental and
can emerge from local bosonic model. Here we will discuss how those ideas were
developed historically.

Let us first consider gauge bosons.  In the standard picture of gauge theory,
the gauge potential $a_\mu$ is viewed as a geometrical object -- a connection
of a fibre bundle.  However, there is another point of view about the gauge
theory.  Many thinkers in theoretical physics were not happy with the
redundancy of the gauge potential $a_\mu$.  It was realized in the early
1970's that one can use gauge invariant loop operators to characterize
different phases of a gauge theory 
(\cit{W7159}; \cit{W7445}; \cit{KS7595}).  It was later
found that one can formulate the entire gauge theory using closed strings
(\cit{BMK7793}; \cit{GRV7978}; \cit{M7991}; \cit{P7947}; \cit{F7987}; \cit{S8053}). Those studies reveal the
intimate relation between gauge theories and closed-string theories --- a
point of view very different from the geometrical notion of vector potential.

In a related development, it was found that gauge fields can emerge from a
local bosonic model, if the bosonic model is in certain quantum phases.  This
phenomenon is also called dynamical generation of gauge fields.  The emergence
of gauge fields from local bosonic models has a long and complicated history.
Emergent $U(1)$ gauge \emph{field} has been introduced in quantum disordered
phase of 1+1D $CP^N$ model \cite{DDL7863,W7985}. In condensed matter physics,
the $U(1)$ gauge \emph{field} have been found in the slave-boson approach to
spin liquid states 
(\cit{BA8880}; \cit{AM8874}).  
The slave-boson approach not only has a $U(1)$ gauge
field, it also has gapless fermion \emph{fields}.

It
is well known that the compact $U(1)$ gauge theory is confining in $1+1$ and
$1+2$D \cite{P7582}.  
The concern about
confinement  led to an opinion that the $U(1)$ gauge field and the gapless
fermion fields are just a unphysical artifact of the ``unreliable''
slave-boson approach.  Thus the key to find emergent gauge bosons and emergent
fermions is not to write down a Lagrangian that contain \emph{gauge fields}
and \emph{Fermi fields}, but to show that gauge \emph{bosons} and
\emph{fermions} actually appear in the physical low energy spectrum.  
However, only when the dynamics of gauge field is such that the gauge field is
in the deconfined phase can the gauge boson appear as a low energy
quasiparticle.  Thus after the initial finding of \Ref{DDL7863}; \Ref{W7985};
\Ref{BA8880}; \Ref{AM8874}, many researches have been trying to find the
deconfined phase of the gauge field.

One way to obtain deconfined phase is to give gauge boson a mass.  In 1988, it
was shown that if we break the time reversal symmetry in a 2D spin-1/2 model,
then the $U(1)$ gauge field from the slave-boson approach can be in a
deconfined phase due to the appearance of Chern-Simons term (\cit{WWZcsp};
\cit{KW8983}). The deconfined phase correspond to a spin liquid state of the
spin-1/2 model \cite{KL8795} which is called chiral spin liquid.  The chiral
spin state contains neutral spin-1/2 excitations that carry fractional
statistics.  A second deconfined phase was found by breaking the $U(1)$ or
$SU(2)$ gauge structure down to a $Z_2$ gauge structure. Such a phase contains
a deconfined $Z_2$ gauge theory \cite{RS9173,Wsrvb} 
and is called $Z_2$ spin liquid (or short ranged RVB state).\footnote{The
$Z_2$ state obtained in \Ref{RS9173} breaks the $90^\circ$ rotation symmetry
while the $Z_2$ state in \Ref{Wsrvb} has all the lattice symmetries.} The
$Z_2$ spin state also contains neutral spin-1/2 excitations.  But now the
spin-1/2 excitations are fermions and bosons.  

The above $Z_2$ spin liquids have a finite energy gap for their neutral
spin-1/2 excitations. In \Ref{BFN9833}, a spin liquid with gapless spin-1/2
excitations was constructed by studying quantum disordered $d$-wave
superconductor.  Such a spin liquid was identified as a $Z_2$ spin liquid
using a $Z_2$ slave-boson theory \cite{SF0050}.  The mean-field ansatz is
given by  
\begin{align}
\label{Z2lCs}
U_{\v i,\v i+{\v x}} &= \chi \tau^3 +\eta \tau^1, & 
U_{\v i,\v i+{\v y}} &= \chi \tau^3 -\eta \tau^1,
\\
a_0^3 \neq &0,  \ \ \ \ a^{1,2}_0=0,  
&
U_{\v i,\v i+{\v x}+\v y} &= 
U_{\v i,\v i+{\v x}-\v y} = \ga \tau^3.
\nonumber 
\end{align}
The diagonal hopping breaks particle-hole symmetry and breaks the $U(1)$
symmetry of the $a^3_0=0$ $d$-wave pairing ansatz down to $Z_2$.  We will call
such an ansatz $Z_2$-gapless ansatz.
The ansatz describes a symmetric spin liquid, since it is invariant under the
combined transformations $(G_{T_x}T_x, G_{T_y}T_{y},
G_{P_x}P_x, G_{P_y}P_y, G_{P_{xy}}P_{xy}, G_0) $ with
\begin{align}
\label{GsZ2lCs}
 G_{T_x}=&\tau^0, &
 G_{T_y}=&\tau^0, &
 G_0=&-\tau^0, \nonumber\\
 G_{P_x}=&\tau^0, &
 G_{P_y}=&\tau^0, &
 G_{P_{xy}}=&i\tau^3 .
\end{align}
The fermion excitations are gapless only at four $\v k$ points with a linear
dispersion.  

The $Z_2$-gapped state and the $Z_2$-gapless state are just two $Z_2$ states
among over 100 $Z_2$ states that can be constructed within the $SU(2)$
slave-boson theory \cite{Wqoslpub}.  The chiral spin liquid and the $Z_2$ spin
liquids provide examples of emergent gauge structure and emergent fermions (or
anyons).  However, those results were obtained using slave-boson theory, which
is not very convincing to many people. 

In 1997, an exact soluble spin-1/2 model \cite{K032}
\begin{equation*}
 H_\text{exact}=16g \sum_{\v i} S^y_{\v i} S^x_{\v i+\hat{\v x}} S^y_{\v
i+\hat{\v x}+\hat{\v y}} S^x_{\v i+\hat{\v y}}
\end{equation*}
was found. The $SU(2)$ slave-boson theory turns out to be exact for such a model
\cite{Wqoexct}. That is by choosing a proper $SU(2)$ mean-field ansatz, the
corresponding mean-field state give rise to an exact eigenstate of $H_\text{exact}$
after the projection. In fact all the eigenstates of $H_\text{exact}$ can be
obtained this way by choosing different mean-field ansatz.  The exact solution
allows us to show the excitations of $H_\text{exact}$ to be fermions and $Z_2$
vortices. This confirms the results obtained from the slave-boson theory.

More exactly soluble or quasi-exactly soluble models were find for dimmer
model \cite{MS0181}, spin-1/2 model on Kagome lattice \cite{BFG0212}, boson
model on square lattice \cite{SM0204}, and Josephson junction array
\cite{IFI0203}.  A model of electrons coupled to pairing fluctuations, with a
local constraint which results in a Mott insulator  that obeys the spin
$SU(2)$, symmetry was also constructed \cite{MS0204}.  Those models realize
the $Z_2$ states.  A boson model that realize $Z_3$ gauge structure
\cite{M0308} and $U(1)$ gauge structure \cite{SM0204,Walight} were also found.
15 years after the slave-boson approach to the spin liquids, now it is easy to
construct (quasi-)exactly soluble spin/boson models that have emergent gauge
bosons and fermions.

We would like to point out that the spin liquids are not the first example of
emergent fermions from local bosonic models.  The first example of emergent
fermions, or more generally, emergent anyons is given by the FQH states.
Although \Ref{ASW8422} only discussed how anyons can emerge from a fermion
system in magnetic field, the same argument can be easily generalized to show
how fermions and anyons  can emerge from a boson system in magnetic field.
Also in 1987, in a study of resonating-valence-bond (RVB) states, emergent
fermions (the spinons) were proposed in a nearest neighbor dimer model on
square lattice (\cit{KRS8765}; \cit{RK8876}; \cit{RC8933}). But, according to
the deconfinement picture, the results in \Ref{KRS8765} and \Ref{RK8876} 
are valid only
when the ground state of the dimer model is in the $Z_2$ deconfined phase. It
appears that the dimer liquid on square lattice with only nearest neighbor
dimers is not a deconfined state (\cit{RK8876}; \cit{RC8933}), and thus it is
not clear if the nearest neighbor dimer model on square lattice \cite{RK8876}
has the deconfined quasiparticles or not \cite{RC8933}.  However, on
triangular lattice, the dimer liquid is indeed a $Z_2$ deconfined state
\cite{MS0181}.  Therefore, the results in \Ref{KRS8765} and \Ref{RK8876} are valid for
the triangular-lattice dimer model and deconfined quasiparticles do emerge in
a dimer liquid on triangular lattice.  

All the above models with emergent fermions are 2D models, where the emergent
fermions can be understood from binding flux to a charged particle
\cite{ASW8422}. Recently, it was pointed out in \Ref{LWsta} that the key to
emergent fermions is a string structure. Fermions can generally appear as ends
of open strings in any dimensions, if the ground state has a condensation of
closed strings. The string picture allows a construction of a 3D local bosonic
model that has emergent fermions \cite{LWsta}.  According to this picture, all
the models with emergent fermions contain closed-string condensation in their
ground states.  Since the fluctuations of condensed closed strings are gauge
fluctuations \cite{BMK7793,S8053,Walight}, this explains why the model with
emergent fermions also have emergent gauge structures. Since the gauge charges
are ends of open strings, this also explains why the emergent fermions always
carry gauge charges.

The second way to obtain deconfined phase is to simply go to higher
dimensions.  In 3+1 dimension, the gapless $U(1)$ fluctuations do not generate
confining interactions. In 4+1 dimensions and above, even non-Abelian gauge
theory can be in a deconfined phase.  So it is not surprising that one can
construct bosonic models on cubic lattice that have emergent gapless photons
($U(1)$ gauge bosons) (\cit{Wlight}; \cit{MS0204}; \cit{Walight}).  

The third way to obtain deconfined phase is to include  gapless excitations
which carry gauge charge. The charged gapless excitations can screen the gauge
interaction to make it less confining.  We would like to remark that the
deconfinement in this case has a different behavior than the previous two
cases.  In the previous two cases, the charged particles in the deconfined
phases become non-interacting quasiparticles at low energies.  In the present
case, the deconfinement only means that those gapless charged particles remain
to be gapless.  Those particles may not become non-interacting quasiparticles
at low energies.  The spin liquids obtained from the sfL ansatz and the uRVB
ansatz (given by \Eq{sF} with $\Del=0$) belong to this case. Those spin liquid
are gapless.  But the gapless excitations are not described by free fermionic
quasiparticles or free bosonic quasiparticles at low energies.  The uRVB state
(upon doping) leads to strange metal states \cite{LN9221} with large Fermi
surface. We will discuss the spin liquid obtained from the sfL ansatz in
subsections \ref{asl} and \ref{screen}.

Finally, we remark that what is common among these three ways to get
deconfinement is that instantons are irrelevant and a certain gauge flux is a
conserved quantity.  We shall exploit this property in section XII.E.

\subsection{The projective symmetry group and quantum order}

The $Z_2$-gapped ansatz \Eq{Z2gA} and the $Z_2$-gapless ansatz \Eq{Z2lCs},
after the projection, give rise to two spin liquid states. The two states have
exactly the same symmetry. The question here is that whether there is a
way to classify these as distinct phases.   According to Landau's symmetry
breaking theory, two states with the same symmetry belong to the same phase.
However, after the discovery of fractional quantum Hall states, we now know
that Landau's symmetry breaking theory does not describe all the phases.
Different quantum Hall states have the same symmetry, but yet they can belong
to different phases since they contain different topological orders
\cite{Wtoprev}. So it is possible that the two $Z_2$ spin liquids contain
different orders that cannot be characterized by symmetry breaking and local
order parameters. The issue here is to find a new set of universal quantum
numbers that characterize the new orders.

To find a new set of universal quantum numbers, we note that although 
the projected wavefunctions of the two
$Z_2$ spin liquids have the same symmetry, their ansatz are invariant under
the same set of symmetry transformations but followed by different gauge transformations (see
\Eq{GsZ2gA} and \Eq{GsZ2lCs}). So the invariant group of the mean-field ansatz
for the two spin liquids are different. The invariant group is called the
Projective Symmetry Group (PSG).  The PSG is generated by the combined
transformations $(G_{T_x}T_x, G_{T_y}T_y, G_{P_x}P_x, G_{P_y}P_y,
G_{P_{xy}}P_{xy}) $ and $G_0$.  We note that the PSG is the symmetry group of
the mean-field Hamiltonian.  Since the mean-field fluctuations in the $Z_2$
states are weak and perturbative in nature, those fluctuations cannot change
the symmetry group of the mean-field theory. Therefore, the PSG of an ansatz
is a universal property, at least against perturbative fluctuations.  The PSG
can be used to characterize the new order in the two $Z_2$ spin liquids
\cite{Wqoslpub}. Such order is called the quantum order.  The two $Z_2$ spin
liquids belong to two different phases since they have different PSG's and
hence different quantum orders.

We know that the symmetry characterization of phases (or orders) have some
important applications. It allows us to classify all the 230 crystal orders in
three dimensions. The symmetry also produces and protects gapless collective
excitations -- the Nambu-Goldstone bosons.  The PSG characterization of
quantum orders has similar applications.  Using PSG, we can classify over 100
different 2D $Z_2$ spin liquids that all have the same symmetry
\cite{Wqoslpub}.  Just like the symmetry group, PSG can also produce and
protect gapless excitations.  However, unlike the symmetry group, PSG can
produce and protects gapless gauge bosons and fermions
\cite{Wqoslpub,Wlight,WZqoind}. 

\section{ $SU(2)$ slave-boson theory of doped Mott insulators}

In order to apply the $SU(2)$ slave-boson theory to high $T_c$ superconductors
we need to first generalize the $SU(2)$ slave-boson to the case with finite
doping.  Then we will discuss how to use the $SU(2)$ slave-boson theory to
explain some of those properties in detail.

\subsection{$SU(2)$ slave-boson theory at finite doping}

The $SU(2)$ slave-boson theory can be generalized to describe doped spin
liquids (\cit{WLsu2}; \cit{LNNWsu2}).  The generalized  $SU(2)$ slave-boson
theory involves two $SU(2)$ doublets $\psi_{\v i}$ and $h_{\v i} = \left
({b_{1\v i}\atop b_{2\v i}}\right) $.  Here $b_{1\v i}$ and $b_{2\v i}$ are
two spin-0 boson fields.  The additional boson fields allow us to form $SU(2)$
singlet to represent the electron operator $c_{\v i}$:
\begin{align}
\label{cpsib}
c_{\up\v i} 
=& {1\over \sqrt 2} h^\dagger_{\v i}\psi_{\v i}
= {1\over \sqrt 2} \left (b^\dagger_{1\v i} f_{\up\v i}
+ b^\dagger_{2\v i}f^\dagger_{\down\v i}\right ) 
\nonumber\\
c_{\down\v i}  
=& {1\over \sqrt 2} h^\dagger_{\v i} \bar \psi_{\v i}
= {1\over \sqrt 2} \left ( b^\dagger_{1\v i}f_{\down\v i}
- b^\dagger_{2\v i}f^\dagger_{\up\v i}\right ) 
\end{align}
where $\bar \psi = i\tau^2 \psi^*$ which is also an $SU(2)$ doublet.  The
$t$-$J$ Hamiltonian 
\begin{align*}
H_{tJ}=\sum_{\<\v i\v j\>} \left [ J \left (\v S_{\v i}\cdot \v S_{\v j}
- {1\over 4} n_{\v i} n_{\v j} \right ) -t(c^\dagger_{\alpha \v i} c_{\alpha \v j} + h.c.)
\right ]
\end{align*}
can now be written in terms of our fermion-boson fields.  The Hilbert space of
the fermion-boson system is larger than that of the $t$-$J$ model.  However,
the local $SU(2)$ singlets satisfying $\left ( \psi^\dagger_{\v i} \v\tau
\psi_{\v i} + h^\dagger_{\v i} \v\tau h_{\v i} \right ) |{\rm phys} \rangle =
0$ form a subspace that is identical to the Hilbert space of the $t$-$J$
model.  On a given site, there are only three states that satisfy the above
constraint.  They are $f^\dagger_1 |0\rangle$, $f^\dagger_2 |0\rangle$, and
${1\over \sqrt 2} \left (b^\dagger_1 + b^\dagger_2 f^\dagger_2 f^\dagger_1
\right ) |0\rangle$ corresponding to a spin up and down electron, and a
vacancy respectively.  Furthermore, the fermion-boson Hamiltonian $H_{tJ}$, as
a $SU(2)$ singlet operator, acts within the subspace, and has same matrix
elements as the $t$-$J$ Hamiltonian.

 We note that just as in eq.~(\ref{Eq.36}), our treatment of the ${1\over 4}
n_{\v i}n_{\v j}$ term introduces a nearest neighbor boson attraction term which we
shall ignore from now on.\footnote{
\Ref{LS0101} have introduced a
slightly different formulation where the combination $\left( \v{S}_{\v i}\cdot
\v{S}_{\v j} - {1\over 4} n_{\v i}n_{\v j} \right)$ is written as $
- {1 \over 2} \left | \left( f_{\v i\up}^\dagger f_{\v j\down}^\dagger -
  f_{\v i\down}^\dagger f_{\v j\up}^\dagger \right) \right |^2 (1 - h_{\v i}^\dagger h_{\v i})
(1 - h_{\v j}^\dagger h_{\v j}) $.  The last two factors are the boson projections which
are needed to take care of the case when both sites $i$ and $j$ are occupied
by holes.  While the formulations are equivalent, the mean field phase diagram
is a bit different in that a nearest-neighbor attraction term may lead to
boson pairing.  The competition between boson condensation and boson pairing
needs further studies but we will proceed without the boson interaction term.}
Now the partition function $Z$ is given by
\begin{equation*}
Z = \int D\psi D\psi^\dagger DhDa^1_0Da^2_0Da^3_0D U\exp
\left(
-\int^\beta_0 d\tau L_2
\right)
\end{equation*}
with the Lagrangian taking the form
\be
\label{128}
L_2 &=& \tilde{J} \sum_{\<\v i\v j\>} \Tr \left[ U_{\v i\v j}^\dagger
U_{\v i\v j} \right] + \tilde{J} 
 \sum_{<\v i\v j>} \left( \psi_{\v i}^\dagger
U_{\v i\v j} \psi_{\v j} +
c.c. \right) \nonumber \\
&+& \sum_{\v i} \psi_{\v i}^\dagger \left(
\partial_\tau - ia_{0\v i}^\ell \tau^\ell \right) \psi_{\v i} \nonumber \\
&+& \sum_{\v i} h_{\v i}^\dagger \left(
\partial_\tau - ia_{0\v i}^\ell \tau^\ell + \mu \right) h_{\v i} \nonumber \\
&-& {1\over 2}\sum_{\<\v i\v j\>} t_{\v i\v j} 
\left(
\psi_{\v i}^\dagger
h_{\v i}h_{\v j}^\dagger
 \psi_{\v j} + c.c.
\right)
\,\,\, . \nonumber \\
\en

Following the standard approach with the choice $\tilde{J} = {3\over 8} J$, we
obtain the following mean-field Hamiltonian
for the fermion-boson system, which is an extension of \Eq{sl2a.8} to the
doped case:
\begin{align}
\label{5}
H_\text{mean} =& \sum_{\<\v i \v j\>} \frac{3}{8}
 J \left[ 
\frac12 \Tr (U_{\v i \v j}^\dag~U_{\v i \v j}) +(\psi^\dag_ i
U_{\v i \v j}
\psi_ j +~h.c.)
\right]  
\nonumber\\
&-{1\over 2}\sum_{\<\v i \v j\>} t (h^\dagger_{\v i} U_{\v i\v j} h_{\v j} + {\rm h.c.} )
\nonumber\\
&
- \mu \sum_{\v i} h^\dagger_{\v i}h_{\v i}  
+\sum_{\v i}  a_0^l (
\psi_{\v i}^\dag \tau^l \psi_{\v i} 
+h_{\v i}^\dag \tau^l h_{\v i} )
\end{align}
The value of the chemical potential $\mu$ is chosen such that the total boson
density (which is also the density of the holes in the $t$-$J$ model) is
\begin{equation*}
\langle h^\dagger_{\v i} h_{\v i}\rangle =
\langle b^\dagger_{1\v i} b_{1\v i} +  b^\dagger_{2\v i} b_{2\v i}\rangle = x.  
\end{equation*}
The values of $a^l_{0}(\v i)$ are chosen such that 
\begin{equation*}
\langle \psi^\dagger_{\v i} \tau^l \psi_{\v i} +
h^\dagger_{\v i} \tau^l h_{\v i}\rangle = 0.
\end{equation*}
For $l = 3$ we have
\begin{equation}
\langle f^\dagger_{\alpha\v i} f_{\alpha\v i} +
b^\dagger_{1\v i} b_{1\v i} - b^\dagger_{2\v i} b_{2\v i}\rangle = 1 \label{6}
\end{equation}
We see that unlike the $U(1)$ slave-boson theory, the density of the fermion
$\langle f^\dagger_{\alpha \v i} f_{\alpha \v i}\rangle$ is not necessarily
equal $1 - x$.  This is because a vacancy in the $t$-$J$ model may be
represented by an empty site with a $b_1$ boson, or a doubly occupied site
with a $b_2$ boson.  


\subsection{ The mean-field phase diagram}

\begin{figure}
\centerline{
\includegraphics[width=3in]{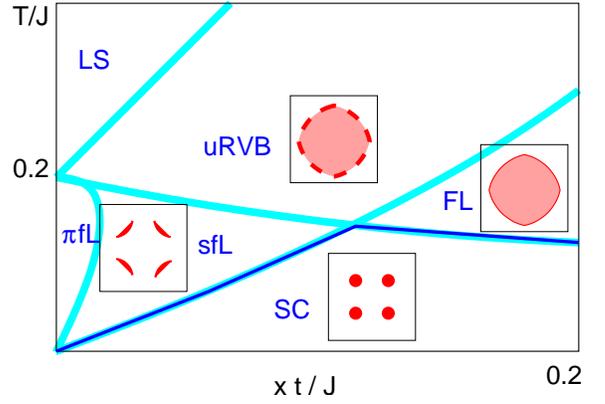}
}
\caption{
$SU(2)$ mean-field  phase diagram for $t/J=1$. The phase diagram for $t/J=2$
is quantitatively very similar to the $t/J=1$ phase diagram, when
plotted in terms of the scaled variable $xt/J$, except the
$\pi$fL phase disappears at a lower scaled doping concentration.
We also plotted the Fermi surface, the Fermi arcs, or the Fermi points in some
phases. \cite{WLsu2} 
}
\label{su2phase}
\end{figure}

To obtain the mean-field phase diagram, we have searched the minima of the
mean-field  free energy for the mean-field ansatz with translation, lattice
and spin rotation symmetries.  We find a phase diagram with six different
phases (see Fig. \ref{su2phase}) \cite{WLsu2}.

{(1)}  The $d$-wave superconducting (SC) phase is described by the following
mean-field ansatz
\begin{align}
\label{dSC}
U_{\v i, \v i+ \hat x} = & -\chi\tau^3 + \Delta\tau^1, \nonumber\\
U_{\v i, \v i+ \hat y} = & -\chi\tau^3 - \Delta\tau^1,
\nonumber\\
a^3_{0}  \neq & 0,\ \ \ \ \ \
a^{1,2}_{0}  =  0,
\nonumber\\
\langle b_1\rangle \ne & 0 ,\ \ \ \ \ \ 
\langle b_2\rangle = 0.
\end{align}
Notice that the boson condenses in the SC phase despite the fact that in our
mean-field  theory the interactions between the bosons are ignored.  
In the SC phase, the fermion and boson dispersion are given by $\pm E_f$ and
$\pm E_b-\mu$, where
\begin{align}
\label{EfEb}
E_f = & \sqrt{(\epsilon_f+a_0^3)^2 + \eta^{2}_f},  \nonumber\\
\epsilon_f  =& -{3J\over 4} (\cos k_x + \cos k_y ) \chi,  \nonumber\\ 
\eta_f  =& -{3J\over 4} (\cos k_x - \cos k_y ) \Delta,  \nonumber\\
E_b = & \sqrt{(\epsilon_b+a_0^3)^2 + \eta^{2}_b},  \nonumber\\
\epsilon_b  =& -2t (\cos k_x + \cos k_y ) \chi,  \nonumber\\ 
\eta_b  =& -2t (\cos k_x - \cos k_y ) \Delta.
\end{align}

{(2)}  The Fermi liquid (FL) phase is similar to the SC phase except that
there is no fermion pairing ($\Delta = 0$).

{(3)} Staggered flux liquid (sfL) phase:
\begin{align}
\label{sFans}
U_{\v i, i + \hat x} & = - \tau^3 \chi - i (-)^{\v i} \Delta ,
\nonumber\\
U_{\v i, i + \hat y} & = - \tau^3 \chi + i (-)^{\v i} \Delta ,
\nonumber\\
a^l_{0}  = & 0, \ \ \ \ \ \ 
\langle b_{1,2}\rangle = 0.
\end{align}
The $U$ matrix is the same as that of the staggered flux phase in the $U(1)$
slave-boson theory, which breaks transition symmetry.  Here the breaking of
translational invariance is a gauge artifact.  In fact, a site dependent
$SU(2)$ gauge transformation $W_{\v i} = e^{-i\pi\tau^1/4} e^{-i\pi (i_x + i_y
) (\tau^1/2+1)}$ maps the sfL ansatz to the $d$-wave pairing ansatz:
\begin{align}
\label{8}
U_{\v i, \v i+ \hat x} = & -\chi\tau^3 + \Delta\tau^1, \nonumber\\
U_{\v i, \v i+ \hat y} = & -\chi\tau^3 - \Delta\tau^1,
\nonumber\\
a^l_{0}  = & 0, \ \ \ \ \ \ 
\langle b_{1,2}\rangle = 0.
\end{align}
which is explicitly translation invariant.  However, the staggered flux
representation of eq.~(\ref{sFans}) is more convenient because the gauge
symmetry is immediately apparent.  Since this $U$ matrix commutes with
$\tau^3$, it is clearly invariant under $\tau^3$ rotation, but not $\tau^1$
and $\tau^2$, and the gauge symmetry has been broken from $SU(2)$ down to
$U(1)$, following the discussion in section X.F.  For this reason we shall
refer to this state as the staggered flux liquid (sfL).

In the sfL phase, the fermion and boson dispersion are still given by $\pm
E_f$ and $\pm E_b-\mu$ with $E_f$ and $E_b$ in \Eq{EfEb}, but now $a_0^3=0$.
Since $a^3_0 = 0$ we have $\langle f^\dagger_{\alpha \v i} f_{\alpha \v i}
\rangle = 1$ and $\langle b^\dagger_1b_1\rangle = \langle
b^\dagger_2b_2\rangle = {x\over 2}$. 

{(4)}  The $\pi$-flux liquid ($\pi$fL) phase is the same as the sfL phase
except here $\chi = \Delta$.

{(5)}  The uniform RVB (uRVB) phase is described by eq. (\ref{sFans}) with
$\Delta = 0$.

{(6)}  A localized spin (LS) phase has $U_{\v i\v j} = 0$ and $a^l_{0\v i} = 0$, where
the fermions cannot hop.


\subsection{Simple properties of the mean-field  phases}

Note that the topology of the phase diagram is similar to that of $U(1)$ mean
field theory shown in Fig.~\ref{U1}.  The uRVB, sfL, $\pi$fL and LS phases
contain no boson condensation and correspond to unusual metallic states.  
As temperature is lowered, the uRVB phase changes into the sfL or $\pi$fL
phases.  A gap is opened at the Fermi surface near $(\pi,0)$ which reduces the
low energy spin excitations.  Thus the sfL and $\pi$fL phases correspond to
the pseudo-gap phase.  

The FL phase contains boson condensation.  In this case the electron Green's
function $\<c^\dag c\>=\<(\psi^\dag h)(h^\dag \psi)\>$ is proportional to the
fermion Green's function $\<\psi^\dag\psi\>$.  Thus the electron spectral
function contain $\del$-function peak in the FL phase.  Therefore, the low
energy excitations in the FL phase are described by electron-like
quasiparticles and the FL phase corresponds to a Fermi liquid phase of
electrons. 

The SC phase contains both the boson and the fermion-pair condensations and
corresponds to a $d$-wave superconducting state of the electrons.  Just like
the $U(1)$ slave-boson theory, the superfluid density is given by
$\rho_s=\frac{\rho_s^b \rho_s^f}{ \rho_s^b +\rho_s^f}$ where $\rho_s^b$ and
$\rho_s^f$ are the superfluid density of the bosons and the condensed
fermion-pairs, respectively. We see that in the low doping limit, $\rho_s \sim
x$ and one need the condensation of both the bosons and the fermion-pairs to
get a superconducting state.  

We would like to point out that the different mean-field  phases contain
different gapless gauge fluctuations at classical level. \ie the gauge groups
for gapless gauge fluctuations are different in different mean-field phases.
The uRVB and the $\pi$fL phases have trivial $SU(2)$ flux and the gapless
gauge fluctuations are $SU(2)$ gauge fluctuations.
In the sfL phase, the collinear $SU(2)$ flux break the $SU(2)$ gauge structure
to a $U(1)$ gauge structure.  In this case the gapless gauge fluctuations are
$U(1)$ gauge fluctuations.  In the SC and FL phases, $\langle b_a\rangle \neq
0$. Since $b_a$ transform as a $SU(2)$ doublet, there is no pure $SU(2)$ gauge
transformation that leave mean-field ansatz $(U_{\v i\v j}, a^l_0, b_a)$
invariant. Thus the invariant gauge group (IGG) is trivial.  As a result, the
$SU(2)$ gauge structure is completely broken and there is no low energy gauge
fluctuations.

\subsection{Effect of gauge fluctuations: enhanced $(\pi,\pi)$ spin
fluctuations in pseudo-gap phase} \label{asl}

The pseudo-gap phase has a very puzzling property which seems hard to explain.
As the doping is lowered, it was found experimentally that both the pseudo-gap
and the antiferromagnetic (AF) spin correlation in the normal state increase.
Naively, one expects the pseudo-gap and the AF correlations to work against
each other. That is the larger the pseudo-gap, the lower the single particle
density of states, the fewer the low energy spin excitations, and the weaker
the AF correlations.

It turns out that the gapless $U(1)$ gauge fluctuations present in the sfL
phase play a key role in resolving the above puzzle \cite{KL9930,RWspin}.  Due
to the $U(1)$ gauge fluctuations, the AF spin fluctuations in the sfL phase
are as strong as those of a nested Fermi surface, despite the presence of the
pseudo-gap. 

To see how the $U(1)$ gauge fluctuation in the sfL phase enhance the AF spin
fluctuations, we map the lattice effective theory for the sfL state onto a
continuum theory.  In the low doping limit, the bosons do not affect the spin
fluctuations much. So we will ignore the bosons and effectively consider the
undoped case. In the sfL phase, the low energy fermions only appear near $\v
k=(\pm \frac{\pi}{2},\pm \frac{\pi}{2})$ Since the fermion dispersion is
linear near $\v k=(\pm \frac{\pi}{2},\pm \frac{\pi}{2})$, those fermions are
described by massless Dirac fermions in the continuum limit:
\begin{eqnarray}
\label{Dirac}
S&=&\int d^{3}x \sum_\mu \sum_{\al=1}^N\bar{\Psi}_{\al}
v_{\al,\mu} \partial_{\mu}\gamma_{\mu}\Psi_{\al}
\end{eqnarray}
where $v_{\al, 0}=1$ and $N=2$, but in the following we will treat $N$ as
an arbitrary integer, which gives us a large $N$ limit of the sfL state.  In
general $v_{\al, 1}\neq v_{\al,2}$. However, for simplicity we will
assume $v_{\al,i}=1$ here.  The Fermi field $\Psi_{\al}$ is a $4\times
1$ spinor which describes lattice fermions $f_{\v i}$ with momenta near $(\pm
\pi/2, \pm \pi/2)$.  The $4\times4$ $\gamma_{\mu}$ matrices form a
representation of the
$\{\gamma_{\mu},\gamma_{\nu}\}=2\delta_{\mu\nu}$ ($\mu,\nu = 0,1,2$) and are
taken to be
\begin{eqnarray}
\gamma_{0}&=&\bpm \sigma_{3}&0\\ 0&-\sigma_{3} \epm, \quad
\gamma_{1}=\bpm \sigma_{2}&0\\ 0&-\sigma_{2}   \epm, \\
\gamma_{2}&=&\bpm\sigma_{1}&0\\  0&-\sigma_{1}\epm
\end{eqnarray}
with $\sigma_{\mu}$ the Pauli matrices.  Finally note that
$\bar{\Psi}_{\sigma} \equiv \Psi^{\dag}_{\sigma} \gamma_{0}$.

The fermion field $\Psi$ couples to the $U(1)$ gauge field in the sfL phase.
To determine the form of the coupling, we note that the $U(1)$ gauge
transformation takes the following form
\begin{equation*}
 f_{\v i} \to e^{i\th_{\v i}} f_{\v i}
\end{equation*}
if we choose the  ansatz \Eq{sFans} to describe the sfL phase.  By requiring
the $U(1)$ gauge invariance of the continuum model, we find the continuum
Euclidean action to be
\begin{eqnarray}
\label{QED3a}
S&=&\int d^{3}x \sum_\mu \sum_{\sigma=1}^N\bar{\Psi}_{\sigma}
v_{\sigma,\mu} (\partial_{\mu}-ia_{\mu})\gamma_{\mu}\Psi_{\sigma}
\end{eqnarray}

The dynamics for the $U(1)$ gauge field arises solely due to the screening by
bosons and fermions, both of which carry gauge charge. In the low doping
limit, however, we will only include the screening by the fermion fields.
After integrating out $\Psi$ in \Eq{QED3a}, we obtain the following effective
action for the $U(1)$ gauge field \cite{KL9930}
\begin{eqnarray}
\cal Z&=&\int Da_{\mu}\exp\Big( -\frac{1}{2}\int\frac{d^3q}{(2\pi)^3}a_{\mu}
(\v{q})\Pi_{\mu\nu}a_{\nu}(-\v{q})\Big) \nonumber \\
\Pi_{\mu\nu}&=&\frac{N}{8}\sqrt{\v{q}^2}\Big(\delta_{\mu\nu}
- \frac{q_{\mu}q_{\nu}}{\v{q}^{2}}\Big)
\label{Pi}
\end{eqnarray}
By simple power counting we can see that the above polarizability makes the
gauge coupling $a_\mu j^\mu$ a marginal perturbation at the free fermion fixed
point.  
Since the conserved current $j^\mu$ cannot have any anomalous dimension, this
interaction is an \emph{exact} marginal perturbation protected by current
conservation.

For $N=2$, the spin operator with momenta near $\v q=(0,0)$, $(\pi,\pi)$, and
$(\pi,0)$ has different form when expressed in terms of $\Psi_\al$.  Near $\v
q=(0,0)$ 
\begin{equation*}
 \v S_u(\v x)=\frac12 \bar \Psi_\al \ga^0 \v \si_{\al\bt} \Psi_\bt
\end{equation*}
Near $\v q=(\pi,\pi)$ 
\begin{equation*}
 \v S_s(\v x)=\frac12 \bar \Psi_\al \v \si_{\al\bt} \Psi_\bt
\end{equation*}
Near $\v q=(\pi,0)$ 
\begin{equation*}
 \v S_{(\pi,0)}(\v x)=\frac12 \bar \Psi_\al 
\bpm 0 & \sigma_{1}\\ \sigma_{1} & 0 \epm
\v \si_{\al\bt} \Psi_\bt
\end{equation*}

\begin{figure}[tb]
\begin{center}
\includegraphics[width=3in]{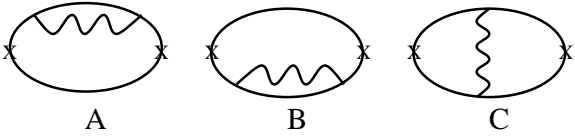}
\caption{Non-zero leading $1/N$ corrections to the staggered spin
correlation function. The ${\v x}$ denotes the vertex which is the $4 \times
4$ unit matrix in  the case of interest.}
\label{Pol1N}
\end{center}
\end{figure}

At the mean-field level, all the above three spin operators have algebraic
correlations $1/r^4$ with decay exponent $4$.  The effect of gauge
fluctuations can be included at $\frac{1}{N}$ order by calculating the
diagrams in Fig. \ref{Pol1N}.  We find that (\cit{RWspin}; \cit{FPS0308}) these three spin
correlaters still have algebraic decays, indicating that the gauge interaction
is indeed marginal.  The decay exponents of the spin correlation near $\v
q=(0,0)$ and $\v q=(\pi,0)$ are not changed and remain to be $4$.  This result
is expected for the spin correlation near $\v q=(0,0)$ since $S_u(\v x)$ is
proportional to the conserved density operator that couple to the $U(1)$ gauge
field.  Therefore $S_u(\v x)$ cannot have anomalous dimension.  
$ \v S_{(\pi,0)}(\v x)$ does not have any anomalous dimension either (at $1/N$
order).  In fact, this result holds to all orders in $1/N$ for the case of
isotropic velocities, due to an $SU(4)$ symmetry 
\cite{HS04}.
Thus the spin fluctuations near $(\pi,0)$ is also not enhanced by the gauge
interaction. This may explain why it is so hard to observe any spin
fluctuations near $(\pi,0)$ in experiments.

$S_s(\v x)$ is found to have a non-zero anomalous dimension.  The spin
correlation near $\v q=(\pi,\pi)$ is found to be $1/r^{4-2\al}$ with
\begin{equation}
\label{eqnu}
\al=\frac{32}{3\pi^2 N}
\end{equation}
In the $\om$-$\v k$ space, the imaginary part of the spin susceptibility near
$(\pi,\pi)$ is given by
\begin{align}
\label{sspin}
& \Im\chi(\omega,\mathbf{q})
\equiv\Im\langle S^{+}(\omega,\mathbf{q}+\v Q)
S^{-}(-\omega,-\mathbf{q}+\v Q) \rangle \nonumber \\ 
=&
\frac{C_s}{2}\sin(2\al\pi)\Gamma(2\al-2)
\Theta(\omega^2-\mathbf{q}^2)\left(\omega^2-\mathbf{q}^2\right)^{1/2-\al} 
\end{align}
where $C_{s}$ is a constant depending on the physics at the lattice scale.

\begin{figure}[tb]
\begin{center}
\includegraphics[width=70mm]{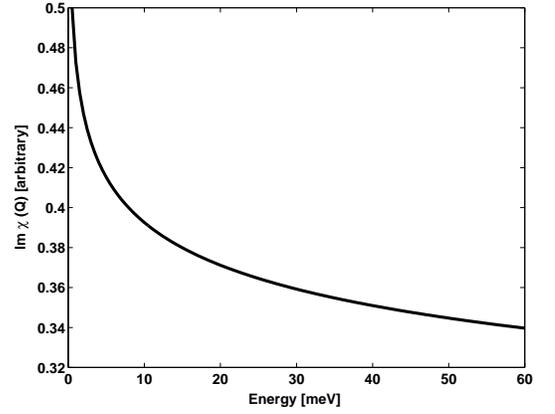}
\caption{Imaginary part of the spin susceptibility at $(\pi,\pi)$. 
Note the divergence at small $\omega$. (from \Ref{RWspin}}
\label{chi_Q_ASL}
\end{center}
\end{figure}

From \Eq{sspin} it is clear that the gauge fluctuations have reduced the
mean-field exponent. If we boldly set $N=2$ which is the physically relevant
case we find $\al =0.54 > 1/2$ which signals the divergence of $\chi(\omega = 0,
q = 0)$.  Thus, after including the gauge fluctuations, the $(\pi,\pi)$ spin
fluctuations are enhanced in the sfL phase despite the pseudo-gap.  In Fig.
\ref{chi_Q_ASL}, we plot the imaginary part of the spin susceptibility at
$(\pi,\pi)$. The $\om$ dependence of the spin susceptibility at $(\pi,\pi)$ is
similar to the one from a nested Fermi surface.

The enhancement of the staggered spin correlation follows the trend found in
Gutzwiller projection of the staggered flux (or equivalently the $d$-wave
pairing) state.  
\Ref{I00} and 
\Ref{PRT0404} reports a
power law decay of the equal time staggered spin correlation function as
$r^{-\nu}$ where $\nu = 1.5$ for the undoped case and $\nu = 2.5$ for
5\% doping, which are considerably slower than the $r^{-4}$ behavior before
projection.  


We remark that with doping, Lorentz invariance is broken by the presence of
bosons. In this case the Fermi velocity receives an logarithmic correction
which enhances the specific heat $\ga$  coefficient and the uniform
susceptibility \cite{KLW9709}.

\subsection{Electron spectral function} \label{specG}

One of the striking properties of the high $T_c$ superconductor is the
appearance of the pseudo-gap in electron spectral function for underdoped
samples, even in the non-superconducting state. To understand this property
within the $SU(2)$ slave-boson theory, we like to calculate the physical
electron Green function.  Since the non-superconducting state for small $x$ is
described by the sfL phase in the $SU(2)$ slave-boson theory, so we need to
calculate the electron Green function in the sfL phase.

\subsubsection{Single
hole spectrum}

The electron Green's function is given by
\begin{equation*}
 G_e(\v x)=\<h^\dag(\v x)\psi(\v x) h(0)\psi^\dag(0)\>
\end{equation*}
If we ignore the gauge interactions between the bosons and the fermions, the
electron Green's function can be written as
\begin{equation*}
 G_{e0}=\<h^\dag h\>_0 \<\psi \psi^\dag\>_0
\end{equation*}
where the subscript $0$ indicates that we ignore the gauge fluctuations when
calculating $\<...\>_0$.

\begin{figure}[t]
\begin{center}
\includegraphics[scale=0.5]{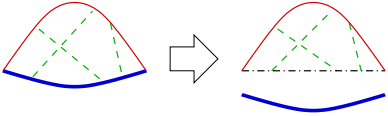}
\end{center}
\caption{
The thick line represents the boson world line,
the thin line represents the fermion world line, and the dash line represent
the gauge interaction. The dash-dot line is the straight return path.
The $U(1)$ gauge interaction is caused by the extra phase term
$e^{i\oint d\v x\cdot \v a}$ due to the $U(1)$ flux through the loop
formed by the boson and the fermion world lines. 
Such flux can be approximated by
the flux through the loop
formed by the fermion world line and the straight return path.
}
\label{bsnfrm}
\end{figure}

The effect of the $U(1)$ gauge fluctuations is an extra phase term $e^{i\oint
d\v x\cdot \v a}$ determined by the $U(1)$ flux through the loop formed by the
boson and the fermion world lines (see Fig. \ref{bsnfrm}).  Since the fermion
has a linear dispersion relation, the area between the boson and the fermion
world lines is of order $|\v x|^2$, where $|\v x|$ is the separation between
the two points of the Green's function.  Such an area is about the same as the
area between the fermion world line and the straight return path (see Fig.
\ref{bsnfrm}).  So we may approximate the effect of $U(1)$ gauge
fluctuations as the effect caused by the $U(1)$ flux through the fermion world
line and the straight return path \cite{RW0140}. This corresponds to
approximate the electron Green's function as
\begin{equation*}
 G_e(\v x)=\<h^\dag(\v x) h(0)\>_0 
\<\psi(\v x) \psi^\dag(0) e^{i\int_0^{\v x} d\v x \cdot \v a}\>
\end{equation*}
where $\int_0^{\v x}d\v x$ is the integration along the straight return path
and $\<...\>$ includes integrating out the gauge fluctuations.

First, let us consider the fermion Green's function.  At the leading
order of a large-$N$ approximation, it was found that \cite{RW0171,RW0140}
\footnote{ 
Note that the usual fermion Green's function $\<\psi(\v x)
\psi^\dag(0)\>$ is not gauge invariant.  As a result, the Green's function is
not well defined and depends on the choices of gauge-fixing conditions
\cite{FT0103,FTV0235,K0206,Y0217}.  If one incorrectly identifies $\<\psi(\v
x) \psi^\dag(0)\>$ as the the electron Green's function, then the electron
Green's function will have different decay exponents for different
gauge-fixing conditions.  In contrast, the combination $\<\psi(\v x)
\psi^\dag(0) e^{i\int_0^{\v x} d\v x \cdot \v a}\>$ is gauge invariant and
well defined.  The resulting electron Green's function does not depend on
gauge-fixing conditions.
}
\begin{equation}
\label{fermG}
 \<\psi(\v x) \psi^\dag(0) e^{i\int_0^{\v x} d\v x \cdot \v a}\>
\propto (\v x^2)^{-(2-\al)/2}  .
\end{equation}
where $\al$ is given in \Eq{eqnu}.  
We note that the above becomes the Green's function for free
massless Dirac fermion when $\al=0$.  The finite $\al$ is the effect of gauge
fluctuations.

\begin{figure}[tb]
\begin{center}
\includegraphics[scale=0.65]{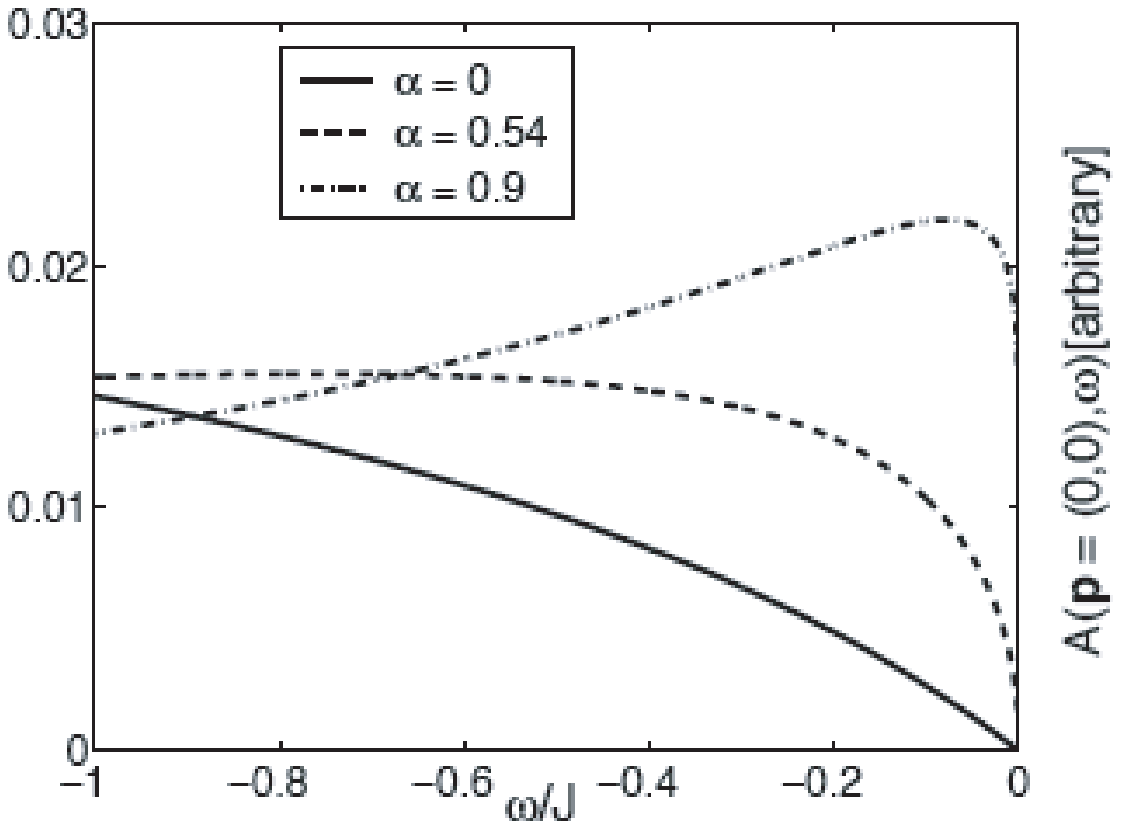}
\end{center}
\caption{
The single hole spectral function at $\left(  {\pi \over 2},{\pi \over 2}    \right)$.  Increasing $\alpha$ corresponds to increasing attraction between fermion and boson due to gauge field fluctuations. (from \Ref{RW0140})
}
\label{specG2}
\end{figure}

For a single hole, the boson Green's function is simply that of a classical
particle.  The electron Green's function $G_e(\v{r},\tau)$ is readily
calculated using \Eq{fermG} and its Fourier transform yields the electron
spectral function.  The result at the nodal position $\left( {\pi \over 2},
{\pi \over 2}   \right)$ is shown in Fig.~(\ref{specG2}).  The $\alpha = 0$
curve is the result without gauge fluctuation.  It is the convolution of the
fermion and Bose spectra and is extremely broad.  The gauge field leads to an
effective attraction between the fermion and boson in order to minimize the
gauge flux enclosed by the fermion on boson vortex lines as shown in
Fig.~(\ref{bsnfrm}).  The result is a piling up of a spectral weight at low
energy with increasing $\alpha$.  Still, the one-hole spectrum remains
incoherent, as is appropriate for a deconfined $U(1)$ spin liquid state.  This
calculation can be extended to finite hole density, which requires making
certain assumptions about the boson Green's function 
(\cit{RW0140}; \cit{FT0103}).
Under certain  conditions they obtain power-like type spectral functions
similar to those of the Luttinger liquid.

\subsubsection{Finite hole density: pseudo-gap and Fermi arcs}
\label{specmean}

Here we will consider the mean-field electron Green's function $G_0$
at finite doping.
Using the expression
of $c_\alpha$ in \Eq{cpsib},
the mean-field electron Green's function is given by the product of the
fermion and boson Green functions.  So the electron spectral function is a
convolution of the boson spectral function and the fermion spectral function.

Let us consider a region of the pseudogap above $T_c$ but at a temperature
which is not too high.  The boson can be considered nearly condensed. The
boson spectral function contain a sharp peak at $\om=0$ and $\v k=0$ and $\v
k=(\pi,\pi)$.  The weight of the peak is of order $x$ and the width is of
order $T$.  At high energies, the boson Green function is given by the
single-boson Green function $G_{b}^{s}$ as if no other bosons are present.  So
the boson spectral function also contain a broad background which extends the
whole band width of the boson band.  The resulting  mean-field electron Green
function has a form (\cit{WLsu2}; \cit{LNNWsu2})
\begin{align}
\label{G0SC}
G_0 
=& {x\over 2} \left ( {u_{\v k}^2\over \omega - E_f} 
+ {v_{\v k}^2\over \omega + E_f}\right ) + G_{in}
\end{align}
where $u$ and $v$ are the coherent factors: 
\begin{align*}
u_{\v k} = &\sqrt{{E_f + \epsilon_f\over 2E_f}} {\rm sgn} (\eta_f ),
\nonumber\\
v_{\v k} = &\sqrt{{E_f - \epsilon_f\over 2E_f}} .  
\end{align*}
The second term $G_{in}$ gives rise to a broad background in the electron. It
comes from the convolution of the background part of the boson spectral
function  and the fermion spectral function.  The first term is the coherent
part since its imaginary part is a peak of width $T$, which is approximated by
a $\delta$-function here.  The quasiparticle dispersion is given by $\pm E_f$.
The peak in the electron spectral function crosses zero energy at four points
at $\v k=(\pm \frac{\pi}{2}, \pm \frac{\pi}{2})$.  Thus the mean-field sfL
phase has four Fermi points.  Also, in the sfL phase, Im$G_{in}$ is non-zero
only for $\omega<0$ and contributes $1/2$ to a total spectral weight which is
$(1+x)/2$.



From the dispersion relation of the peak $\om=E_f(\v k)$ and the fact that
Im$G_{in}\approx 0$ when $\omega<-E_f(\v k)$, we find that the electron
spectral function contain the gap of order $\Del$ at $(0,\pi)$ and $(\pi,0)$
even in the non-superconducting state.  So the mean-field electron spectral
function of the $SU(2)$ slave-boson theory can explain the pseudo-gap in the
underdoped samples.  However, if we examine the mean-field electron spectral
function more closely, we see that the Fermi surface of the quasiparticles is
just four isolated points $(\pm \pi/2,\pm \pi/2)$.  This property does not
agree with experiments. 

\begin{figure}
\includegraphics[width=1.5in]{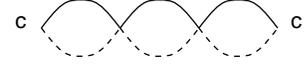}
\caption{
A diagram for renormalized electron Green function.
The solid (dash) line is the fermion (boson) propagator.
}
\label{su2longf2}
\end{figure}

\begin{figure}
\includegraphics[width=3in]{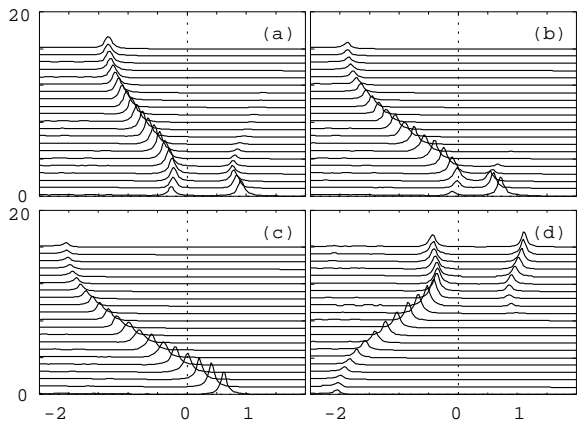}
\caption{
The electron spectral function for, from top down, 
(a) $k=(-\pi/4, \pi/4) \to (\pi/4, 3\pi/4)$,
(b) $k=(-\pi/8, \pi/8) \to (3\pi/8, 5\pi/8)$, 
(c) $k=(0, 0) \to (\pi/2, \pi/2)$, and
(d) $k=(0, \pi) \to (0,0)$.
We have chosen $J=1$.
The paths of the four momentum scans are shown in Fig. \ref{su2longf7}.
}
\label{su2longf3}
\end{figure}

In reality there is a strong attraction between the boson and the fermions due
to the fluctuation around the mean-field state.  The dominant effect comes
from the gauge fluctuations which attempt to bind the bosons and the fermions
into electrons.  This corresponds to an effective attraction between the
bosons and the fermions.  In the case of a single hole, the interaction with
gauge fields can be treated as discussed in the last section.  Here we proceed
more phenomenologically.  One way to include this effect is to use the diagram
in Fig. \ref{su2longf2} to approximate the electron Green function, which
corresponds to an effective short range interaction of form
\begin{equation}
-\frac{V}{2} (\psi^\dagger h) (h^\dagger \psi )=-V  c^\dagger c
\end{equation}
with $V < 0$.  We get
\begin{equation}
G = \frac{1}{\left( G_0 \right)^{-1} + V}
\label{GV}
\end{equation}
The first contribution to $V$ come from the fluctuations of $a_{0}^{\ell}$
which induces the following interaction between the fermions and the bosons:
\begin{equation}
 \psi^{\dagger} \v{\tau} \psi \cdot h^{\dagger}
\v{\tau}h
\end{equation}
The second one (whose importance was pointed out by 
\Ref{L9527}) is
the fluctuations of $|\chi_{\v i\v j}|$ which induces
\begin{equation}
-t (\psi^\dagger h)_{\v j} (h^\dagger \psi )_{\v i} =-2t c^\dagger_{\v j} c_{\v i}
\end{equation}
This is nothing but the original hopping term. We expect the coefficient $t$
to be reduced due to screening, but in the following we adopt the form
\begin{equation}
V(\v k)=U + 2t(\cos k_x +\cos k_y)
\end{equation}
for $V$ in eq. (\ref{GV}). 
In Fig. \ref{su2longf3} and \ref{su2longf3a} we plot the electron spectral
function calculated from eq. (\ref{GV}) (\cit{WLsu2}; \cit{LNNWsu2}).

\begin{figure}
\includegraphics[width=2in]{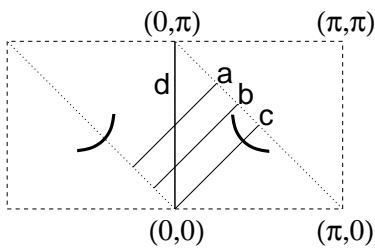}
\caption{
The solid line a, b, c, and d are paths of the four momentum scans
in Fig. \ref{su2longf3}.
The solid curves are schematic representation of
the Fermi segments where the quasiparticle peak
crosses the zero energy.
}
\label{su2longf7}
\end{figure}  

\begin{figure}
\includegraphics[width=3in]{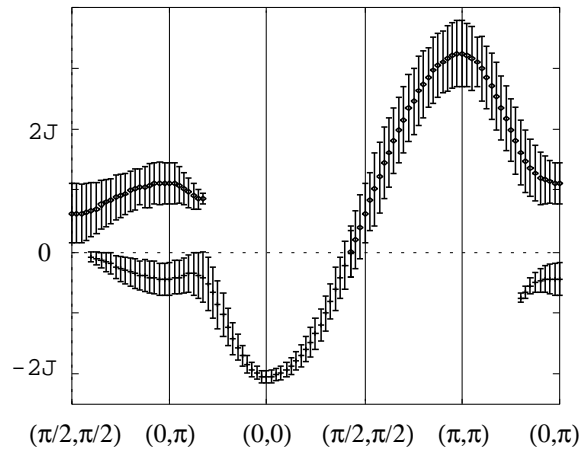}
\caption{
The points describe the dispersion of the quasi-particle peaks for the
s-flux liquid phase in Fig. \ref{su2longf3a}.  The vertical bars are proportional
to the peak values of Im$G_U$ which are proportional to
the quasi-particle weight.
}
\label{su2longf3a}
\end{figure}

We have chosen 
$t=2J$, $\chi = 1$, $\Delta/\chi = 0.4$, $x = 0.1$, and $T = 0.1 J$. 
The value of $U$ is determined from requiring the renormalized electron Green
function to satisfies the sum rule
\begin{equation}
\int_0^\infty \frac{d\omega }{2\pi}\int \frac{d^2k}{(2\pi)^2}
 \Im G= x
\end{equation}

We find that the gap near $(0, \pm \pi)$ and $(\pm \pi,0)$ survives the
binding potential $V(\v k)$.
However spectral functions near $(\pm \frac{\pi}{2}, \pm \frac{\pi}{2})$ are
modified.  The Fermi point at $(\frac{\pi}{2}, \frac{\pi}{2})$ for the  mean
field electron Green function $G_0$ is stretched into a Fermi segment as shown
in Fig.  \ref{su2longf7}.  As we approach the uRBV phase, $\Delta$ decreases
and the Fermi arcs are elongated.  Eventually the arcs join together to form a
large closed Fermi surface.

While the phenomenological binding picture successfully produces Fermi arcs,
the results are not as satisfactory for the anti-nodal points.  While an
energy gap is produced near $(0,\pi)$, the theory gives a rather sharp
structure at the gap and we see from Fig. \ref{su2longf3a} that the gap above
and below the Fermi energy is not symmetric.  

This exposed a serious weakness of the slave-boson gauge theory approach. With
finite hole density, the bosons tend to condense at the mean field level. In
reality the holes are strongly coupled to gauge fluctuations which tend to
suppress the Bose condensation. While the fermions are also coupled to gauge
fields, the Fermi statistics allow us to approach the problem perturbatively
by introducing artificial expansion parameters such as $\frac{1}{N}$ . In
contrast, the problem of bosons coupled to gauge fields is much less
understood. Furthermore the gauge fields mediate strong attraction between
fermions and bosons and, in the case of $SU(2)$ theory, between $b_1$ and $b_2$
bosons which carry opposite gauge charges. In the phenomenological approach
outlined above, the bosons are treated as almost condensed (\ie a narrow peak
in the spectral function is assumed) and bind with a fermion. The assumption
of ``almost Bose condensation'' leads to sharp hole spectra at both the nodal
and anti-nodal points and the latter disagrees with experiment. 
Furthermore it can be shown that the assumption of Bose condensation leads to
a decoupling of the electron to the electromagnetic field, and as a result, the
current carried by the quasiparticles $j = e\frac{dE_{\v k}(\v A)}{d\v A}$ 
is strongly reduced from $ev_F$ which disagrees with experiments
(see IX.B). 

\Ref{WL9893} took a first step towards addressing this problem by assuming
that the binding between the bosons and the fermions and/or between the
$b_1$ and $b_2$ bosons prevents single-boson
condensation.  The superconducting state characterized by $\< cc\>\neq 0$
contains only boson pair condensation, \ie $\< b_1b_2 \> \neq 0$ while
$\<b\>=0$. They show that with this assumption the quasiparticle current can
be a finite fraction of $ev_F$ , \ie the $\al$ parameter in \Eq{Eq.6} does
not have to go as $x$. The competition between fermion-boson binding,
boson-boson binding and Bose condensation is a complicated problem which is
still poorly understood at present.

STM experiment reveals a rather
broad structure for both particle and hole excitations 
\cite{HLK04}
and ARPES measurements, which can measure only the occupied states, show
a reduction of the density of states over a broad energy range 
\cite{RSK0301}.  These lineshapes are more reminiscent of those shown in Fig.
\ref{specG2} for intermediate $\alpha$.  It appears that the assumption of
almost Bose condensation and simple binding via a
short range potential do not capture the subtlety of gauge fluctuation
effects near $(0,\pi)$ where fermions and bosons appear to be closer to being
deconfined.  This dichotomy between nodal and anti-nodal electronic structure
is an important issue which remains open for further theoretical work.

\subsection{Stability of algebraic spin liquids} \label{screen}

The sfL mean-field ansatz leads to an gapless spin liquid. We will call this
the $U(1)$ spin liquid, and it is an example of a class which we call
algebraic spin liquid (ASL) since all the spin correlations have algebraic
decay.  We would like to stress that the ASL is a phase of matter, not a
critical point at a phase transition between two phases.

ASL has a striking property: its low energy excitations interact with each
other even down to zero energy. This can be seen from the correlation
functions at low energies which always contain branch cut without any poles.
The lack of poles implies that we cannot use free bosonic or free fermionic
quasiparticles to describe the low energy excitations.  For all other commonly
know gapless states, such as solids, superfluids, Fermi liquids, \etc, the
gapless excitations are always described by free bosons or free fermions. The
only exception is the 1D Luttinger liquid.  Thus the ASL can be view as an
example of Luttinger liquids beyond one dimension.

We know that interactions tend to open up energy gaps.  From this point of
view, one might have thought that the only self consistent gapless excitations
are the ones  described by free quasiparticles. Knowing the gapless
excitations in the ASL interact down to zero energy, we may wonder does ASL
really exist?  Have we overlooked some effects which open up energy gap and
make ASL unstable?

Indeed, in the above calculation, we have overlooked two effects. Both of them
can potentially destabilize the ASL.  First, the self-energy in Fig.
\ref{Pol1N}A,B, contains a cut-off dependent term which gives the fermion
$\Psi$ a cut-off dependent mass $m(\La)\bar \Psi\Psi$. In the above
calculation, we have dropped such a term. If such a cut-off-dependent term was
kept, the fermions would gain a mass which would destabilize the ASL. 

Second, we have overlooked the effects of instantons described by the
space-time monopoles of the $U(1)$ gauge field.  After integrating out the
massless fermions, the effective action of the $U(1)$ gauge field has a form
\Eq{LaNL}.  Unlike the Maxwell term discussed in section IX.D, which produced
a $1/r$ potential, in this case the interaction of the  space-time monopoles
is described by a $\log (r)$ potential.  That is the action of the pair of
space-time monopoles separated by a distance $r$ is given by $ C\log (r)$.
Just like the Coulomb gas in 2D, if the coefficient $C$ is larger than 6, then
the instanton effect is an irrelevant perturbation and the inclusion of the
instantons will not destabilize the ASL \cite{IL8988}. If the coefficient $C$
is less then 6, then the instanton effect is a relevant perturbation and the
inclusion of the instantons will destabilize the ASL.

Recently, it was argued in \Ref{HS0301} and \Ref{HSS0310} that the instanton
effect always represent a relevant perturbation due to a screening effect of
3D Coulomb gas, regardless the value of $C$.  This led to a conclusion in
\Ref{HS0301} that the ASL described by the sfL state does not exist.  The easiest way
to understand the screening effect of the 3D Coulomb gas is to note that the
partition function of the Coulomb gas can be written as a path integral
\begin{align}
 &\int \prod d^3\v x_i e^{-C \sum q_i q_j\log |\v x_i-\v x_j|}
\nonumber\\
=&\int \cD\phi e^{-\int d^3\v x \frac{2\pi}{C}
\prt \phi \sqrt{-\prt^2} \prt \phi -g \cos(\phi)}
\label{monopole}
\end{align}
If we integrate out short distance fluctuations of $\phi$, a counter term
$K(\prt \phi)^2$ can be generated. The counter term changes the long distance
interaction of the space-time monopoles from $\log (r)$ to $1/r$.  The
space-time monopoles with $1/r$ interaction always represent a relevant
perturbation, which will destabilize the ASL.  Physically, the change of the
interaction from $\log (r)$ to $1/r$ is due to the screening effect of
monopole-anti-monopole pairs. Thus the counter term $K(\prt \phi)^2$
represents the screening effect.

The issue of the stability of the ASL has been examined by \Ref{Wqoslpub},
\Ref{RWspin} and more carefully by 
\Ref{HSF0451} using an argument based on PSG. They came to the conclusion that the $U(1)$ spin liquid is
stable for large enough $N$, if the $SU(2)$ spin symmetry is generalized to
$SU(N)$.  They showed that there is no relevant operator which can destabilize
the deconfined fixed point which consists of $2N$ two-component Dirac fermions
coupled to {\em noncompact} $U(1)$ gauge fields, for $N$ sufficiently large.
\Ref{HSF0451} also pointed out the fallacy of the monopole screening argument.
We summarize some of the salient points below.

The operators which perturb the noncompact fixed point can be classified into
two types, those which preserve the flux and those which change the flux by
$2\pi$.  The latter are instanton creation operators which restore the
compactness of the gauge field.  Among the first type there are four fermion
terms which are readily seen to be irrelevant, but as mentioned earlier, the
dangerous term is the quadratic fermion mass term.  The important point is
that the mass terms are forbidden by the special symmetry described by PSG.
The discrete symmetry (such as translation and rotation) of the sfL state
defined on the lattice imposes certain symmetry on the continuum Dirac field
which forbids the mass term.  Another way of seeing this is that after
integrating out the short distance fluctuations, if a mass term is generated
it can be described in the lattice model as a deformation of the mean-field
ansatz $\del U_{\v i\v j}$. Since the short distance fluctuations are
perturbative in nature, the deformation $\del U_{\v i\v j}$ cannot change the
symmetry of the ansatz $\bar  U_{\v i\v j}$ that describe the ground state,
\ie if $\bar  U_{\v i\v j}$ is invariant under a PSG, $\del U_{\v i\v j}$ must
be invariant under the same PSG. One can show that for all the possible
deformations that are invariant under the sfL PSG described by \Eq{GssF}, none
of them can generate the mass term for the fermions. Thus the masslessness of
the fermions are protected by the sfL PSG.

As for the second type of operators which change the flux, 
\Ref{HSF0451} appeal to a result in conformal field theory which relates the
scaling dimension of such operators to the eigenvalues of states on a sphere
with a magnetic flux through the surface 
\cite{BKW0249}.  This
is easily bound by the ground state energy of $2N$ component Dirac fermions on
the sphere which clearly scale as $N$.  Thus the creation of instantons is
also irrelevant for sufficiently large $N$.

As far as the monopole screening argument goes, the fallacy is that in that
argument the fermions are first integrated out completely in order to derive
an effective action for the field $\phi$ shown in eq.~\ref{monopole}.  Then
renormalization group arguments generate a $K(\partial \phi)^2$ term.  However
implicit in this procedure is the assumption that the fermions are rapidly
varying variables compared with the monopoles.  The fact that the fermions are
gapless makes this procedure unreliable.  (One could say that the screening
argument implicitly assumes mass generation for the fermions.)  A better
approach is to renormalize the monopoles and the fermions on the same footing,
\ie let the infrared cutoff length scale for the fermion $(L_f)$ and the
monopoles $(L_m)$ to approach infinity with a fixed ratio, e.g.  $L_f/L_m =
1$.  In this case integrating out the fermions down to scale $L_f$
will produce an effective action for the $U(1)$ gauge field of the form

\begin{equation*}
{g(L_f) \over 16 \pi} f_{\mu v}f^{\mu v}
\end{equation*}
where the running coupling constant $g(L_f) \sim L_f$.  This in turn generates an
interaction between two monopoles separated by a distance $r$ which is of
order $g(L_f)/r$.  To calculate such an interaction, we should integrate out all
the fermions with wavelength less than $r$.  We find the interaction to be
${g(r) \over r} \sim r^0$, indicating a logarithmic interaction between
monopoles.   Thus the logarithmic interaction is constantly being rejuvenated
and cannot be screened.  This can be cast into a normalization group language
and we can see that the flow equation for the coupling constant $g$ is
modified from the form used by \cite{HS0301}.  The extra term leads to the
conclusion that the instanton fugacity scales to zero and the instanton
becomes irrelevant for $N$ larger than a certain critical value.  

To summarize, the ASL derived from the sfL ansatz contain a quantum order
characterized by the sfL PSG \Eq{GssF}. The sfL PSG forbids the mass term of the
fermions. To capture such an effect, we \emph{must} drop the mass term in the
self-energy in our calculation in the continuum model \cite{RWspin}. Ignoring
the mass term is a way to include the effects of PSG into the continuum model.
Similarly,  we must ignore the screening effect described by the $K(\prt
\phi)^2$ term when we consider instantons.  We are then assured that instanton
effects are irrelevant in the large $N$ limits.  So the ASL exists and is
stable at least in the large $N$ limit.  The interacting gapless excitations
in the ASL are protected by the sfL PSG.  It is well known that the symmetry
can protect gapless Nambu-Goldstone modes.  The above example shows that the
PSG and the associated quantum order can also protect gapless excitations
\cite{Wqoslpub,WZqoind,RWspin}.


\section{Application of gauge theory to the high $T_c$ superconductivity
problem}

\begin{figure}
\includegraphics[width=2in]{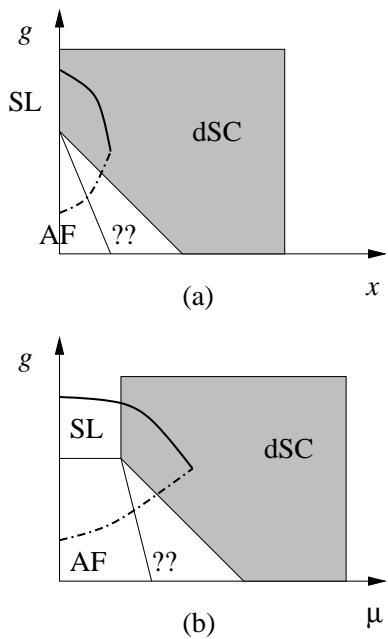}
\caption{
a) Schematic zero temperature phase diagram showing the route between the
antiferromagnetic Mott insulator and the $d$-wave superconductor.  The
vertical axis is labeled by a parameter $g$ which may be taken as a measure
of the frustration in the interaction between the spins in the Mott insulator.
AF represents the antiferromagnetically ordered state. SL is a spin liquid
insulator that could potentially be reached by increasing the frustration.
The path taken by the cuprate materials as a function of doping $x$ is shown
in a thick  dashed-dot line. The question marks represent regions where the
physics is not clear at present. Doping the spin liquid naturally leads to the
dSC state. The idea behind the spin liquid approach is to regard the
superconducting system at non-zero $x$  as resulting from doping the spin
liquid as shown in the solid line, though this is not the path actually taken
by the material.  b) Same as in Fig. \protect\ref{sldsc1}(a) but as a function
of chemical potential rather than hole doping.
}
\label{sldsc1}
\end{figure}

Now we summarize how the gauge theory concepts we have described may be
applied to the high $T_c$ problem.  The central observation is that high $T_c$
superconductivity emerges upon doping a Mott insulator.  The antiferromagnetic
order of the Mott insulator disappears rather rapidly and is replaced by the
superconducting ground state.  The ``normal'' state above the superconducting
transition temperature exhibits many unusual properties which we refer to as
pseudogap behavior.  How does one describe the simultaneous suppression of
N\'{e}el order and the emergence of the pseudogap and the superconductor from
the Mott insulator? The approach we take is to first understand the nature of
a possible nonmagnetic Mott state at zero doping, the spin liquid state, which
naturally becomes a singlet superconductor when doped.  This is the central
idea behind the RVB proposal \cite{A8796} and is summarized in Fig.
\ref{sldsc1}.  The idea is that doping effectively frustrates the N\'{e}el
order so that the system is pushed across the transition where the N\'{e}el
order is lost.  In the real system the loss of N\'{e}el order may proceed
through complicated states, such as incommensurate charge and spin order,
stripes or inhomogeneous charge segregation 
\cite{CEK03}.
However, in this direct approach the connection with superconductivity is not
al all clear.  Instead it is conceptually useful to arrive at the
superconducting state via a different path, starting from a spin liquid state.
Recently, 
\Ref{SL0466} have elaborated upon this point of view which
we summarize below.

\subsection{Spin liquid, quantum critical point and the pseudogap} It is
instructive to consider the phase diagram as a function of the chemical
potential rather than the hole doping as shown in Fig. \ref{sldsc1}(b). 

\begin{figure}
\includegraphics[width=2.4in]{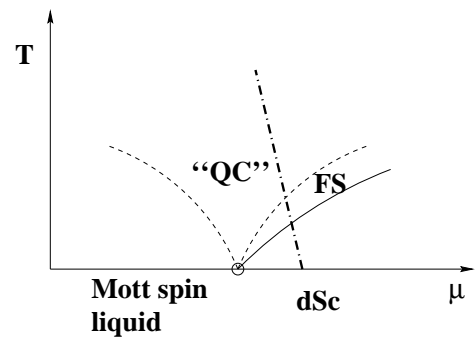}
\caption{Schematic phase diagram for a doping induced Mott transition between
a spin liquid insulator and a $d$-wave superconductor.  The bold dot-dashed
line is the path taken by a system at hole density $x$ that has a
superconducting ground state. The region marked FS represents the fluctuation
regime of the superconducting transition. The region marked QC is the quantum
critical region associated with the Mott critical point. This region may be
identified with the high temperature pseudogap phase in the experiments. }
\label{mqc}
\end{figure}

Consider any spin liquid Mott state that when doped leads to a $d$-wave
superconductor. As a function of chemical potential, there will then be a 
zero temperature phase transition where the holes first enter the system. For
concreteness we will simply refer to this 
as the Mott transition.
The associated quantum critical fixed point will control 
the physics in a finite non-zero range of parameters. The various crossovers
expected near such transitions are well-known and are shown in Fig.
\ref{mqc}.

Sufficiently close to this zero temperature critical point many aspects of the
physics will be universal. The regime in which such universal behavior 
is observed will be limited by `cut-offs' determined by microscopic
parameters. In particular we may expect that the cutoff scale is provided by
an energy of 
a fraction of $J$ (the exchange energy for the spins in the Mott insulator).
We note that this corresponds to a reasonably high temperature scale.

Now consider an underdoped cuprate material at fixed doping $x$. Upon
increasing the temperature this will follow a path in Fig.\ref{mqc} that is
shown schematically. 
The properties of the system along this path may be usefully discussed in
terms of the various crossover regimes.  In particular it is clear that the
`normal' state above the superconducting transition is to be understood
directly as the finite temperature `quantum critical'
region associated with the Mott transition. Empirically this region
corresponds to the pseudogap regime. Thus our assertion is that the pseudogap
regime 
is controlled by the unstable zero temperature fixed point associated with the
(Mott) transition to a Mott insulator.

What are the candidates for the spin liquid phase?  There have been several
proposals in the literature.  One proposal is the dimer phase  
\cite{S0313}.  Strictly speaking, this is a valence bond solid and not a spin liquid:
it is a singlet state which breaks translational symmetry.  It has been shown
by 
\Ref{RS9068} that within the large $N$ Schwinger boson approach
the dimer phase emerges upon disordering the N\'{e}el state.  Sachdev and
collaborators have shown that doping the dimer state produces a $d$-wave
superconductor \cite{VS9916}.  However, such a superconductor also inherits
the dimer order and has a full gap to spin excitations, at least for low
doping.  As we have seen in this review, there are strong empirical evidence
for gapless nodal quasiparticles in the superconducting state.  In our view,
it is more natural to start with translation invariant spin liquid states
which produce $d$-wave superconductors with nodal quasiparticles when doped.

We see from Section X that the spin liquid states are rather exotic beasts in
that their excitations are conveniently described in terms of fractionalized
spin 1/2 ``spinon'' degrees of freedom.  We discussed in Section X.G that spin
liquids are characterized by their low energy gauge group.  Among spin liquids
with nodal fermionic spinons, two versions, the $Z_2$ and the $U(1)$ spin
liquids have bee proposed.  The $Z_2$ gauge theory was advocated by
\cite{SF0050}.  It can be considered as growing out of the fermion pairing
phase of the $U(1)$ mean field phase diagram shown in Fig.~\ref{U1}.  The
pairing of fermions $\Delta_{\v i\v j} = \langle   f_{\v i\up} f_{\v i\down} - f_{\v i\down}
f_{\v i\up}     \rangle$ breaks the $U(1)$ gauge symmetry down to $Z_2$, \ie
only $f \rightarrow -f$ remains unbroken.  One feature of this theory is that
in the superconducting state $hc/e$ vortices tend to have lower energy than
$hc/2e$ vortices, particularly at low doping.  We saw in section IX.C that
$hc/2e$ vortices involve suppression of the pairing amplitude $|\Delta_{\v i\v j}|$
at the center and cost a large energy of order $J$.  On the other hand, one
can form an $hc/e$ vortice by winding the boson phase by $2\pi$, leaving the
fermion pairing intact inside the core.  Another way of describing this from
the point of view of $Z_2$ gauge theory is that the $hc/2e$ vortex necessarily
involves the presence of a $Z_2$ gauge flux (called a vison by Senthil and
Fisher) in its core.  The finite energy cost of the $Z_2$ flux dominates in
the low doping limit and raises the energy of the $hc/2e$ vortices.
Experimental proposals were made 
\cite{SF0192} to provide for a
critical test of such a theory by detecting the vison excitation or by
indirectly looking for signatures of stable $hc/e$ vortices.  To date, all
such experiments have yielded negative results and provided fairly tight
bounds on the vison energy 
\cite{BWG0187}.   

We are then left with the $U(1)$ spin liquid as the final candidate.  The mean
field basis of this state is the staggered flux liquid state of the $SU(2)$
mean field phase diagram (Fig. \ref{su2phase}).  The low energy theory of this
state consists of fermions with massless Dirac spectra (nodal quasiparticles)
interacting with a $U(1)$ gauge field.  Note that this $U(1)$ gauge field
refers to the low energy gauge group and is not to be confused with the $U(1)$
gauge theory in section IX, which refers to the high energy gauge group, in
the nomenclature of section X.G.  This theory was treated in some detail in
Section XI.  This state has enhanced ($\pi,\pi$) spin fluctuations but no long
range N\'{e}el order, and the ground states becomes a $d$-wave superconductor
when doped with holes.  As we shall see, a low energy $hc/2e$ vortex can be
constructed, thus overcoming a key difficulty of the $Z_2$ gauge theory.
Furthermore, an objection in the literature about the stability of the $U(1)$
spin liquid has been overcome, at least for sufficiently large $N$ (see
section IX.F)  It has also been argued by Senthil and Lee, 2004 that even if
the physical spin 1/2 case does not possess a stable $U(1)$ liquid phase, it
can exist as a critical state separating the N\'{e}el phase from a $Z_2$ spin
liquid and may still have the desired property of dominating the physics of
the pseudogap and the superconducting states.
An example of deconfinement appearing at the critical point between two ordered
phases is recently pointed out by \cite{SVB0490}.

In the next section we shall further explore the properties of the $U(1)$ spin
liquid upon doping.  We approach the problem from the low temperature limit
and work our way up in temperature.  This regime is conveniently described by
a nonlinear $\sigma$-model effective theory.

\subsection{$\sigma$-model effective theory and new collective modes in the
superconducting state}

Here we attempt to reduce the large number of degrees of freedom in the
partition function in \Eq{128} to the few which dominate the low energy
physics.  We shall ignore the amplitude fluctuations in the fermionic degree
of freedom which are gapped on the scale of $J$.  The bosons tend to Bose
condense.  We shall ignore the amplitude fluctuation and assume that its phase
is slowly varying on the fermionic scale, which is given by $\xi =
\epsilon_F/\Delta$ in space.  In this case we can have an effective field
theory ($\sigma$-model) description where the local boson phases are the slow
variables and the fermionic degrees of freedom are assumed to follow them.  We
begin by picking a mean field representation $U_{\v i\v j}^{(0)}$.  The choice of
the staggered flux state $U_{\v i\v j}^{SF}$ given by \Eq{EfEb} is most convenient
because $U_{\v i\v j}^{SF}$ commutes with $\tau^3$, making explicit the residual
$U(1)$ gauge symmetry which corresponds to a $\tau^3$ rotation.  Thus we
choose $ U_{\v i\v j}^{(0)} = U_{\v i\v j}^{SF} e^{ia_{\v i\v j}^3\tau^3} $ and replace the
integral over $U_{\v i\v j}$ by an integral over the gauge field $a_{\v i\v j}^3$.  It
should be noted that any $U_{\v i\v j}^{(0)}$ which are related by $SU(2)$ gauge
transformation will give the same result.  At the mean field level, the bosons
form a band with minima at $Q_0$.  Writing $h = \tilde{h} e^{i\v{Q}_0\cdot
\v{r}}$, we expect $\tilde{h}$ to be slowly  varying in space and time.  We
transform to the radial gauge, 
\ie we write
\be
\tilde{h}_{\v i} = g_{\v i}
\bpm b_{\v i} \\
0 
\epm 
\; , 
\label{eq.154}
\en
where $b_{\v i}$ can be taken as real and positive and $g_{\v i}$ is an $SU(2)$ matrix
parametrized by
\be
g_{\v i}=
\bpm z_{\v i1} & -z_{\v i2}^\ast \\
z_{\v i2} & z_{\v i1}^\ast 
\epm
\; 
\label{eq.155}
\en
where
\be
z_{\v i1}  = e^{i\alpha_{\v i}} e^{-i{\phi_{\v i}\over 2}} \cos {\theta_{\v i} \over 2}
\label{eq.156}
\en
and
\be
z_{\v i2}  = e^{i\alpha_{\v i}} e^{i{\phi_{\v i}\over 2}} \sin {\theta_{\v i} \over 2} \,\,\, .
\label{eq.157}
\en
We ignore the boson amplitude fluctuation and replace $b_{\v i}$ by a constant
$b_0$.

An important feature of eq. (\ref{128}) is that $L_2$ is invariant under the 
$SU(2)$ gauge transformation
\be
\tilde{h}_{\v i} &=& g_{\v i}^\dagger h_{\v i}   \label{eq.158} \\ 
\tilde{\psi}_{\v i} &=& g_{\v i}^\dagger \psi_{\v i}  \label{eq.159}\\
\tilde{U}_{\v i\v j} &=& g_{\v i}^\dagger U_{\v i\v j}^{(0)} g_{\v j}
\label{eq.160}
\en
and
\be
\tilde{a}_{0\v i}^\ell \tau^\ell = g^\dagger a_{0\v i}^\ell \tau^\ell g - g \left(
\partial _\tau g^\dagger \right) \,\,\, .
\label{eq.161}
\en
Starting from eq. (\ref{128}) and making the above gauge transformation, the
partition function is
integrated over $g_{\v i}$ instead of $h_{\v i}$ and the Lagrangian
takes the form
\begin{align}
L_2^\prime &= {\tilde{J} \over 2} \sum_{<\v i\v j>} \Tr \left(
\tilde{U}_{\v i\v j}^\dagger \tilde{U}_{\v i\v j} \right) +
\tilde{J} \sum_{<\v i\v j>} \psi_{\v i}^\dagger
\tilde{U}_{\v i\v j}\psi_{\v j} + c.c. \nonumber \\
&+ \sum_{\v i} \psi_{\v i}^\dagger  \left( \partial_\tau - ia_{0\v i}^\ell \tau^\ell  \right)
 \psi_{\v i}\nonumber + \sum_{\v i}  
 \left(- ia_{0\v i}^3 + \mu_B \right)
 b_0^2 \nonumber \\
  &- \sum_{\v i\v j,\sigma} \tilde{t}_{\v i\v j} b_0^2 f_{\v j\sigma}^\dagger f_{\v i\sigma}
\label{eq.162} 
\end{align}
We have removed the tilde from $\tilde{\psi}_{\v i\sigma}$,
$\tilde{f}_{\v i\sigma}$, $\tilde{a}_0^\ell$ because these are integration
variables and $\tilde{t}_{\v i\v j} = t_{\v i\v j}/2$.  
Note that $g_{\v i}$ appears only in $\tilde{U}_{\v i\v j}$.  For every configuration
$\{g_{\v i}(\tau),  a^3_{\v i\v j}(\tau)\} $ we can, in principle, integrate out the
fermions and $a_0^\ell$ to obtain an energy functional.  This will constitute
the $\sigma$-model description.  In practice, we can make the slowly varying
$g_{\v i}$ approximation and solve the local mean field equation for
$a_{0\v i}^\ell$.  This is the approach taken by 
\Ref{LNNWsu2}.
Note that since $\{g_{\v i}\}$ appears only in the fermionic Lagrangian via
$\tilde{U}_{\v i\v j}$ in eq. (\ref{eq.162}), the resulting energy functional is
entirely fermionic in origin, and no longer has any bosonic contribution.

\begin{figure}
\includegraphics[width=2.0in]{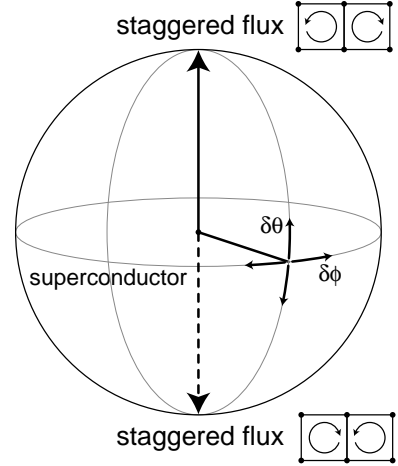}
\caption{The quantization axis $\v{I}$ in the $SU(2)$ gauge theory.  The north
and south poles correspond to the staggered flux phases with shifted orbital
current patterns.  All points on the equators are equivalent and correspond to
the $d$-wave superconductor.  In the superconducting state one particular
direction is chosen on the equator.  There are two important collective modes.
The $\theta$ modes correspond to fluctuations in the polar angle
$\delta\theta$ and the $\phi$ gauge mode to a spatially varying fluctuation in
$\delta\phi$.}
 \label{fig.26}
\end{figure}

The $\si$-model depends on $\{ g_{\v i}(\tau), a^3_{\v i\v j}(\tau) \}$,
\ie it is characterized by $\al_{\v i}$, $\th_{\v i}$, $\phi_{\v i}$ and the gauge field $a_{\v i\v
j}^3$. $\al_{\v i}$ is the familiar overall phase of the electron operator which
becomes half of the pairing phase in the superconducting state. To help
visualize the remaining dependence of freedom, 
it is useful to introduce the local quantization axis
\be \v{I}_{\v i} = z_{\v i}^\dagger \v{\tau} z_{\v i} = ( \sin \theta_{\v i} \cos \phi_{\v i}, \sin
\theta_{\v i} \sin \phi_{\v i}, \cos \theta_{\v i}) \label{eq.163} 
\en 
Note that $\v{I}_{\v i}$ is
independent of the overall phase $\alpha_{\v i}$, which is the phase of the
physical electron operator.  Then different orientations of $\v{I}$ represent
different mean field states in the $U(1)$ mean field theory.  This is shown in
Fig. \ref{fig.26}.  For example, $\v{I}$ pointing to the north pole
corresponds to $g_{\v i} = I$ and the staggered flux state.  This state has $a_0^3
\neq 0$, $a_0^1 = a_0^2 = 0$ and has small Fermi pockets.  It also has orbital
staggered currents around the plaquettes.  $\v{I}$ pointing to the south pole
corresponds to the degenerate staggered flux state whose staggered pattern is
shifted by one unit cell.  On the other hand, when $\v{I}$ is in the equator,
it corresponds to a $d$-wave superconductor.  Note that the angle $\phi$ is a
gauge degree of freedom and states with different $\phi$ anywhere along the
equator are gauge equivalent.  A general orientation of $\v{I}$ corresponds to
some combination of $d$-SC and $s$-flux.

At zero doping, all orientations of $\v{I}$ are energetically the same.  This
symmetry is broken by doping, and the $\v{I}$ vector has a small preference to
lie on the equator.  At low temperature, there is a phase transition to a
state where $\v{I}$ lies on the equator, \ie the $d$-SC ground state.  It is
possible to carry out a small expansion about this state and work out
explicitly the collective modes 
\cite{LN0316}.  In an ordinary
superconductor, there is a single complex order parameter $\Delta$ and we
expect an amplitude mode and a phase mode.  For a charged superconductor the
phase mode is pushed up to the plasma frequency and one is left with the
amplitude mode only.  In the gauge theory we have in addition to $\Delta_{\v i\v j}$
the order parameter $\chi_{\v i\v j}$.  Thus it is natural to expect additional
collective modes.  From Fig. \ref{fig.26} we see that two modes are of special
interest corresponding to small $\theta$ and $\phi$ fluctuations.  Physically
the $\theta$ mode corresponds to local fluctuations of the $s$-flux states
which generate local orbital current fluctuations.  These currents generate a
small magnetic field (estimated to be $\sim10$ gauss) which couples to
neutrons. 
\Ref{LN0316} predict a peak in the neutron scattering
cross-section at $(\pi,\pi)$, at energy just below $2\Delta_0$, where
$\Delta_0$ is the maximum $d$-wave gap.  This is {\em in addition} to the
resonance mode discussed in section III.B which is purely spin fluctuation in
origin.  The orbital origin of this mode can be distinguished from the spin
fluctuation by its distinct form factor 
(\cit{HMA9166}; \cit{CKN0240})

The $\phi$ mode is more subtle because 
$\phi$ is the phase of a Higgs field, \ie it is part of
the gauge degree of freedom.  It
turns out to correspond to a relative oscillation of the {\em amplitudes} of
$\chi_{\v i\v j}$ and $\Delta_{\v i\v j}$ and is again most prominent at $(\pi,\pi)$.
Since $|\chi_{\v i\v j}|$ couples to the bond density fluctuation, inelastic Raman
scattering is the tool of choice to study this mode, once the technology
reaches the requisite 10~meV energy resolution.  
\Ref{LN0316} point
out that due to the special nature of the buckled layers in LSCO, this mode
couples to photons and may show up as a transfer of spectral weight from a
buckling phonon to a higher frequency peak.  Such a peak was reported
experimentally \cite{KTM0304}, but it is apparently not unique to LSCO as the
theory would predict, and hence its interpretation remains unclear at this
point.

From Fig. \ref{fig.26} it is clear that the $\sigma$-model representation of
the $SU(2)$ gauge theory is a useful way of parameterizing the myriad $U(1)$
mean field states which become almost degenerate for small doping.  The low
temperature $d-SC$ phase is the ordered phase of the $\sigma$-model, while in
the high temperature limit we expect the $\v{I}$ vector to be disordered in
space and time, to the point where the $\sigma$-mode approach fails and one
crosses over to the $SU(2)$ mean field description.  The disordered phase of
the $\sigma$-model then corresponds to the pseudogap phase.  How does this
phase transition take place?  It turns out that the destruction of
superconducting order proceeds via the usual route of BKT proliferation of
vortices.  To see how this comes about in the $\sigma$-model description, we
have to first understand the structure of vortices. 

\subsection{Vortex structure} 

\begin{figure}
\includegraphics[width=2.5in]{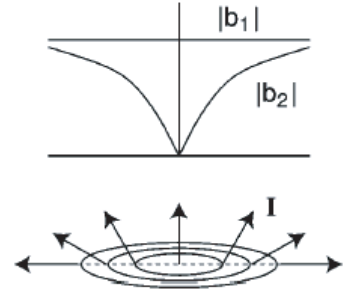}
\caption{Structure of the superconducting vortex.  Top: $b_1$ is constant
while $b_2$ vanishes at the center and its phase winds by $2\pi$.  Bottom: The
isospin quantization axis points to the north pole at the center and rotates
towards the equatorial plane as one moves out radially.  The pattern is
rotationally symmetric around the $\hat{z}$ axis.}
 \label{vortex}
\end{figure}

The $\sigma$-model picture leads to a natural
model for a low energy $hc/2e$ vortex 
\Ref{LW0117}.  It takes advantage
of the existence of two kinds of bosons $b_1$ and $b_2$ with opposite gauge
charges but the same coupling to electromagnetic fields.  Far away from the
vortex core, $|b_1| = |b_2|$ and $b_1$ has constant phase while $b_2$ winds
its phase by $2\pi$ around the vortex.  As the core is approached $|b_2)$ must
vanish in order to avoid a divergent kinetic energy, as shown in Fig.
\ref{vortex}(top).
The quantization axis $\v{I}$ provides a nice way to visualize this structure
[Fig. \ref{vortex}(bottom)].  It smoothly rotates to the north pole at the vortex core,
indicating that at this level of approximation, the core consists of the
staggered flux state.  The azimuthal angle winds by $2\pi$ as we go around the
vortex. It is important to remember that $\v{I}$ parameterizes  only the
internal gauge degrees of freedom $\theta$ and $\phi$ and the winding of
$\phi$ by $2\pi$ is different from the usual winding of the overall phase
$\alpha$ by $\pi$ in an $hc/2e$ vortex.  To better understand the phase
winding we write down the following continuum model for the phase $\theta_1,
\theta_2$ of $b_1$ and $b_2$, valid far away from the core.
\be
D = \int d^2x {K\over 2} 
\left[ \rule{0in}{.15in}
\left( \v{\nabla} \theta_1 -\v{a} - \v{A} \right) \right.
&+&
\left.
\left(
\nabla \theta_2 + \v{a} - \v{A}
\right)^2
\right] \nonumber \\
&& + \cdots
\label{eq.164}
\en
where $\v{a}$ stands for the continuum version of $a^3_{\v i\v j}$ in the last
section, and $\v{A}$ is the electromagnetic field ($e/c$ has been set to be
unity).  We now see that the $hc/2e$ vortex must contain a half integer vortex
of the $\v{a}$ gauge flux with an opposite sign.  Then $\theta_1$ sees zero
flux while $\theta_2$ sees $2\pi$ flux, consistent with the windings chosen in
Fig.\ref{vortex}.  This vortex structure has low energy for small $x$ because
the fermion degrees of freedom remain gapped in the core and one does not pay
the fermionic energy of order $J$ as in the $U(1)$ gauge theory.  Physically,
the above description takes advantage of the states with almost degenerate
energies (in this case the staggered flux state) which is guaranteed by the
$SU(2)$ symmetry near half filling.  There is direct evidence from STM
tunneling that the energy gap is preserved in the core 
\cite{MRE9554,PHG0036}.  This is in contrast to theoretical
expectations for conventional $d$-wave vortex cores, where a large resonance
is expected to fill in the gap in the tunneling spectra 
\cite{WM9576}.

We can clearly reverse the roles of $b_1$ and $b_2$ to produce another vortex
configuration which is degenerate in energy.  In this case $\v{I}$ in Fig.
\ref{vortex}
points to the south pole.  These configurations are sometimes referred to
merons (half of a hedgehog) and the two halves can tunnel to each other via
the appearance of instantons in space-time.  The time scale of the tunneling
event is difficult to estimate, but should be considerably less than $J$.
Depending on the time scale, the orbital current of the staggered flux state
in the core generates a physical staggered magnetic field which may be
experimentally observable by NMR (almost static), $\mu$SR (intermediate time
scale) and neutron (short time scale).  The experiment must be performed in a
large magnetic field so that a significant fraction of the area consists of
vortices and the signal of the staggered field should be proportional to $H$.
A $\mu$SR experiment on underdoped YBCO has detected such a field dependent
signal with a local field of $\pm$18 gauss 
\cite{MKB0202}.
However $\mu$SR is not able to determine whether the field has an orbital or
spin origin and this experiment is only suggestive, but by no means
definitive, proof of orbital currents in the vortex core.  In principle,
neutron scattering is a more definitive probe, because one can use the form
factor to distinguish between orbital and spin effects.  However, due to the
small expected intensity, neutron scattering has so far not yielded any
definite results.

As discussed in section XI.E, we expect enhanced $(\pi, \pi)$ fluctuations to
be associated with the staggered flux liquid phase.  Indeed, the $s$-flux
liquid state is our route to N\'{e}el order and if gauge fluctuations are
large, we may expect to have quasi-static N\'{e}el order inside the vortex
core.  Experimentally, there are reports of enhanced spin fluctuations in the
vortex core by NMR experiments 
(\cit{CMH0073}; \cit{MSB0105}; \cit{MSH0303}; \cit{KKM0203}).  There
are also reports of static incommensurate spin order forming a halo around the
vortex in the LSCO family 
(\cit{KYS0077}; \cit{LAC0159}; \cit{LRC0299}; \cit{KLE0228}.  One possibility is that these halos are
the condensation of pre-existing soft incommensurate modes known to exist in
LSCO, driven by quasi-static N\'{e}el order inside the core.  We emphasize the
$s$-flux liquid state is our way of producing antiferromagnetic order starting
from microscopies and hence is fully consistent with the appearance of static
or dynamical antiferromagnetism in the vortex core.  Our hope is that gauge
fluctuations (including instanton effects) are sufficiently reduced in doped
systems to permit a glimpse of the staggered orbital current.  The detection
of such currently fluctuations will be a strong confirmation of our approach.

Finally, we note that orbital current does not show up directly in STM
experiments, which are sensitive to the local density of states.  However,
\Ref{KLW0226} have considered the possibility of interference
between Wannier orbitals on neighboring lattice sites, which could lead to
modulations of STM signals {\em between} lattice positions.  STM experiments
have detected $4 \times 4$ modulated patterns in the vortex core region and
also in certain underdoped regions.  Such patterns appear to require density
modulations which are in addition to our vortex model.

\subsection{Phase diagram} 

\begin{figure}
\includegraphics[width=2.5in]{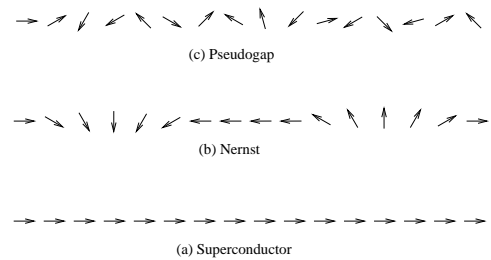}
\caption{
Schematic picture of the quantization axis $\v{I}$ in different parts of the
phase diagram shown in Fig.~\ref{Nernst}. (a)~In the superconducting phase
$\v{I}$ is ordered in the $x$-$y$ plane.  (b)~In the Nernst phase, $\v{I}$
points to the north or south pole inside the vortex core. (c)~The pseudogap
corresponds to a completely disordered arrangement of $\v{I}$. ($\v{I}$ is a
three dimensional vector and only a two dimensional projection is shown.)
}
\label{isospin}
\end{figure}

We can now construct a phase diagram of the
underdoped cuprates starting from the $d$-wave superconductor ground state at
low temperatures.  The vortex structure allows us to unify the $\sigma$-model
picture with the conventional picture of the destruction of superconducting
order in two dimensions, ie., the BKT transition via the unbinding of
vortices.  The $\sigma$-model 
contains in addition to the pairing phase $2\al$, the phases
$\th$ and $\phi$. However, we saw in section XII.C that a particular
configuration of $\th$ and $\phi$ is favored in side the vortex core.
The $SU(2)$ gauge
theory provides a  mechanism for cheap vortices which are necessary for a BKT
description, as discussed in section V.B.  If the core energy is too large,
the system will behave like a superconductor on any reasonable length scale
above  $T_{\rm BKT}$, which is not in accord with experiment.  On the other
hand, if the core energy is small compared with $T_c$, vortices will
proliferate rapidly. They overlap and lose their identity.  As discussed
section V.B, there is strong experimental evidence that vortices survive over
a considerable temperature range above $T_c$.  Taken as a whole, these
experiments require the vortex core energy to be cheap, but not too cheap,
\ie of the order of $T_c$.  
\Ref{HL0401} have attempted a
microscopic modeling of the proliferation of vortices.  They assume an
$s$-flux core and estimate the energy from projected wavefunction
calculations.  They indeed found that there is a large range of temperature
above the BKT transition where vortices grow in number but still maintain
their identity.  This forms a region in the phase diagram which may be called
the Nernst region shown in Fig.~\ref{Nernst}.  The corresponding picture of
the $\v{I}$ vector fluctuation is shown in Fig.~\ref{isospin}.  Above the
Nernst region the $\v{I}$ vector is strongly fluctuating and is almost
isotopic.  This is the strongly disordered phase of the $\sigma$-model. The
vortices have lost their identity and indeed the $\sigma$-model description
which assumes well defined phases of $b_1$ and $b_2$ begin to break down.
Nevertheless, the energy gap associated with the fermions remains.  This is
the pseudogap part of the phase diagram in Fig.~\ref{Nernst}.  In the $SU(2)$
gauge theory this is understood as the $U(1)$ spin liquid.  There is no order
parameter in the usual sense associated with this phase, as all fluctuations
including staggered orbital currents and $d$-wave pairing become short range.
Is there a way to characterize this state of affairs other than the term spin
liquid?  This question is addressed in the next section.

\subsection{Signature of the spin liquid}

\Ref{SL0466}
pointed out that if the pseudogap region is controlled
by the $U(1)$ spin liquid fixed point, it is possible to characterize this
region in a certain precise way.  The spin liquid is a de-confined state,
meaning that instantons are irrelevant.  Then the $U(1)$ gauge flux is a
conserved quantity.  Unfortunately, it is not clear how to couple to this
gauge flux using conventional probes.  We note that the flux associated with
the $\v{a}^3$ gauge field is {\em different} from the $U(1)$ gauge flux
considered in section IX, which had the meaning of spin chirality.  In the
case where the bosons are locally condensed and their local phase well
defined, it is possible to identify the gauge flux in terms of the local 
phase variables.
The gauge magnetic field $\cal{B}$ is given by
\be
{\cal B} & =& (\v \nabla \times \v{a}^3)_z \nonumber \\
& = & \frac{1}{2} \hat{n} \cdot \partial_x \hat{n} \times \partial_y \hat{n}
\label{eq.165}
\en
where
\be
\hat{n} = (\sin \theta \cos \alpha, \sin \theta \sin \alpha, \cos \theta) \nonumber \;\; .
\en
with $\th$ and $\al$ defined in \Eq{eq.156}.
Note that the azimuthal angle associated with $\hat{n}$ is now the pairing
phase $\alpha$, in contrast with the vector $\v{I}$ we considered earlier.
The gauge flux is thus related to the local pairing and $s$-flux order as
\be
{\cal B} = \frac{1}{2} 
\left( \v \nabla  \hat{n}_z  \times \v \nabla \alpha \right)_z
\label{eq.166}
\en
and it is easily checked that the vortex structure described in section XII.C
contains a half integer gauge flux.

In the superconducting state the gauge flux is localized in the vortex core
and fluctuations between $\pm$ half integer vortices are possible via
instantons, because the instanton action is finite.  The superconductor is in
a confined phase as far as the $U(1)$ gauge field is concerned.  As the
temperature is raised towards the pseudogap phase this gauge field leaks out
of the vortex cores and begins to fluctuate more and more homogeneously.


The asymptotic conservation of the gauge flux at the Mott transition fixed
point
potentially provides some possibilities for its detection. At non-zero
temperatures in the non-superconducting regions, the flux 
conservation is only approximate (as the instanton fugacity is small but
non-zero). Nevertheless at low enough temperature the conserved flux will 
propagate diffusively over a long range of length and time scales. Thus there
should be an extra diffusive mode that is present at low temperatures in the 
non-superconducting state. It is however not clear how to design a probe that
will couple to this diffusive mode at present.

Alternately the vortex structure described above provides a useful way to
create and then detect the gauge flux in the non-superconducting
normal state. We will first describe this by ignoring the
instantons completely in the normal state. The effects of
instantons will then be discussed.

\begin{figure}[tb]
\centering
\includegraphics[width=2.0in]{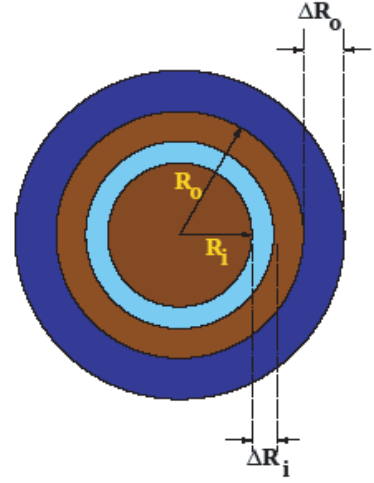}
\caption{Structure of the sample needed for the proposed experiment. The outer annulus (in dark blue) has the highest $T_c$. The inner annulus 
(in light blue) has a smaller $T_c$. The rest of the sample (in brown) has even smaller $T_c$. 
 } 
\label{u1ft1}
\end{figure}

Consider first a large disc of
cuprate material which is such that the doping level changes as a function of
the radial distance from the center as shown in Fig. \ref{u1ft1}.
The outermost annulus has the largest doping $x_1$. The inner annulus has a
lower doping level $x_2$. The rest of the sample is at a doping level
$x_3< x_2 < x_1$. The corresponding transition temperatures $T_{c1,2,3}$
will be such that $T_{c3} < T_{c2} < T_{c1}$. We also imagine that
the thickness $\Delta R_o, \Delta R_i$ of the outer and inner  annuli
are both much smaller than the penetration
depth for the physical vector potential $A$. The penetration depth
of the internal gauge field $a$ is expected to be small and we
expect it will be smaller than $\Delta R_o, \Delta R_i$.  
We also imagine that the radius of this inner annulus $R_i$ is a substantial
fraction of the radius $R_o$
of the outer annulus.

 Now consider the following set of operations on such a sample.

(i) First cool in a magnetic field to a temperature $T_{in}$ such that
$T_{c2} < T_{in} < T_{c1}$. The outer ring will then go superconducting while
the rest of the sample stays normal.
In the presence of the field the outer ring will condense into a state in
which there is a net vorticity on going around the ring.
We will be interested in the case where this net vorticity is an odd multiple
of the basic $hc/2e$ vortex. If as assumed the
physical penetration depth is much bigger than the thickness $\Delta R_o$ then
the physical magnetic flux enclosed by the ring will not be quantized.

(ii) Now consider turning off
the external magnetic field. The vortex present in the outer superconducting
ring will stay (manifested as a small circulating persistent current)
and will give rise to a small magnetic field. As explained above if the
vorticity is odd, then it must be associated with a flux of the
internal gauge field that is $\pm \pi$. This internal gauge flux must
essentially all be in the inner `normal' region of the sample with
very small penetration into the outer superconducting ring. It will spread out
essentially evenly over the full inner region.

We have thus managed to create a configuration with a non-zero  internal gauge
flux
in the non-superconducting state.

(iii) How do we detect the presence of this internal gauge flux? For that
imagine now cooling the sample further to a temperature
$T_{fin}$ such that $T_{c3} < T_{fin} < T_{c2}$. Then the inner ring will also
go superconducting. This is to be understood as the condensation
of the two boson species $b_{1,2}$. But this condensation occurs in the
presence of some internal gauge flux. When the bosons $b_{1,2}$ condense in
the inner ring,
they will do so in a manner that quantizes the internal gauge flux enclosed by
this inner ring into an integer multiple of $\pi$. If as
assumed the inner radius is a substantial fraction of the outer radius then
the net internal gauge flux will prefer the quantized values $\pm \pi$ rather
than
be zero (see below). However configurations of the inner ring that enclose
quantized internal gauge flux of $\pm \pi$ also necessarily
contain a physical vortex that is an odd multiple of $hc/2e$. With the
thickness of the inner ring being smaller than the physical penetration depth, 
most of the physical magnetic flux will escape. There will still be a small
residual physical flux due to the current in the inner ring associated with
the 
induced vortex. This residual physical magnetic flux can then be detected.

Note that the sign of the induced physical flux is independent of the sign of
the initial magnetic field. Furthermore
the effect obtains only if the initial vorticity in the outer ring is odd. If
on the other hand the initial vorticity is
even the associated internal gauge flux is zero, and there will be no induced
physical flux when the inner ring goes superconducting.

The preceding discussion ignores any effects of instantons. In
contrast to a bulk vortex in the superconducting state the
vortices in the set-up above have macroscopic cores. The internal
gauge flux is therefore distributed over a region of macroscopic
size. Consequently if instantons are irrelevant at long scales in
the normal state, their rate may be expected to be small. At any
non-zero temperature (as in the proposed experiment) there will be
a non-zero instanton rate which will be small for small
temperature.

When such instantons are allowed then the internal gauge flux
created in the sample after step (ii) will fluctuate between the
values $+\pi$ and $-\pi$. However so long as the time required to
form the physical vortex in step (iii), which we expect to be short electronic
time scale, is much shorter than the
inverse of the instanton rate we expect that the effect will be seen.
Since the cooling is assumed slow enough that the system always stays in
equilibrium, the outcome of the experiment is determined by thermodynamic
considerations.  
\Ref{SL0466} estimated the energies of the various
stages of the operation and concluded that for sample diameters under a micron
and sufficiently low temperatures (= 10 K), such an experiment may be
feasible.

\section{Summary and outlook}

In this review we have summarized a large body of work which views high
temperature superconductivity as the problem of doping of a Mott insulator.
We have argued that the $t$-$J$ model, supplemented by $t^\prime$ terms,
contains the essence of the physics.  We offer as evidence numerical work
based on the projected trial wavefunctions, which correctly predicts the
$d$-wave pairing ground state and a host of properties such as the superfluid
density and the quasiparticle spectral weight and dispersion.  Analytic theory
hinges on the treatment of the constraint of no double occupation.  The
redundancy in the representations used to enforce the constraint naturally
leads to various gauge theories.  We argue that with doping, the gauge theory
may be in a deconfined phase, in which case the slave-boson and fermion
degrees of freedom, which were introduced as mathematical devices, take on a
physical meaning in that they are sensible starting points to describe
physical phenomena.  However, even in the deconfined phase, the coupling to
gauge fluctuations is still of order unity and approximation schemes (such as
large $N$ expansion) are needed to calculate physical properties such as spin
correlation and electron spectral function.  These results qualitatively
capture the physics of the pseudogap phase, but certainly not at a
quantitative level.  Nevertheless, our picture of the vortex structure and how
they proliferate gives us a reasonable account of the phase diagram and the
onset of $T_c$.

One direction of future research is to refine the treatment of the low energy
effective model, \ie fermions and bosons coupled to gauge fields, and
attempt more detailed comparison with experiments such as photoemission
lineshapes, \etc.  On the other hand, it is worthwhile to step back and take a
broader perspective.  What is really new and striking about the high
temperature superconductors is the strange ``normal'' metallic state for
underdoped samples.  The carrier density is small and the Fermi surface is
broken up by the appearance of a pseudogap near $(0,\pi)$ and $(\pi, 0)$,
leaving a ``Fermi arc'' near the nodal points.  All this happens without
doubling of the unit cell via breaking translation or spin rotation symmetry.
How this state comes into being in a lightly doped Mott insulator is the crux
of the problem.  We can distinguish between two classes of answers.  The
first, perhaps the more conventional one, postulates the existence of a 
symmetry-breaking state
which gaps the Fermi surface, and further assumes that thermal fluctuation
prevents this state from ordering.  A natural candidate for the state is the
superconducting state itself.  However, it now appears that phase fluctuations
of a superconductor can explain the pseudogap phenomenon only over a
relatively narrow temperature range, which we called the Nernst regime.
Alternatively, a variety of competing states which have nothing to do with
superconductivity have been proposed, often on a phenomenological level, to
produce the pseudogap.  We shall refer to this class of theory as ``thermal''
explanation of the pseudogap.

A second class of answer, which we may dub the ``quantum'' explanation,
proposes that the pseudogap is connected with a fundamentally new quantum
state.  Thus, despite its appearance at high temperatures, it is argued that
it is a high frequency phenomenon which is best understood quantum
mechanically.  The gauge theory reviewed here belongs to this class, and views
the pseudogap state as derived from a new state of matter, the quantum spin
liquid state.  The spin liquid state is connected to the N\'{e}el state at
half filling by confinement.  At the same time, with doping a $d$-wave
superconducting ground state is naturally produced.  We argue that rather than
following the route taken by the cuprate in the laboratory of evolving
directly from the antiferromagnet to the superconductor, it is better
conceptually to start from the spin liquid state and consider how AF and
superconductivity develop from it.  In this view the pseudogap is the
closest we can get to obtaining a glimpse of the spin liquid which up to now
is unstable in the square lattice $t$-$J$ model. 

Is there a ``smoking gun'' signature to prove or disprove the validity of this
line of theory?  Our approach is to make specific predictions as much as
possible in the hope of stimulating experimental work.  This is the reason we
make special emphasis on the staggered flux liquid with its orbital current
fluctuations, because it is a unique signature which may be experimentally
detectable.  Our predictions range from new collective modes in the
superconducting state, to quasi-static order in the vortex core.
Unfortunately the physical manifestation of the orbital current is a very weak
magnetic field, which is difficult to detect, and to date we have not found
experimental verification.  Besides orbital current, we also propose an
experiment involving flux generation in a special geometry. This experiment
addresses the fundamental issue of the quantum spin liquid as the origin of
the pseudogap phase.

The pseudogap metallic state is so strange that at the beginning, it is not
clear if a microscopic description is even possible. So the microscopic
description provided by the $SU(2)$ slave-boson theory, although still
relatively qualitative, represents  important progress and leads to some deep
insights.  A key finding is that the parent spin liquid is a new state of
matter that cannot be described by Landau's symmetry breaking theory.  The
description of the parent spin liquid, such as the $SU(2)$ slave-boson theory,
must involve gauge theory.  Even if one starts with an ordered phase and later
uses quantum fluctuations to restore the symmetry, the resulting description
of the symmetry restored  state, if found, appears to always contain gauge
fields \cite{WMW9846}.  Thus the appearance of the gauge field in the quantum
description of the  pseudogap metal is not a mathematical artifact of the
slave-boson theory. It is a consequence of a new type of correlations in those
states. The new type of correlations represents a new type of order
\cite{Wqoslpub}, which make those states different from the familiar states
described by Landau's symmetry breaking theory.

From this perspective, the study of high temperature superconductivity may
have a much broader and deeper impact than merely understanding high
temperature superconductivity. Such a study is actually a study of new states
of matter. It represents our entry into a new exciting world that lies beyond
Landau's world of symmetry breaking.  Hopefully the new states of matter may
be discovered in some materials other than high temperature superconductors.
The slave-boson theory and the resulting gauge theory developed for high
temperature superconductivity may be useful for these new states of matter
once they are discovered in experiments. [Examples of these new states of
matter have already been discovered in many theoretical toy models
(\cit{K032}; \cit{MS0181}; \cit{BFG0212}; \cit{Wqoexct}).] At the moment,
gauge theory is the only known language to describe this new state of
affairs.  We believe the introduction of this subject to condensed matter
physics has enriched the field and will lead to may interesting further
developments.

\begin{acknowledgments}
P.A.L. acknowledges support by NSF grant number
DMR-0201069.
X.G.W. acknowledges support by NSF Grant No. DMR--01--23156,
NSF-MRSEC Grant No. DMR--02--13282, and NFSC no. 10228408.
\end{acknowledgments}


\end{document}